%% file: mnras_template.tex
\documentclass[fleqn,usenatbib]{mnras}

\usepackage{newtxtext,newtxmath}

\usepackage[T1]{fontenc}

\DeclareRobustCommand{\VAN}[3]{#2}
\let\VANthebibliography\thebibliography
\def\thebibliography{\DeclareRobustCommand{\VAN}[3]{##3}\VANthebibliography}

\usepackage{graphicx}	
\usepackage{amsmath}	
\usepackage{caption}
\usepackage{subcaption}
\usepackage{xcolor}
\usepackage[dvipsnames]{xcolor}
\usepackage{nicefrac}
\usepackage[none]{hyphenat}
\usepackage{placeins}
\usepackage[nobiblatex]{xurl}
\usepackage{orcidlink}
\usepackage{subfiles}

\newcommand{\gaia}{\textit{Gaia}}
\newcommand{\ngts}{NGTS}
\newcommand{\tess}{\textit{TESS}}
\newcommand{\astep}{ASTEP}
\newcommand{\coralie}{CORALIE}
\newcommand{\harps}{HARPS}
\newcommand{\soar}{SOAR}
\newcommand{\gemini}{Gemini}

\newcommand{\allesfitter}{\textsc{Allesfitter}}
\newcommand{\ariadne}{\textsc{Ariadne}}
\newcommand{\ldtk}{\textsc{ldtk}}
\newcommand{\lightkurve}{\textsc{Lightkurve}}
\newcommand{\specmatch}{\textsc{SpecMatch-Emp}}
\newcommand{\pysynphot}{\textsc{PySynphot}}

\newcommand{\objonengts}{NGTS-34}
\newcommand{\objonetic}{TIC-147277741}
\newcommand{\objtwo}{TOI-4940}
\newcommand{\objthreetoi}{TOI-6669}
\newcommand{\objthreengts}{NGTS-35}

\newcommand{\objoner}{3.65\pm0.22}
\newcommand{\objonem}{19.1_{-4.5}^{+4.9}}
\newcommand{\objonep}{43.12655_{-0.00017}^{+0.00012}}

\newcommand{\objtwor}{6.61\pm0.37}
\newcommand{\objtwom}{<89}
\newcommand{\objtwomwhole}{41\pm16}
\newcommand{\objtwop}{25.867811_{-0.000056}^{+0.000058}}

\newcommand{\objthreer}{10.90\pm0.65}
\newcommand{\objthreem}{152_{-19}^{+22}}
\newcommand{\objthreep}{25.241192\pm0.000022}
\newcommand{\objthreee}{0.192_{-0.033}^{+0.037}}
\newcommand{\objthreet}{451_{-10}^{+11}}
\newcommand{\objthreetround}{450}

\title{A 43 day transiting Neptune and two 25 day Saturns from TESS, NGTS and ASTEP}

\author[A. Kendall et al.]{
Alicia Kendall $^{\orcidlink{0009-0006-0719-9229}}$,$^{1}$\thanks{E-mail: ak842@leicester.ac.uk}
Solène Ulmer-Moll $^{\orcidlink{0000-0003-2417-7006}}$,$^{2,3,4}$
Samuel Gill $^{\orcidlink{0000-0002-4259-0155}}$,$^{5,6}$
Matthew R. Burleigh $^{\orcidlink{0000-0003-0684-7803}}$,$^{1}$
\newauthor
Michael R. Goad $^{\orcidlink{0000-0002-2908-7360}}$,$^{1}$
David R. Anderson $^{\orcidlink{0000-0001-7416-7522}}$,$^{7}$
Edward M. Bryant $^{\orcidlink{0000-0001-7904-4441}}$,$^{5,8}$
Baptiste Lavie $^{\orcidlink{0000-0001-8884-9276}}$,$^{3}$
\newauthor
Maddalena Bugatti,$^{3}$
Javier A. Acevedo Barroso $^{\orcidlink{0000-0002-9654-1711}}$,$^{9}$
Michal Steiner $^{\orcidlink{0000-0003-3036-3585}}$,$^{3}$
Diana Dragomir $^{\orcidlink{0000-0003-2313-467X}}$,$^{10}$
\newauthor
Steven Villanueva Jr. $^{\orcidlink{0000-0001-6213-8804}}$,$^{11}$
Daniel J. Stevens $^{\orcidlink{0000-0002-5951-8328}}$,$^{12}$
Arvind F. Gupta $^{\orcidlink{0000-0002-5463-9980}}$,$^{13}$
Scott Gaudi $^{\orcidlink{0000-0003-0395-9869}}$,$^{14}$
\newauthor
Guoyou Sun $^{\orcidlink{0000-0003-3162-3350}}$,$^{15}$
Alastair Claringbold $^{\orcidlink{0000-0003-1309-5558}}$,$^{5,6}$
Lauren Doyle $^{\orcidlink{0000-0002-9365-2555}}$,$^{5,6}$
Tristan Guillot $^{\orcidlink{0000-0002-7188-8428}}$,$^{16}$
Olga Suarez $^{\orcidlink{0000-0002-3503-3617}}$,$^{16}$
\newauthor
Djamel Mékarnia $^{\orcidlink{0000-0001-5000-7292}}$,$^{16}$
Amaury H.M.J. Triaud $^{\orcidlink{0000-0002-5510-8751}}$,$^{17}$
Philippe Bendjoya $^{\orcidlink{0000-0002-4278-1437}}$,$^{16}$
Carl Ziegler $^{\orcidlink{0000-0002-0619-7639}}$,$^{18}$
\newauthor
Andrew W. Mann $^{\orcidlink{0000-0003-3654-1602}}$,$^{19}$
Steve B. Howell $^{\orcidlink{0000-0002-2532-2853}}$,$^{20}$
Sergio B. Fajardo-Acosta $^{\orcidlink{0000-0001-9309-0102}}$,$^{21}$
Colin Littlefield $^{\orcidlink{0000-0001-7746-5795}}$,$^{22}$
\newauthor
Douglas A. Caldwell $^{\orcidlink{0000-0003-1963-9616}}$,$^{20,23}$
Michelle Kunimoto $^{\orcidlink{0000-0001-9269-8060}}$,$^{24}$
Pamela Rowden $^{\orcidlink{0000-0002-4829-7101}}$,$^{25}$
Veselin Kostov $^{\orcidlink{0000-0001-9786-1031}}$,$^{11,23}$
\newauthor
Jesus Noel Villaseñor,$^{26}$
Douglas Alves $^{\orcidlink{0000-0002-5619-2502}}$,$^{27,28}$
Ioannis Apergis $^{\orcidlink{0009-0004-7473-4573}}$,$^{5,6}$
David J. Armstrong $^{\orcidlink{0000-0002-5080-4117}}$,$^{5,6}$
\newauthor
Matthew P. Battley $^{\orcidlink{0000-0002-1357-9774}}$,$^{3,29}$
Daniel Bayliss $^{\orcidlink{0000-0001-6023-1335}}$,$^{5,6}$
François Bouchy $^{\orcidlink{0000-0002-7613-393X}}$,$^{3}$
Sarah L. Casewell $^{\orcidlink{0000-0003-2478-0120}}$,$^{1}$
\newauthor
Maximilian N. Günther $^{\orcidlink{0000-0002-3164-9086}}$,$^{30}$
George T. Harvey $^{\orcidlink{0009-0009-1823-7501}}$,$^{1}$
Faith Hawthorn $^{\orcidlink{0000-0002-8675-182X}}$,$^{5,6}$
James S. Jenkins $^{\orcidlink{0000-0003-2733-8725}}$,$^{31,28}$
\newauthor
Monika Lendl $^{\orcidlink{0000-0001-9699-1459}}$,$^{3}$
James McCormac $^{\orcidlink{0000-0003-1631-4170}}$,$^{5,6}$
Maximilano Moyano $^{\orcidlink{0000-0002-7927-9555}}$,$^{7}$
Louise D. Nielsen $^{\orcidlink{0000-0002-5254-2499}}$,$^{32}$
\newauthor
Ares Osborn $^{\orcidlink{0000-0002-5899-7750}}$,$^{33}$
Toby Rodel $^{\orcidlink{0009-0009-2175-7284}}$,$^{34}$
Suman Saha $^{\orcidlink{0000-0001-8018-0264}}$,$^{28,31}$
Stephane Udry $^{\orcidlink{0000-0001-7576-6236}}$,$^{3}$
Jose I. Vines  $^{\orcidlink{0000-0002-1896-2377}}$,$^{7}$
\newauthor
Peter J. Wheatley $^{\orcidlink{0000-0003-1452-2240}}$,$^{5,6}$
Tafadzwa Zivave $^{\orcidlink{0009-0001-8055-995X}}$,$^{5,6}$
\\\\
The authors' affiliations are listed at the end of the paper.}
\date{Accepted XXX. Received YYY; in original form ZZZ}

\pubyear{2025}

\begin{document}
\label{firstpage}
\pagerange{\pageref{firstpage}--\pageref{lastpage}}
\maketitle

\begin{abstract}
Beyond orbital periods of 10 days, there is a dearth of known transiting gas giants. On longer orbits, planets are less affected by their host star, and become ideal probes of planet formation, migration and evolution. We report the discovery of a long period Neptune and two Saturns, each initially identified as single transits in the \tess\ photometry, and solved through additional transits from ground-based follow-up photometric observations by \ngts\ and \astep. High-resolution radial velocity mass measurements using \coralie\ and \harps\ confirm their planetary nature. From joint modelling of the photometric and spectroscopic data, we determine an orbital period of $\objonep~$days, radius of $\objoner~\mathrm{R_{\oplus}}$, and mass of $\objonem~\mathrm{M_{\oplus}}$ for \objonengts\,b, making it one of the longest period well-characterized transiting Neptunes. Orbiting a late F-type star, bright in the K-band (Kmag$~\simeq7.9$), it is amenable for cool atmosphere studies using JWST or Ariel. \objtwo\,b is a small Saturn on a $\objtwop~$day orbit with a radius of $\objtwor~\mathrm{R_{\oplus}}$ and an upper mass limit $\objtwom~\mathrm{M_{\oplus}}$. \objthreengts\,b(=\objthreetoi\,b) is a larger Saturn on a $\objthreep~$day, moderately eccentric orbit ($e = \objthreee$), with a radius of $\objthreer~\mathrm{R_{\oplus}}$ and a mass of $\objthreem~\mathrm{M_{\oplus}}$. With an assumed albedo $A=0.3$, each of these planets has an equilibrium temperature below 700K, with \objthreengts\,b especially cold at $\objthreetround~$K. These three giants add to the small but growing population of long period planets that can further our understanding of planet formation mechanisms.

\end{abstract}




\section{Introduction}
\label{sec:intro}
Most gas giants with well measured parameters are Hot Jupiters (HJs), where hot implies that they orbit close to their host star, with short orbital periods of less than 10 days. Being at a close orbital separation increases the probability of transiting and a shorter period provides more opportunities to observe transits and take radial velocity data across all phases, leading to clearer phase-folded transit photometry and radial velocity signals. Additionally, in terms of practicality for observing a HJ, larger masses and radii place them well above the typical telescope signal-to-noise ratio limits. These factors add up to make them more easily detectable, to the point of many early exoplanet detections being HJs \citep[][amongst others]{mayorqueloz}. Planets with radii greater than $0.5R_J$ and orbital periods shorter than 10 days accounted for over half of the cumulative confirmed exoplanets with a measurement for both orbital period and radius up until 2013. However, despite their abundance of detections, it is challenging to investigate their history. For HJs and other short-period gas giants with such close proximity to their host star, effects such as photoevaporation~\citep{photoevap} or inflation~\citep{millerandfortney} become important, which can lead to complex evolution of their orbital, atmospheric and planetary properties.

Further out from their host stars, there is a more sparse population of known exoplanets. At longer periods, it has long been noticed that few gas giants exist in the so-called ``period valley'' (around $10 < P < 100~$days, with varying estimates in the literature, such as~\citealt{udryperiodvalley} and~\citealt{jonesperiodvalley}). Equivalently, in terms of semi-major axes, there is a a dearth between 0.2 and 0.6 AU, as noted by \cite{cummingperiodvalley}. 

In their simulations, \cite{idalin1,idalin2} also describe a planet desert, this time a paucity of planets intermediate in mass between the populations of HJs and Neptunes ($10-100~M_\oplus$) and with semi major axes less than 3~AU. This is due to planet masses growing rapidly as they pass through this range during gas accretion, so they rarely finish at these masses. However, in this mass range, \cite{wittenmyerperiodvalley} conclude the period valley arises as a result of selection effects caused by the low masses and long periods. For masses $\gtrsim 100~M_{\oplus}$, they find that the gap is indeed real. 

These longer period warm giants orbit at a greater distance from and are less irradiated by their host star, so are less impacted by the effects experienced by HJs. Targeted searches in this parameter space could yield important candidates to bolster the known population distribution in this region. In particular, those candidates around moderately bright stars are highly amenable for spectroscopic follow-up and atmospheric characterization with instruments such as the Atmospheric Remote-sensing Infrared Exoplanet Large-survey~\citep[Ariel,][]{ariel} or the Near-Infrared Spectrograph~\citep[NIRSpec,][]{nirspec} on JWST. Having a larger sample of characterised long period planets will be highly beneficial for not only exploring evolution, formation and migration mechanisms, but also for gaining a better understanding of colder atmospheres. As equilibrium temperatures drop, various chemical and cloud transitions occur, such as CO to CH\textsubscript{4} at $\simeq850~$K, condensation of KCl at $\simeq700~$K, the transition between N\textsubscript{2} to NH\textsubscript{3} at $\simeq500~$K or condensation of NH\textsubscript{3} at $\simeq150~$K~\citep{transitions}. These approximate transitions are sensitive to many factors other than equilibrium temperature, such as vertical mixing, internal temperature, and metallicity, so more example systems with a variety of planets and stellar hosts are required to further understand these transitions. Studying and classifying colder planets will help us form connections between HJs or other hot gas giants and our Solar System planets, especially in the regime where nitrogen abundance can be measured in the form of ammonia, which could provide constraints on location of planetary formation~\citep{nitrogen}. 

The Transiting Exoplanet Survey Satellite~\citep[\tess,][]{TESS} is a space-based long-timescale photometry mission that covers sectors for $27.4~$days at a time. For short period planets, we see multiple transits per \tess\ sector, leading to easily solved orbital periods. For planets with orbital periods approaching the length of a \tess\ sector and longer, we expect to record at most two transits per sector if the first transit occurs early enough in the sector and the period is sufficiently short. If the star is not observed in chronological sectors, the next observed transit may be separated by a large time-frame when \tess\ eventually returns to that area of sky in a future sector. Hence, the total observed transits will be few in number and the orbital period will not be well defined, making follow-up difficult. The true periods of these candidates will be an integer divisor of the time separation between the transits. In order to solve the period for these planets with single or double transit events, we require further photometry and radial velocity follow-up to target potential period solutions. These are chosen based on likelihood, either confirming or rejecting each possibility, until only one solution remains. These difficulties bias the overall distribution of planets discovered by \tess\ towards those having shorter periods, and many longer periods planets have been missed. Previous works have shown \tess\ has strong potential to find these longer period planets through single transit events~\citep{cooke1,cooke2,villanueva}, with recent yield simulations indicating that beyond 25 days, approximately 75 per cent of planets which have transits in \tess\ data are yet to be discovered~\citep{tiara}. This forms the motivation of the work undertaken by the Next Generation Transit Survey (\ngts) Long Period Planets Programme.

This paper presents three warm giants below 700~K where the orbital periods have been successfully resolved: \objonengts\,b, \objtwo\,b and \objthreengts\,b / \objthreetoi\,b. In Section~\ref{sec:obs}, we describe the data taken and instruments used. In Section~\ref{sec:methods}, we describe the analysis and fitting performed for both the host stars and planets. In Section~\ref{sec:results}, we discuss the results of the joint modelling and place these planets in context of the confirmed planet population, in particular \objonengts\,b as one of the longest period well-defined transiting Neptunes and \objthreengts\,b as a particularly cool Saturn.

\section{Observations}
\label{sec:obs}
\subsection{\tess\ photometry}
\label{sec:tess}
\tess\ observes the sky with 4 cameras, each with an identical 10.5-cm diameter lens which provide a combined field-of-view of 2300 square degrees. The \tess\ bandpass is centred on \textit{I\textsubscript{C}} band, but spans 600 - 1000~nm, with the spectral response shown in \cite{TESS}. 

For \objonengts\,b(=\objonetic\,b), \tess\ has observed 4 transits, one in each of Sectors 9, 10, 36 and 63. However, this candidate does not have a TESS threshold crossing event (TCE), nor has it been alerted as a \tess\ Object of Interest~\citep[TOI,][]{toiscatalog}, likely due to the shallow depth of 0.6~ppt.

\objtwo\,b was initially found by the TESS Faint Star Search~\citep{faintsearch}, using data products from the Quick Look Pipeline~\citep[QLP,][]{qlp1,qlp2}. It was first seen as an event of 2.33~ppt depth and 4.073~hours duration centered at BJD=2458334.071 in the light curve for Sector 1. \tess\ has observed 16 transits in total for \objtwo\,b, but still one per sector with one exception: initially in Sectors 1, 2, 4, 8, 11, 28, 29, 30, 31, 34 and 38 which were sufficient to solve the orbital period, returning in Cycle 5 with two transits in Sector 64 and one transit in each of Sectors 65, 68 and 69 to refine the period further. \objtwo\ was also observed in Sectors 3, 61 and 62, but no transits were seen in these sectors. The TESS Science Office reviewed the data validation reports for the QLP detections up to Sector 38 and issued a TOI alert for \objtwo\,b on 21\textsuperscript{st} December 2021.

\objthreengts\,b / \objthreetoi\,b was initially flagged as a Community TESS Object of Interest (CTOI) on 10\textsuperscript{th} May 2021 by Guoyou Sun as a single transit event centered at BJD=2459283.545 in Sector 36, found through analysis of Science Processing Operations Center~\citep[SPOC,][]{spoc} pipeline data using VStar~\citep{vstar}. There was also a TCE for Sector 36 at BJD=2459283.5 with a depth of 10.36~ppt and a duration of 2.962~hours. The TESS Science Office issued an alert for this object as a TOI on 21\textsuperscript{st} September 2023. A transit from Sector 9 was retrospectively identified. With only 2 transits initially, this was an especially challenging system to solve, requiring further follow up observations. After the period was confirmed with the first successful \ngts\ observation, a third transit was observed in Sector 63. The timelines of \tess\ transits for each object are described in Tables~\ref{tab:TIC147timeseries}, \ref{tab:TIC237timeseries} and \ref{tab:TIC333timeseries}.

For all \tess\ lightcurves in this paper, we use the publicly available data downloaded via \lightkurve~\citep[a package for accessing and analyzing astronomical flux time series data taken by \tess,][]{lightkurve}. The data used are either the processed from 120s cadence target data by the SPOC pipeline~\citep{spoc}, or processed from the Full Frame Image (FFI) data as part of the TESS-SPOC~\citep{tessspoc} project, which produces 1800s cadence lightcurves for the primary mission (S1-26), 600s cadence for the first extended mission (S27-39), and a 200s cadence for the second extended mission and after (past S40). We use Pre-search Data Conditioning Simple Aperture Photometry flux~\citep[PDC-SAP flux,][]{smith12,stumpe12,stumpe14}, where long term systematic trends have been removed and the flux received by the photometric aperture has been corrected for known nearby stars where applicable.

\begin{table}
	\centering
	\caption{Timeline of observations for \objonengts. Dates for photometry are for in-transit data.}
	\label{tab:TIC147timeseries}
	\begin{tabular}{lc}
		\hline
		Instrument & Dates \\
		\hline
		\textbf{Photometry}&\\
		TESS-SPOC (1800s, S9) & 07/03/2019 \\
        TESS-SPOC (1800s, S10) & 19/04/2019 \\
        TESS-SPOC (600s, S36) & 09/03/2021 \\
        \ngts\ &  27/01/2023\\
        TESS-SPOC (200s, S63) & 12/03/2023\\
		\hline
		\textbf{Radial Velocity}&\\
		\coralie\ & 15/02/2022 - 05/07/2022\\
        \harps\ & 10/03/2022 - 30/06/2022\\
		\hline
	\end{tabular}
\end{table}
\begin{table}
	\centering
	\caption{Timeline of observations for \objtwo. Dates for photometry are for in-transit data.}
	\label{tab:TIC237timeseries}
	\begin{tabular}{lc}
		\hline
		Instrument & Dates \\
		\hline
		\textbf{Photometry}&\\
		TESS-SPOC (1800s, S1) & 03/08/2018 \\
        TESS-SPOC (1800s, S2) & 29/08/2018 \\
        TESS-SPOC (1800s, S4) & 20/10/2018\\
        TESS-SPOC (1800s, S8) & 26/02/2019 \\
        TESS-SPOC (1800s, S11) & 15/05/2019\\
        TESS-SPOC (600s, S28) & 22/08/2020 \\
        TESS-SPOC (600s, S29) & 17/09/2020 \\
        TESS-SPOC (600s, S30) & 13/10/2020 \\
        TESS-SPOC (600s, S31) & 08/11/2020 \\
        TESS-SPOC (600s, S34) & 24/01/2021 \\
        TESS-SPOC (600s, S38) & 08/05/2021 \\
        \astep\ &  26/06/2022 \\
        \ngts\ &  18/01/2023\\
        SPOC (120s, S64) & 06/04/2023 \\
        SPOC (120s, S64) & 02/05/2023\\
        SPOC (120s, S65) & 28/05/2023\\
        SPOC (120s, S68) & 14/08/2023\\
        SPOC (120s, S69) & 09/09/2023\\
		\hline
		\textbf{Radial Velocity}&\\
		\coralie\ & 08/01/2022 - 07/03/2022\\
        \harps\ & 02/02/2022 - 21/10/2023\\
		\hline
	\end{tabular}
\end{table}
\begin{table}
	\centering
	\caption{Timeline of observations for \objthreengts. Dates for photometry are for in-transit data.}
	\label{tab:TIC333timeseries}
	\begin{tabular}{lc}
		\hline
		Instrument & Dates \\
		\hline
		\textbf{Photometry}&\\
		TESS-SPOC (1800s, S9) & 09/03/2019 \\
        SPOC (120s, S36) &  10/03/2021\\
        \ngts\ &  12/05/2022\\
        \ngts\ &  25/12/2022\\
        SPOC (120s, S63) & 12/03/2023\\
        \ngts\ &  11/03/2023\\
        \ngts\ &  05/04/2023\\
        \ngts\ &  20/06/2023\\
		\hline
		\textbf{Radial Velocity}&\\
		\coralie\ & 01/06/2021 - 02/05/2022\\
        \harps\ & 28/12/2021 - 29/10/2022\\
		\hline
	\end{tabular}
\end{table}

\subsection{\ngts\ photometry}
\label{sec:ngts}
\begin{table}
	\centering
	\caption{NGTS multicamera in-transit photometry observations for each object, with standard deviation precision achieved for corresponding out of transit data.}
	\label{tab:ngtsphotometry}
	\begin{tabular}{lccc}
		\hline
		Night & Cameras & Coverage & Std. Dev. (ppm) \\
		\hline
		\textbf{\objonengts}&&&\\
		27/01/2023 & 8 & Ingress & 91\\
		\hline
		\textbf{\objtwo}&&&\\
		18/01/2023 & 8 & Ingress & 316\\
		\hline
		\textbf{\objthreengts}&&&\\
        12/05/2022 & 1 & Full & 1165\\
        25/12/2022 & 5 & Ingress, near Full & 1591\\
        11/03/2023 & 7 & Egress & 3406\\
        05/04/2023 & 8 & Ingress, near Full & 2478\\
        20/06/2023 & 7 & Egress, near Full & 786\\
		\hline
	\end{tabular}
\end{table}

The majority of ground-based photometry observations in this paper were taken by the Next Generation Transit Survey (\ngts), based at the ESO Paranal Observatory, Chile. \ngts\ operates an automated array of twelve 20-cm independently mounted and steerable telescopes. This places \ngts\ as one of the few ground-based facilities with the capability to observe multiple \tess\ single-transit objects (or other candidates) simultaneously on a given night. The \ngts\ telescopes have been designed to take high-precision photometry. Matching and complementing \tess, a single telescope is capable of RMS=400~ppm in 30 minutes for all stars brighter than \tess\ magnitude (Tmag) $\simeq12$, and using multiple telescopes, the array is capable of RMS=100\,ppm in 30 minutes for stars brighter than Tmag$\simeq9$, reaching the scintillation limit for bright stars~\citep{ngtsppm,seanscintillation}. \ngts\ photometry is stable night-to-night and is capable of identifying exoplanet transits and stellar variability from night-to-night offsets \citep[e.g.,][]{wasp39b,hip41378f}. Each telescope has a field-of-view of 8 square degrees, which allows suitable reference stars to be chosen for even the brightest TESS candidates. The custom filter used by \ngts\ spans $520-890~$nm and is specifically designed to allow precise photometry of exoplanet transits~\citep{Wheatley}.

All but one of the \ngts\ observations in this paper were taken using multiple cameras simultaneously, with specifics found in Table~\ref{tab:ngtsphotometry}. These observations were reduced using standard aperture photometry routines and detrended for systematics as described in \cite{Wheatley}. The first transit of \objthreengts\ observed by \ngts\ occurred on May 12\textsuperscript{th} 2022, 2 months after alias chasing began on March 10\textsuperscript{th} 2022. This confirmed the 25-day period alias, with each of the 4 subsequent observations refining the period further. For \objonengts\ and \objtwo, the period was already solved through \tess\ transits; however, \ngts\ observed an ingress of each in 2023 to decrease the uncertainty on the ephemerides and confirm the transit from the ground.  To reflect the contributions to refining the parameters for or solving these systems and the fraction of transits observed, we assign \objonetic\,b as \objonengts\,b and \objthreengts\,b alongside the TOI designation \objthreetoi\,b. The order of these observations with respect to the \tess\ photometry can be found in Tables~\ref{tab:TIC147timeseries}, \ref{tab:TIC237timeseries} and \ref{tab:TIC333timeseries}.

\subsection{\astep\ photometry}
\label{sec:astep}
The Antarctic Search for Transiting ExoPlanets~\citep[\astep,][]{astep2,astep3,astep4} is a ground based facility located at Dome C in the Concordia Base, Antarctica, which aims to observe transiting exoplanets. The upgraded \astep + optical 40-cm telescope has the capability to simultaneously observe in red and blue bands~\citep[800~nm and 550~nm respectively,][]{astep4,astep5,astep6}. 

Due to the extremely low data transmission rate at the Concordia station, the data are processed on-site using IDL~\citep{Mekarnia:2016} and Python~\citep{astep5} aperture photometry pipelines. The calibrated light curve is reported via email and the raw light curves of about 1,000 stars are transferred to Europe on a server in Rome, Italy, and are then available for deeper analysis. These data files contain each star’s flux computed through $10$ fixed circular aperture radii so that optimal light curves can be extracted.

For \objtwo, we use one ASTEP egress taken on 26\textsuperscript{nd} June 2022 in R-band with 100s exposures and a FWHM of 6.7$\arcsec$. We used a 9.7$\arcsec$ aperture to extract the photometry. All raw photometry data are shown in the supplementary material.

\subsection{\coralie\ spectroscopy}
\label{sec:coralie}
Many single transit events have turned out to be long period eclipsing binary stars instead of planets \citep[such as][]{lpbinary1,lpbinary2,lpbinary3,lpbinary4,lpbinary5}.  To rule out a stellar companion before higher precision radial velocity follow-up to determine a planetary mass, we use \coralie~\citep{coralie1,coralie2}, a fiber-fed \'{e}chelle spectrograph installed on the 1.2-m Leonard Euler telescope at the ESO La Silla Observatory, Chile. In good conditions and with an exposure time of 40 minutes, \coralie\ can achieve a precision of $5-6~\mathrm{m/s}$ for bright solar-like stars \citep[e.g.][]{coralieprecision}.

For \objonengts, we obtained a total of 11 spectra between the nights of February 15\textsuperscript{th} and July 5\textsuperscript{th} 2022, with a G2 mask, exposure times from 1200s to 1800s and with S/N between 30 and 80 at $550~$nm. We remove one of these points due to large uncertainty and low signal-to-noise ratio. The RMS of the remaining points is $6.5~\mathrm{m/s}$, with a mean uncertainty of $5.6~\mathrm{m/s}$ which is larger than the final fitted semi-amplitude. 

Similarly for \objtwo, we took 8 spectra between the nights of January 8\textsuperscript{th} and March 7\textsuperscript{th} 2022, with a G2 mask, exposure times 1800s and with S/N varying between 5 and 10 at $550~$nm. The RMS of the radial velocity points is $38.2~\mathrm{m/s}$ with a mean uncertainty of $34.3~\mathrm{m/s}$. This is again far larger than the final fitted semi-amplitude.

Finally, for \objthreengts, we observed 12 spectra between the nights of June 1\textsuperscript{st} 2021 and May 2\textsuperscript{nd} 2022, with a K5 mask, exposure times 2400s, reaching S/N between 5 and 10 at $550~$nm. The RMS of the radial velocity points is $32.5~\mathrm{m/s}$. The mean uncertainty ($52.2~\mathrm{m/s}$) is about 30 per cent larger than the final fitted semi-amplitude.

All spectra were reduced using the standard \coralie\ reduction pipeline, and radial velocity measurements derived from standard cross-correlation techniques. None of the systems were found to have a semi-amplitude large enough to have an estimated mass indicative of a stellar companion. Each star was then scheduled for precision radial velocity follow-up, as detailed in the next section.

\subsection{\harps\ spectroscopy}
\label{sec:harps}
To derive the masses of each companion, we used the higher precision \harps\ echelle spectrograph~\citep{harps} on the 3.6\,m telescope, also based at the ESO La Silla Observatory in Chile. The data were taken as part of ongoing long-period transiting giants programmes (IDs 112.261U.001, 109.23J8.001 and 109.22L8.001, PI: Ulmer-Moll).

For \objonengts, 14 observations were made between the nights of March 10\textsuperscript{th} and June 30\textsuperscript{th} 2022.  Observations for \objonengts\ were taken using a G2 mask, and either an exposure time of $1200~$s or of $1500~$s, with S/N ranging between 50 and 110 at $550~$nm. This equates to an RMS of $3.0~\mathrm{m/s}$ and a mean uncertainty of $1.9~\mathrm{m/s}$.

Over a longer timescale for \objtwo, 17 spectra were taken between the nights of February 2\textsuperscript{nd} 2022 and October 21\textsuperscript{st} 2023. Observations for \objtwo\ were taken using a G2 mask and an exposure time of $1800~$s, which reached S/N from 10 to 30 at $550~$nm, with RMS of $12.2~\mathrm{m/s}$ and mean uncertainty of $6.3~\mathrm{m/s}$.

Lastly, for \objthreengts, 22 spectra were taken between the nights of December 28\textsuperscript{th} 2021 and October 29\textsuperscript{th} 2022.  Observations for \objthreengts\ were also taken using an exposure time of $1800~$s but with a K5 mask, and with S/N varying between 8 and 24 at $550~$nm, this time leading to an RMS of $27.7~\mathrm{m/s}$ and mean uncertainty of $9.9~\mathrm{m/s}$. 

\harps\ spectra were reduced using the standard \harps\ reduction pipeline with radial velocity measurements derived using the cross-correlation technique. Both \coralie\ and \harps\ raw radial velocities are shown in the supplementary material.

\subsection{Speckle imaging}
\label{sec:speckle}
Nearby companions can cause false positive transit-like signals or introduce contaminating flux that alters the depth of a real transit~\citep{corot,ciardi2015}. To ensure that these are not unresolved close binaries, the backgrounds of \objtwo\ and \objthreengts\ were checked for nearby companions using high resolution speckle imaging.

\begin{figure}
	\includegraphics[width=\columnwidth]{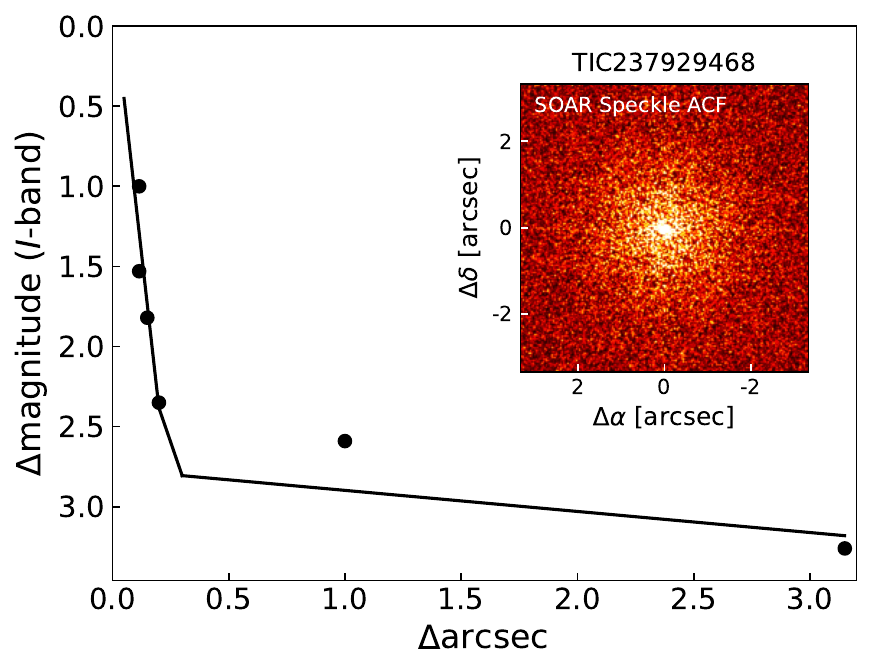}
    \caption{Speckle sensitivity curves for \objtwo\ in the I band (879~nm), taken by HRCam. The reconstructed image is shown in the top right of the figure. There is no evidence for a close stellar companion ($\Delta mag < 2.5$) within separations 0.2 to 3.0 arcseconds.}
    \label{fig:TIC237soar}
\end{figure}

HRCam is a high-resolution speckle imaging instrument mounted on the \soar\ 4.1m telescope in Chile~\citep{hrcam}. HRCam took speckle imaging of \objtwo\ on November 11\textsuperscript{th} 2022 in the I-band (879~nm). As seen in Figure~\ref{fig:TIC237soar}, this shows no evidence of a bright, close stellar companion  ($\Delta mag < 2.5$) from 0.2 to 3.0 arcseconds.

\begin{figure}
	\includegraphics[width=\columnwidth]{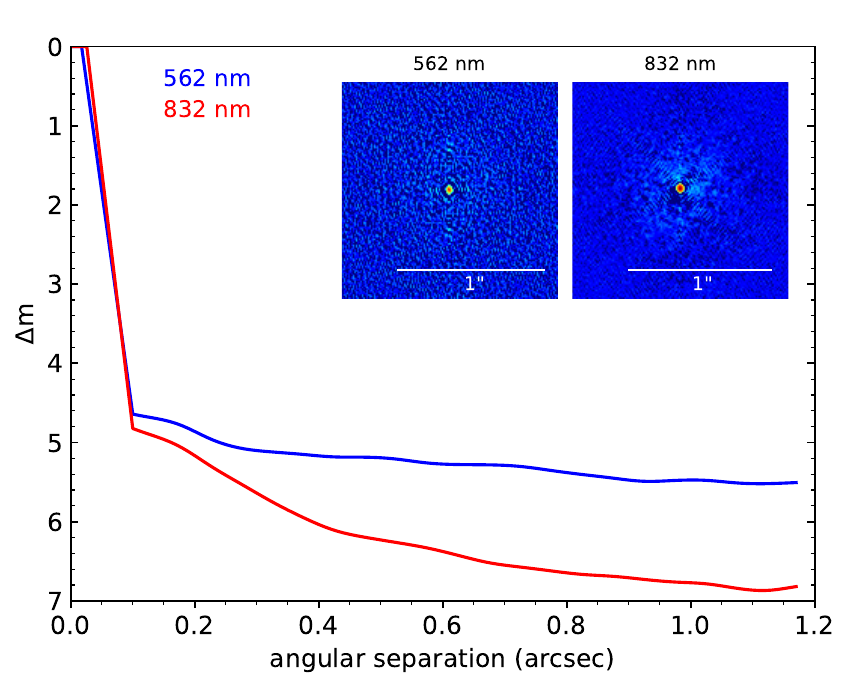}
    \caption{ Speckle sensitivity curves for \objthreengts\ in both 562~nm and 832~nm filters, taken by Zorro. The reconstructed images for both filters are shown in the top right of the figure. There is no evidence for a close stellar companion ($\Delta mag < 4$) within separations 0.1 to 1.2 arcseconds.}
    \label{fig:TIC333gemini}
\end{figure}

Zorro is a dual-channel speckle imaging instrument mounted on the \gemini\ South 8m telescope in Chile~\citep{zorro}. Zorro took speckle imaging of \objthreengts\ on May 18\textsuperscript{th} 2022 in both blue 562~nm and red 832~nm narrowband filters, which was reduced via the standard pipeline described in \cite{howell11}. Shown in Figure~\ref{fig:TIC333gemini}, the imaging data do not detect any bright, close stellar companion ($\Delta mag < 4$) within separations of 0.1 to 1.2 arcseconds. The small features above and below the target in the reconstructed 562~nm image are artifacts commonly seen in speckle imaging for very faint stars. 

\section{Analysis}
\label{sec:methods}
\subsection{Stellar Parameters}
\label{sec:stellarparams}
To have reliable derived parameters from the fit for each object, such as $R_p$ or $M_p$, it is necessary to characterize the host star and have well defined stellar parameters. This subsection describes the analysis to retrieve the properties of each host star. 

\subsubsection{\specmatch}
\label{sec:specmatch}
\begin{table}
	\centering
	\caption{Stellar parameter priors calculated using \specmatch~\protect\citep{specmatch}.}
	\label{tab:specmatchparams}
	\begin{tabular}{lccc}
		\hline
		Parameter & \objonengts & \objtwo & \objthreengts\\
		\hline
        $T_\mathrm{\rm{eff}}$ (K) & $6014\pm110$ & $5556\pm110$ & $4699\pm110$\\
		$\log{g}$ & $4.23\pm0.12$ & $4.22\pm0.12$ & $4.55\pm0.12$\\
		$R_\mathrm{\star}$ ($R_\odot$) & $1.37\pm0.18$ & $1.13\pm0.18$ & $0.78\pm0.10$\\
        $\mathrm{[Fe/H]}$ & $0.240\pm0.090$ & $0.360\pm0.090$ & $0.090\pm0.090$\\
		\hline
	\end{tabular}
\end{table}
To characterise spectra of target stars, \specmatch~\citep{specmatch} compares the combined spectrum to high-resolution spectra taken using KECK/HIRES in the package's empirical library of 404 stars. For our targets, the combined spectrum is from \harps. First, we correct for radial velocity by shifting the target spectrum to the same wavelength scale as a reference spectrum from the library which has the largest median cross-correlation peak in the wavelength range $5120-5220$~\AA. Then, we match our modified target spectrum to each spectra included in the library using the $5140-5200$~\AA\ section of each spectrum, and pick out the best 5 matches from the chi-squared surfaces. We synthesize a linear combination of these to be an even closer match to the target star, from which we derive a weighted average of the stellar parameters from library parameters. \specmatch\ uses a standard error for these parameters, depending on which region on the HR diagram the star is in, since the accuracy of the output of \specmatch\ depends on the spread of the stars in the library and the scatter of their parameters in that region. For example, it adopts an uncertainty $\pm110.0$~K in $T_{\rm{eff}}$, $\pm0.12$ in $\log{g}$, $\pm0.09$ in $\mathrm{[Fe/H]}$ and $\pm16~$per cent in $R_{\mathrm{\star}}$ for $T_{\rm{eff}} \geq 4500~$K. The results of this analysis - the values that will be used as priors for \ariadne~\citep{ariadne} - can be found in Table~\ref{tab:specmatchparams}. 

For \objtwo, due to a lower density of library stars in the parameter space, the chosen five were initially split between two populations: main sequence stars and sub-giants. Certain stars from the library had to be removed from the analysis to force \specmatch\ to choose either population. We opted against sub-giants, since the main sequence linear combination produces a lower chi-squared when compared to the combined target spectrum. Additionally, regardless of whether main sequence or sub-giant priors from \specmatch\ were used, \ariadne\ SED fitting produces a result of $R \simeq 1.2~R_\odot$ and favours the main sequence option in terms of Bayesian evidence.

\subsubsection{SED Fitting with \ariadne}
\label{sec:ariadne}
\begin{figure}
     \centering
     \begin{subfigure}[b]{\columnwidth}
         \centering
         \includegraphics[width=0.9\columnwidth]{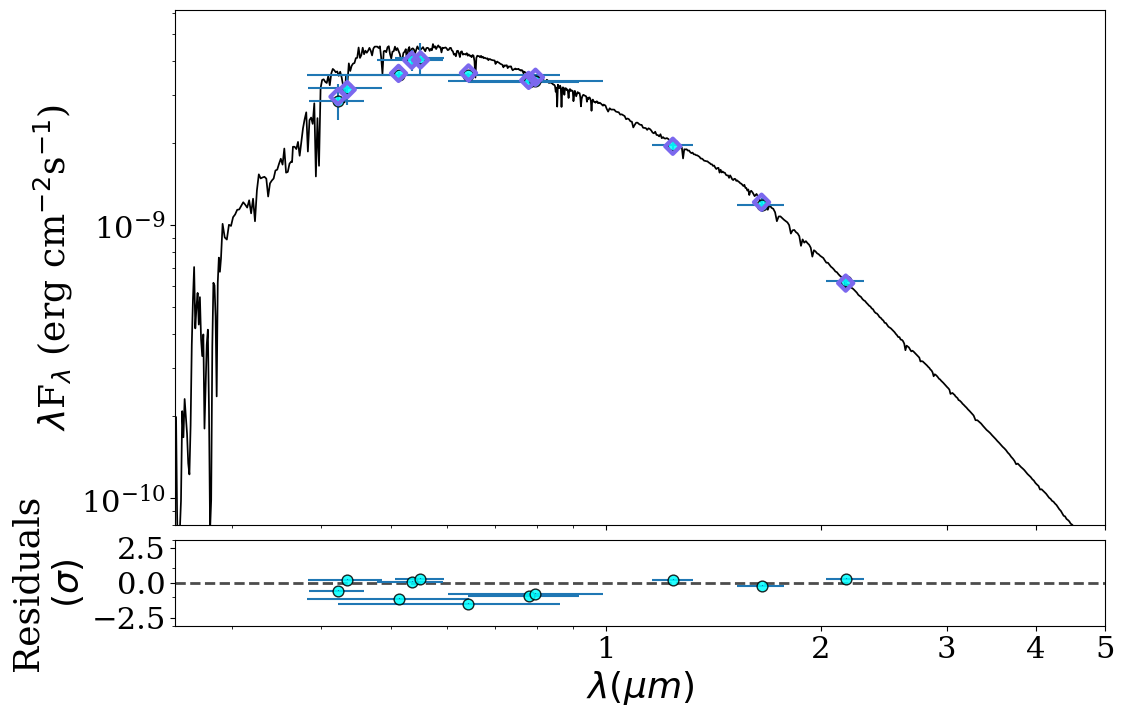}
         \caption{SED fit for \objonengts. The plotted model is the \textsc{ck04}~\citep{castellikurucz} model.}
         \label{fig:tic147sed}
     \end{subfigure}
     \hfill
     \begin{subfigure}[b]{\columnwidth}
         \centering
         \includegraphics[width=0.9\columnwidth]{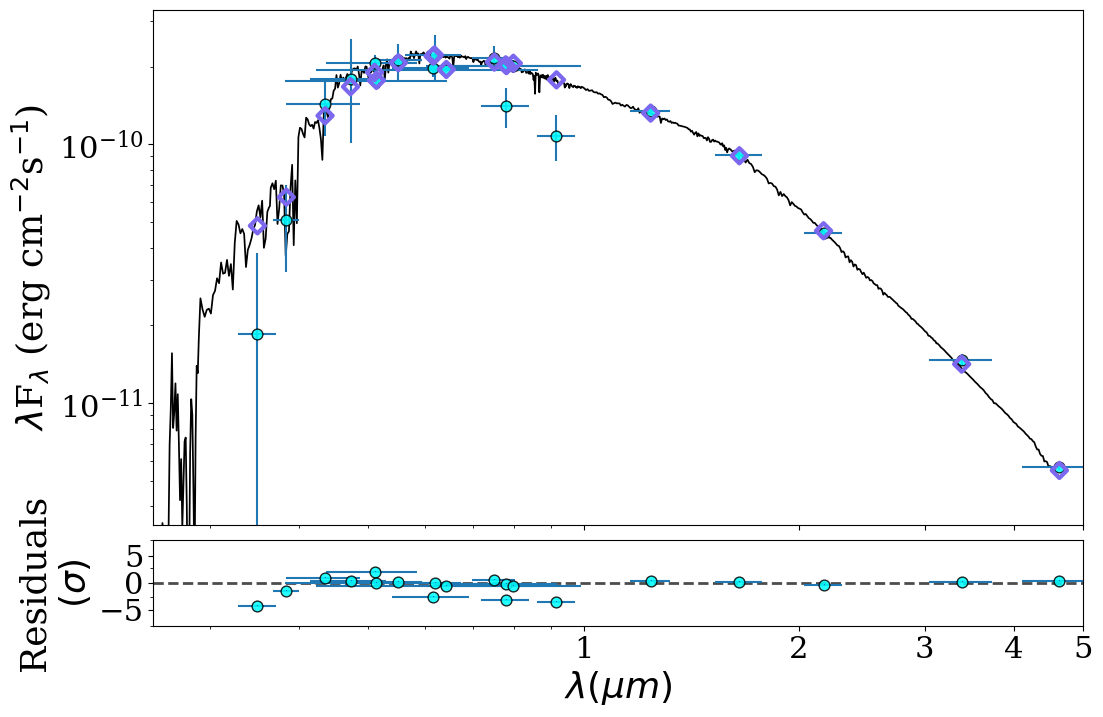}
         \caption{SED fit for \objtwo. The plotted model is the \textsc{ck04}~\citep{castellikurucz} model.}
         \label{fig:tic237sed}
     \end{subfigure}
     \hfill
     \begin{subfigure}[b]{\columnwidth}
         \centering
         \includegraphics[width=0.9\columnwidth]{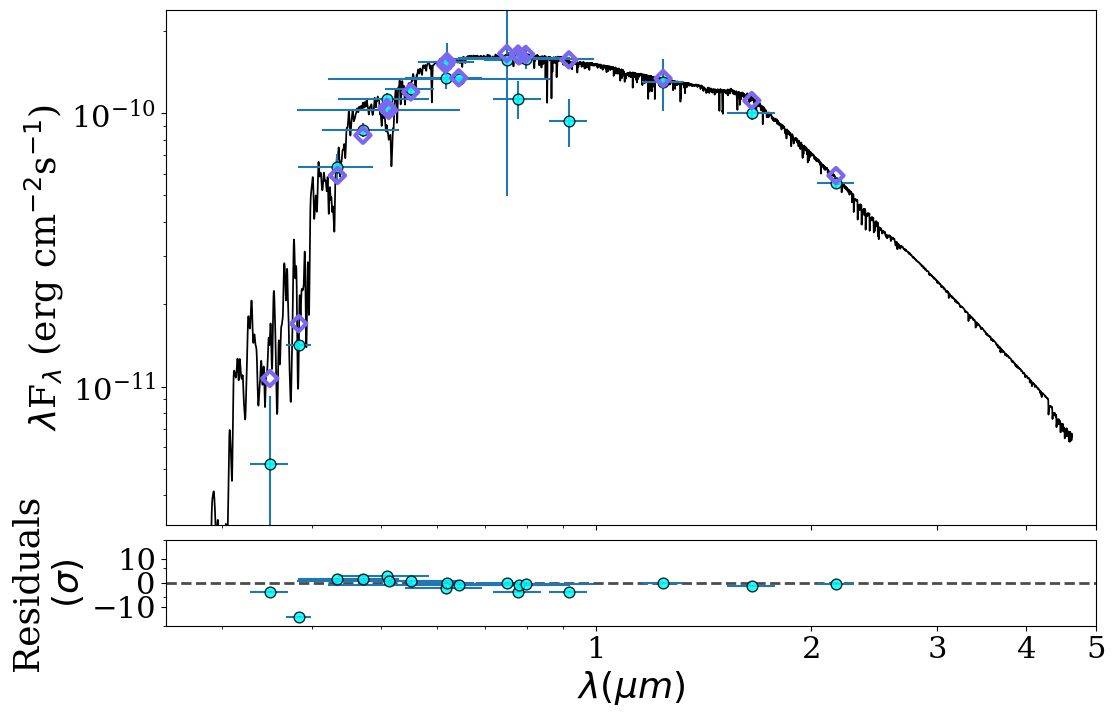}
         \caption{SED fit for \objthreengts. The plotted model is the \textsc{BT-Settl}~\citep{btallard,bthauschildt} model.}
         \label{fig:tic333sed}
     \end{subfigure}
        \caption{SED fits with residuals generated using \ariadne. Blue points are the broadband photometry accessed from the catalogue, and purple diamonds are the synthetic photometry fit. The black line in each plot is the most heavily weighted model included in the Bayesian Model Averaging.}
        \label{fig:SED}
\end{figure}
\begin{table*}
	\centering
	\caption{Top: from \gaia~\citep{gaiapaper,gaiadr3}; Middle: Broadband photometry used in SED fitting; Bottom: Output Stellar Parameters from \ariadne~\citep{ariadne}, where uncertainties for $R_\mathrm{\star}$ and $M_\mathrm{\star}$ for all three and ages for \objtwo\ and \objthreengts\ have been modified to include the systematic uncertainty between stellar models.}
	\label{tab:stellarparams}
	\begin{tabular}{lccc}
		\hline
		Parameter & \objonengts & \objtwo & \objthreengts\\
		\hline
        \multicolumn{4}{l}{\textbf{\gaia\ Astrometric Properties}}\\
		TIC ID & 147277741 & 237929468 & 333736132\\
        TOI &  & 4940 & 6669\\
        \gaia\ Source ID & 5387844681072434944 & 4679529734152652672(DR2) & 3531943236058701824\\
         & & 4679529738448820608(DR3) & \\
		RA & 10:49:59.93 & 03:53:00.28 & 11:21:06.32\\
		Dec & -44:32:52.29 & -62:24:28.68 & -26:03:59.82\\
		pmRA (mas {yr}\textsuperscript{-1}) & $-0.948\pm0.012$ & $0.879\pm0.016$ & $-43.481\pm0.013$\\
		pmDec (mas {yr}\textsuperscript{-1}) & $-66.735\pm0.014$ & $-3.648\pm0.018$ & $-13.347\pm0.012$\\
		Parallax (mas) & $8.310\pm0.014$ & $2.860\pm0.012$ & $5.454\pm0.016$\\
		Distance (pc) & $120.34\pm0.20$ & $349.7\pm1.5$ & $183.35\pm0.53$\\
		\hline
		\multicolumn{4}{l}{\textbf{Broadband Photometry Magnitudes}}\\
		SkyMapper$_u$ & & $14.466\pm0.010$ & $15.8440\pm0.0080$\\
		SkyMapper$_v$ & & $14.0590\pm0.0090$ &$15.4470\pm0.0040$\\
        TYCHO$_{B,MvB}$ & $9.971\pm0.018$ & & \\
        Ground Johnson$_B$ & $9.831\pm0.014$ & $13.196\pm0.012$ &$14.075\pm0.028$\\
        SDSS$_g$ & & $12.778\pm0.018$ &$13.5610\pm0.0080$\\
        SkyMapper$_g$ & & $12.6110\pm0.0020$ & $13.2640\pm0.0020$\\
        \gaia$~{BP}$(DR3) & $9.4183\pm0.0028$ & $12.6755\pm0.0028$ &$13.2579\pm0.0029$\\
        TYCHO$_{V,MvB}$ & $9.299\pm0.014$ & & \\
        Ground Johnson$_V$ & $9.220\pm0.014$ & $12.442\pm0.015$ &$13.036\pm0.019$\\
        SkyMapper$_r$ & & $12.2630\pm0.0030$ & $12.6750\pm0.0050$\\
        SDSS$_r$ & & $12.253\pm0.022$ &$12.644\pm0.030$\\
        \gaia$~G$(DR3) & $9.1407\pm0.0028$ & $12.2938\pm0.0028$  &$12.7016\pm0.0028$\\
        SDSS$_i$ & & $12.068\pm0.017$ &$12.42\pm0.11$\\
        SkyMapper$_i$ & & $12.0990\pm0.0050$ &$12.3360\pm0.0060$\\
        \gaia$~{RP}$(DR3) & $8.6983\pm0.0038$ & $11.7545\pm0.0038$ &$12.0045\pm0.0038$\\
		\tess\ & $8.7504\pm0.0060$ & $11.8153\pm0.0060$ &$12.0742\pm0.0061$\\
		SkyMapper$_z$ & & $12.0830\pm0.0060$ &$12.2330\pm0.0060$\\
        2MASS$_J$ & $8.228\pm0.023$ & $11.146\pm0.023$ &$11.181\pm0.026$\\
        2MASS$_H$ & $7.994\pm0.042$ & $10.784\pm0.022$ &$10.680\pm0.026$\\
        2MASS$_{Ks}$ & $7.925\pm0.021$ & $10.769\pm0.023$ & $10.556\pm0.022$\\
		WISE$_{RSR,W1}$ & & $10.688\pm0.022$ &\\
		WISE$_{RSR,W2}$ & & $10.746\pm0.021$ &\\
		\hline
		\multicolumn{4}{l}{\textbf{\ariadne\ Output Stellar Parameters}}\\
		$T_\mathrm{\rm{eff}}$ (K) & $6134_{-50}^{+55}$ & $5504_{-55}^{+43}$ & $4717_{-84}^{+18}$\\\vspace{1pt plus 1pt minus 1pt} 
		$\log{g}$ & $4.22_{-0.11}^{+0.10}$ & $4.232_{-0.068}^{+0.106}$ & $4.57_{-0.14}^{+0.11}$\\\vspace{1pt plus 1pt minus 1pt} 
		$\mathrm{[Fe/H]}$ & $0.228_{-0.085}^{+0.084}$ & $0.326_{-0.049}^{+0.037}$ & $0.085_{-0.110}^{+0.071}$\\\vspace{1pt plus 1pt minus 1pt} 
		$M_\mathrm{\star}$ ($M_\odot$) & $1.266_{-0.081}^{+0.063}$ & $1.012_{-0.047}^{+0.040}$ & $0.789_{-0.048}^{+0.035}$\\\vspace{1pt plus 1pt minus 1pt} 
		$R_\mathrm{\star}$ ($R_\odot$) & $1.367\pm0.061$ & $1.163_{-0.053}^{+0.054}$ & $0.753_{-0.033}^{+0.051}$\\\vspace{1pt plus 1pt minus 1pt} 
        $L_\mathrm{\star}$ ($L_\odot$) & $2.38\pm0.11$ & $1.116_{-0.064}^{+0.061}$ & $0.253_{-0.018}^{+0.024}$\\\vspace{1pt plus 1pt minus 1pt} 
        $A_v$ & $0.036_{-0.022}^{+0.035}$ & $0.024_{-0.017}^{+0.035}$ & $0.027_{-0.027}^{+0.066}$\\\vspace{1pt plus 1pt minus 1pt} 
		Distance (pc) & $120.34_{-0.18}^{+0.19}$ & $350.19_{-0.71}^{+1.53}$ & $183.42_{-0.43}^{+0.63}$\\\vspace{1pt plus 1pt minus 1pt} 
		Age (Gyr) & $2.7_{-1.8}^{+1.3}$ & $9.8_{-3.0}^{+2.5}$ & $1.5_{-1.5}^{+11.3}$\\\vspace{1pt plus 1pt minus 1pt} 
        Spectral Type & F8V & G8V & K3.5V\\
		\hline
	\end{tabular}
\end{table*}

To produce estimates of the stellar parameters for each star, the results from Section~\ref{sec:specmatch} and the \gaia\ distance calculated from the inverse parallax in Table~\ref{tab:stellarparams} were used as normal priors in \ariadne~\citep[a SED fitting tool,][]{ariadne}, along with each star's right ascension, declination, and \gaia\ DR3 ID.

\begin{figure}
	\includegraphics[width=\columnwidth]{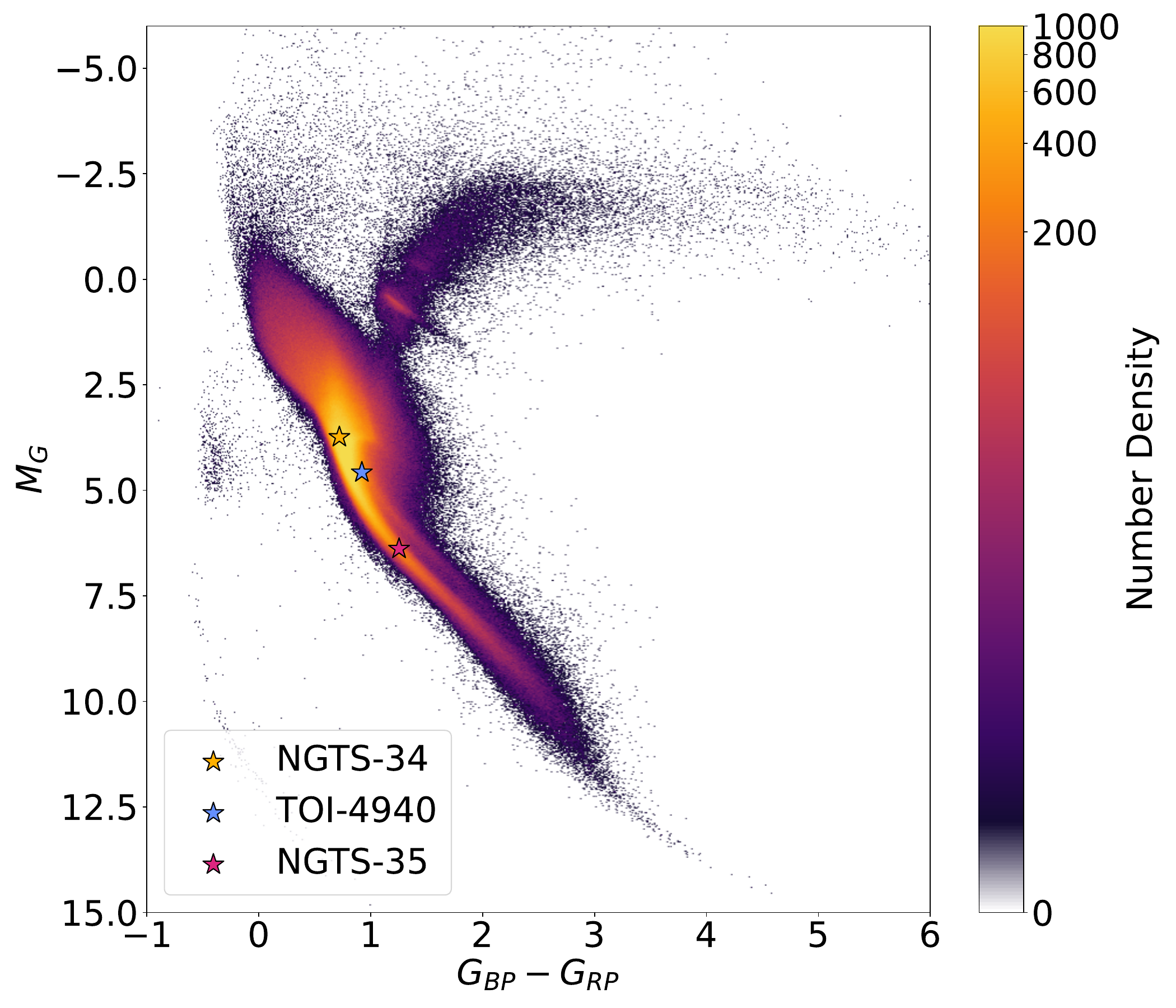}
    \caption{HR diagram of the TESS-SPOC FFI sample described in \citet{laurenHR}, generated using \gaia\ DR3~\citep{gaiadr3} colours and parallax, where the colour bar represents the log density of stars. The three host stars \objonengts, \objtwo\ and \objthreengts\ are represented by yellow, blue and pink stars respectively.}
    \label{fig:HR}
\end{figure}

\ariadne\ first accesses a catalog of broadband photometry plus any user added photometry (Table~\ref{tab:stellarparams}). Atmospheric model grids (\textsc{Phoenix V2}, \citealt{phoenixhusser}: \textsc{BT-Settl, BT-NextGen} and \textsc{BT-Cond}, \citealt{btallard,bthauschildt}; \textsc{ck04}, \citealt{castellikurucz}; \textsc{Kurucz}, \citealt{kurucz}) are interpolated in $T_\mathrm{\rm{eff}}-\log{g}-\mathrm{[Fe/H]}$ space to create model SEDs, with distance, $R_\mathrm{\star}$ and $A_v$ used as additional model parameters. Some models are unused based on predicted effective temperature conditions. Nested sampling~\citep[NS,][]{ns} is performed on each SED using the \textsc{dynesty} package~\citep{dynesty}, which both estimates parameters from the SED and calculates the Bayesian evidence for the model. The last step is Bayesian model averaging, where the relative probabilities of each of the models are used to calculate a weighted average for each of the fitted parameters, which results in a higher degree of precision than using just one of the atmospheric models. The output is a variety of fitted parameters, alongside a mass estimate derived using MIST isochrones~\citep{mist} and a Mamajek spectral type~\citep{pecautmamajek}, which are listed in the lower section of Table~\ref{tab:stellarparams}. The SED fits for each star plotted using the highest weighted model are found in Figure~\ref{fig:SED}. We present these host stars on a HR diagram in Figure~\ref{fig:HR}, using the TESS-SPOC FFI sample for Sectors 1-55 correlated with \gaia\ DR3~\citep{gaiadr3} to retrieve colours and parallaxes, as detailed in \citet{laurenHR}.

\citealt{tayar} showed that analysis of stellar parameters is limited by systematic uncertainties between stellar models. To reflect this, we use a 4.2 per cent error for stellar radii~\citep[as suggested in][]{tayar} and calculate the standard deviation of each star’s mass, age, $\log{g}$, $\mathrm{[Fe/H]}$, and $T_\mathrm{\rm{eff}}$ from different sets of isochrones using the KIAUHOKU package~\citep{kiauhoku1,kiauhoku2}. We add these additional uncertainties in quadrature to the output parameters from \ariadne. For the majority of the parameters for all three stars, including this additional error makes a negligible difference compared to the output parameters from \ariadne\ to 2 significant figures. However, for the radii and masses of the three stars, and the ages of \objtwo\ and \objthreengts\, we substitute the error from this analysis into Table~\ref{tab:stellarparams}. We note that the KIAUHOKU package finds a non-physical lower uncertainty for age for \objthreengts, so we assign the lower limit as 0~Gyr.

\ariadne\ was also used on the neighbour 8.2" away from \objonengts\ (\gaia\ Source ID 5387844676776121472), to calculate input parameters for \pysynphot~\citep[][]{pysynphot} which we use to estimate dilution in Section~\ref{sec:phot}. After using values for $T_\mathrm{\rm{eff}}$, $R_{\star}$ and distance from \gaia\ as normal priors in place of priors estimated by \specmatch, we retrieve $T_\mathrm{\rm{eff}} = 4330_{-80}^{+70}~$K, $\mathrm{[Fe/H]} = -0.13\pm0.19$ and $\log{g} = 4.33_{-0.39}^{+0.47}$. 

\subsubsection{Activity}
\label{sec:activity}
The chromospheric flux ratio ($\log{R'_{hk}}$), an indicator of stellar activity, can be calculated by converting the Mount Wilson S Index ($S_{MW}$) using the star's effective temperature, B-V colour and a conversion factor~\citep{noyesrhk,ruttenrhk}. The majority of stars have a $\log{R'_{hk}}$ between $-5.2$ and $-4.4$. However, the most active stars are considered to have $\log{R'_{hk}}\gtrsim-4.3$~\citep{saikiarhk,henryrhk}. We calculate $\log{R'_{hk}}$ for the stars in this paper using $T_{\rm{eff}}$ from \ariadne, the $S_{MW}$ from each point in the \harps\ data files and the colour difference $B-V$ from the ground Johnson broadband photometry in Table~\ref{tab:stellarparams}. After calculating a weighted average across the points for each object, \objonengts\ has $\log{R'_{hk}} = -4.632 \pm 0.022$, \objtwo\ has $\log{R'_{hk}} = -5.41 \pm 0.28$, and \objthreengts\ has $\log{R'_{hk}} = -4.562 \pm 0.100$. This places \objonengts\ and \objthreengts\ within the common, less active region, and \objtwo\ even less active. However, towards the upper edge of the common region, these stars can still have large activity induced RV variations~\citep[e.g.][]{suarez17}.

Using these values for $\log{R'_{hk}}$ and the activity-rotation relations from \cite{suarez15} and \cite{suarez16}, we estimate the rotation period for each star.  We also search for evidence of the rotation period signals in the SAP-FLUX \tess\ photometry with transit regions masked using normalised generalised Lomb-Scargle periodograms.

For \objonengts, the estimated rotation periods are $12.8~$days using \cite{suarez15} with a metallicity between 0.1 and 0.3, and $5.9~$days for an F-type star in \cite{suarez16}. Searching the photometry for these periods, we do not see any evidence of these periods when combining Sectors 9 and 10. In Sector 36, we see a narrow peak at $6.3~$days and wider peak at $12.8~$days. Similarly, we see peaks at $13.3$ and $6.6~$days in Sector 63. This is closest to $P_{rot}$ and the $P_{rot}/2$ harmonic for $12.8~$days, but could also be related to $5.9~$days.

For \objtwo, the metallicity is outside of the regions for which there are coefficients in \cite{suarez15}. However, taking the closest option of $0.1<[\mathrm{Fe/H]}<0.3$ since 0.3 is within $1\sigma$ of the metallicity derived for \objtwo, we find an estimated rotation period of $85.9~$days. Using \cite{suarez16}, we find $P_{rot}=71.5~d$ for a G-K-M star with $\log{R'_{hk}}<-4.5$ and $P_{rot}=35.5~d$ for G-type star with $\log{R'_{hk}}<-4.6$. Note that the error on $\log{R'_{hk}}$ is largest for this object and these estimated rotation periods could be inaccurate. To search for these large periods, we combine data for TESS-SPOC Sectors 1-4 and 28-31 and SPOC Sectors 61 and 62, 64 and 65, and 68 and 69. All the SPOC combinations contain a wide peak or rise between 32.2-43.1 days, which could include the 35.5-day rotation period. Where four sectors have been combined, we do see a rise towards where the other alternatives could occur, but at longer periods the periodogram is less well-defined. For all but the Sector 1-4 and 61/62 data, there is a recurring peak at $14.3-14.5$~days. In the Sector 1-4 and 61/62 data, an asymmetric 12.2-day peak appears. It is difficult to know what to attribute these signals to, but the closest option is $P_{orb}/2$, or perhaps it is a clue to the true rotation period.

For \objthreengts, we derive rotation periods of $15.1~$days from \cite{suarez15} with a metallicity between -0.1 to 0.1; $15.3~$days from \cite{suarez16} for a G-K-M star with $\log{R'_{hk}}<-4.5$; and $20.2~$days for a K star with $\log{R'_{hk}}>-4.6$. We do not see peaks around any of these periods. However, we do see peaks in all three sectors around $12-13~$days (e.g. $P_{orb}/2$) and in Sectors 9 and 35 peaks between $7-8~$days (e.g. $P_{rot}/2$ if we take the rotation period as $15.3~$days due to lower uncertainty on coefficients).

\subsection{Global Modelling \& Planet Parameters}
\label{sec:planet}
To determine the physical parameters of the system, we utilised \allesfitter~\citep{allesfitter, allesfittercode} to simultaneously fit all photometry and radial velocities from Section~\ref{sec:obs}. \allesfitter\ is publicly available software that acts as an envelope for many common useful packages, such as \textsc{ellc}~\citep[light curve/radial velocity models,][]{ellc} or \textsc{celerite}~\citep[Gaussian Process (GP) models,][]{celerite}. It can effectively model transits of planets, binaries, stellar variability and more, in both photometric and spectroscopic data, using a choice of Monte Carlo Markov Chain (MCMC) sampling through \textsc{emcee}~\citep{emcee} or NS through \textsc{dynesty}~\citep{dynesty}. For our modelling, we used dynamic NS, with 500 live points and a convergence threshold of $\delta\ln{Z} \leq 0.01$, as is default and recommended in \cite{allesfitter} and \cite{dynesty}.\\
\subsubsection{Photometry}
\label{sec:phot}
\begin{figure*}
	\includegraphics[width=2\columnwidth]{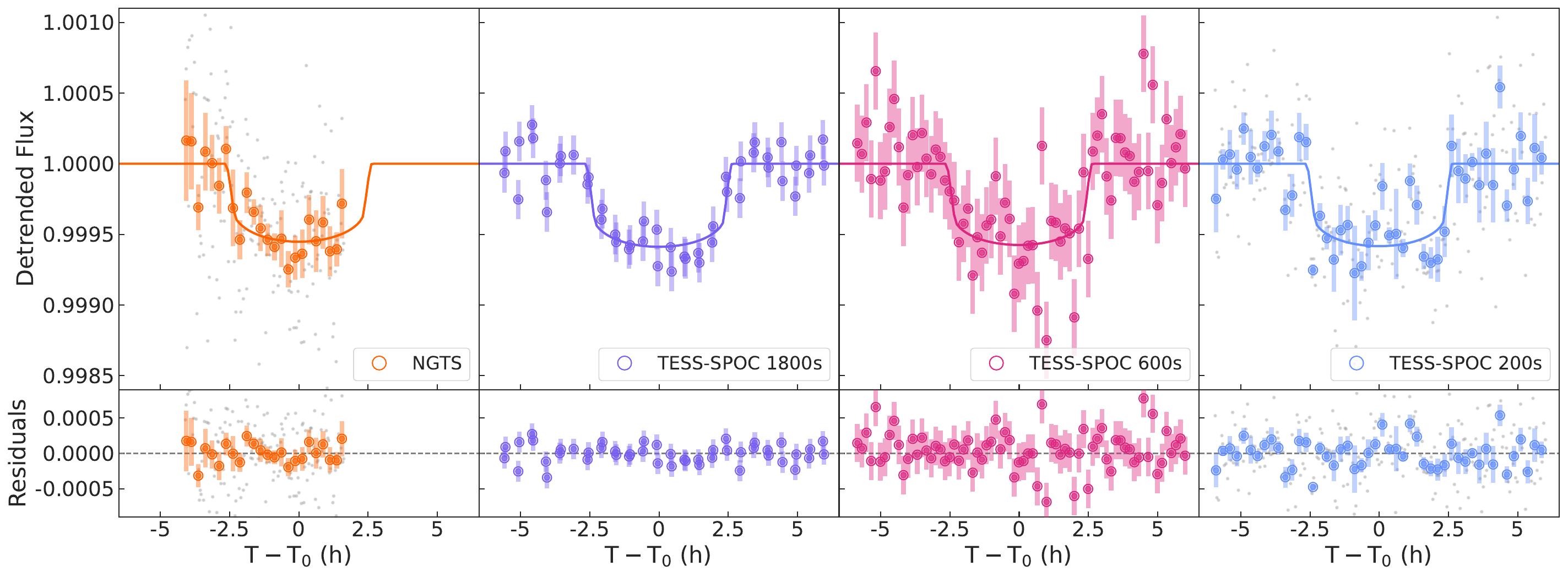}
    \caption{Photometry observations of \objonengts\ grouped by instrument and cadence with residuals, phase-folded on the best fitting period $P = \objonep~$days. From left to right: \ngts\ binned to 2 minutes in grey and binned to 15 minutes in orange, where the red noise was modelled by a hybrid spline; TESS-SPOC 1800s (Sectors 9 and 10) in purple where the red noise was modelled by a GP (Matern-3/2); TESS-SPOC 600s (Sector 36) in pink where the red noise was modelled by a GP (Matern-3/2); and TESS-SPOC 200s (Sector 63) in grey and binned to 15 minutes in blue, where the red noise was modelled by a hybrid spline. The solid lines are the posterior median models for each instrument from the joint fit from \allesfitter.}
    \label{fig:TIC147Phot}
\end{figure*}
\begin{figure*}
	\includegraphics[width=2\columnwidth]{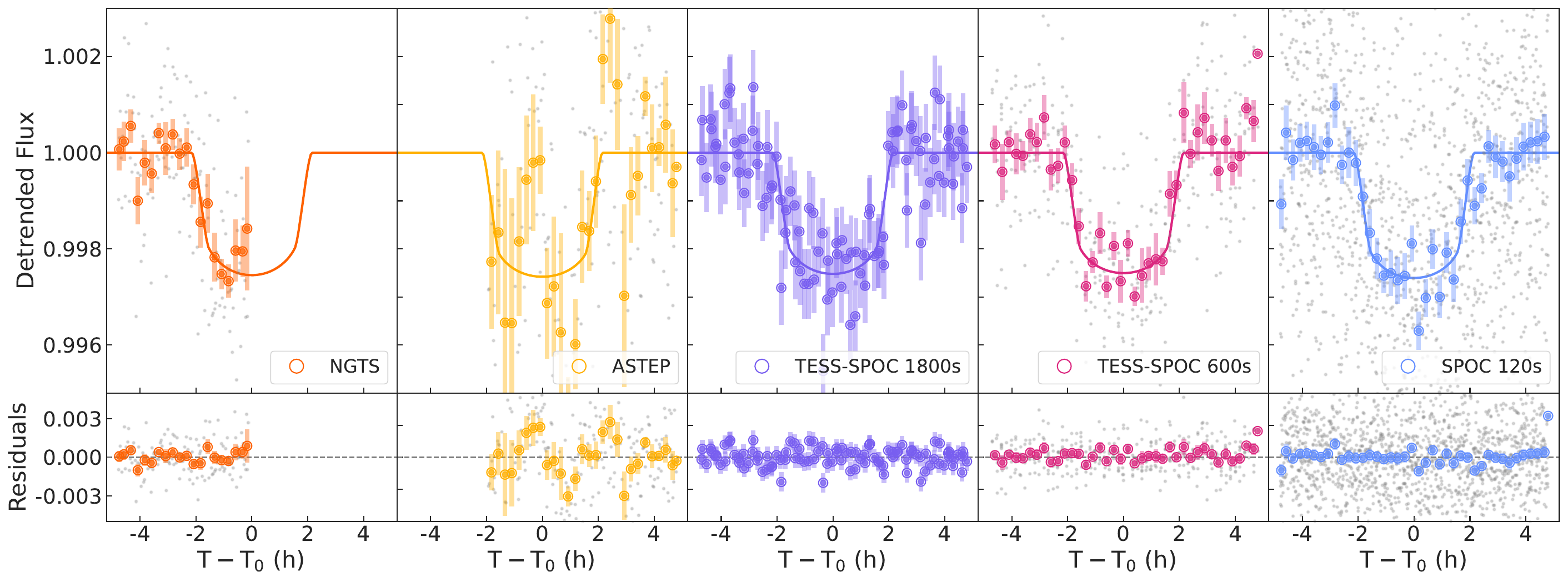}
    \caption{Photometry observations of \objtwo\ grouped by instrument or cadence with residuals, phase-folded on the best fitting period $P = \objtwop~$days. From left to right: \ngts\ binned to 2 minutes in grey and binned to 15 minutes in orange, where the red noise was modelled by a hybrid spline; \astep\ in 100s exposures in grey and binned to 15 minutes in yellow, where the red noise was modelled by a hybrid spline;  TESS-SPOC 1800s (Sectors 1-4, 8 and 11) in purple where the red noise was modelled by a flat offset; TESS-SPOC 600s (Sectors 28-31, 34 and 38) in grey and binned to 15 minutes within the phase-folded lightcurve in pink, where the red noise was modelled by a GP (Matern-3/2); and SPOC 120s (Sectors 61, 62, 64, 65, 68 and 69) in grey and binned to 15 minutes within the phase-folded lightcurve in blue, where the red noise was modelled by a hybrid spline.The solid lines are the posterior median models for each instrument from the joint fit from \allesfitter.}
    \label{fig:TIC237Phot}
\end{figure*}
\begin{figure*}
	\includegraphics[width=2\columnwidth]{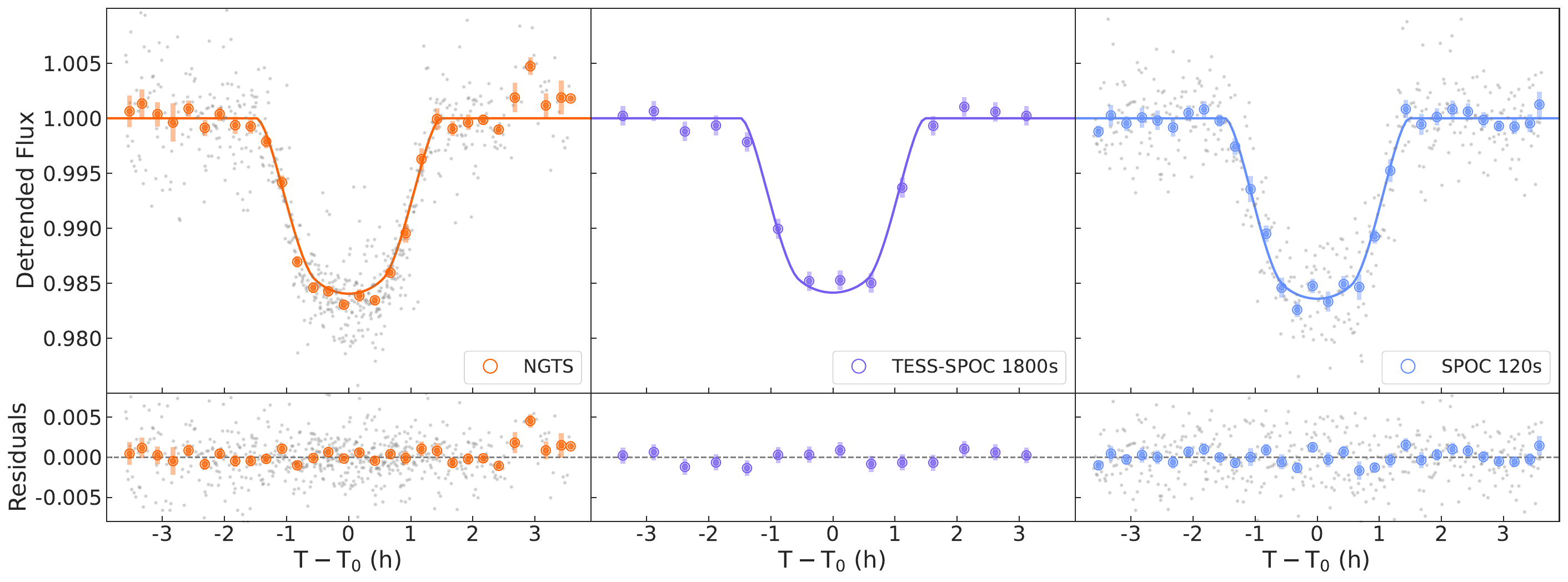}
    \caption{Photometry observations of \objthreengts\ grouped by instrument or cadence with residuals, phase-folded on the best fitting period $P = \objthreep~$days. From left to right: \ngts\ binned to 2 minutes in grey and binned to 15 minutes within the phase-folded lightcurve in orange, where the red noise was modelled by a GP (Matern-3/2); TESS-SPOC 1800s (Sector 9) in purple where the red noise was modelled by a GP (Matern-3/2); and SPOC 120s (Sector 36 and 63) in grey and binned to 15 minutes within the phase-folded lightcurve in pink, where the red noise was modelled by a hybrid spline. The solid lines are the posterior median models for each instrument from the joint fit from \allesfitter.}
    \label{fig:TIC333Phot}
\end{figure*}

To prepare the photometry for NS, we find the relative flux through normalising the raw lightcurves to a baseline of 1, dividing through by the median out of transit flux. To exclude outliers in the \ngts\ data, we apply a 5-sigma clip. To reduce computational time, the \ngts\ data were binned to 2 minutes using \lightkurve~\citep{lightkurve}. As described earlier, the \tess\ photometry data products retrieved from \lightkurve\ keep their cadences of either 1800s, 600s, 200s or 120s depending on the when they were taken and which pipeline was used. We grouped the data by cadence to link related parameters and minimise the number of parameters \allesfitter\ had to fit (i.e. host limb darkening coefficients are the same across \tess\ sectors) which further reduces computational time, and to ensure an extended out of transit baseline over which to fit a GP, while still allowing inclusion of exposure interpolation settings in the fit to avoid bias from longer cadence data.

\begin{table}
	\centering
	\caption{Estimated out of transit values for a GP Matern-3/2 red noise model using \allesfitter~\citep{allesfitter,allesfittercode}}
	\label{tab:GPpriors}
	\begin{tabular}{lcc}
		\hline
		Instrument & $\ln{\sigma}$ & $\ln{\rho}$\\
		\hline
		\textbf{\objonengts}&&\\
		${\mathrm{TESS-SPOC~1800s}}$ & $-9.323_{-0.078}^{+0.075}$ & $-1.62\pm0.16$ \\
		${\mathrm{TESS-SPOC~600s}}$ & $-9.38_{-0.11}^{+0.10}$ & $-2.17_{-0.25}^{+0.26}$ \\
		\hline
		\textbf{\objtwo}&&\\
		${\mathrm{TESS-SPOC~600s}}$ & $-8.65\pm0.10$ & $-2.21_{-0.34}^{+0.32}$ \\
		\hline
		\textbf{\objthreengts}&&\\
		${\mathrm{NGTS}}$ & $-4.82_{-0.11}^{+0.13}$ & $-1.24_{-0.22}^{+0.27}$ \\
		${\mathrm{TESS-SPOC~1800s}}$ & $-7.00_{-0.17}^{+0.20}$ & $0.13\pm0.25$\\
		\hline
	\end{tabular}
\end{table}

For both the photometry and later the radial velocity data, a red noise model is used in the global fit to handle effects from instrument systematics and stellar activity. To decide an appropriate red noise model for each piece of photometry data, flat offsets, sloped linear, hybrid spline and GP Matern-3/2 models were tested by comparing Bayesian evidence between models. Hybrid spline is one of several callable options in Allesfitter~\citep{allesfitter,allesfittercode}. It is a cubic spline, where hybrid means that at each step a least squares minimization determines the cubic spline parameters for the model. For the 1800s cadence TESS-SPOC data for \objtwo, it appears that most variation had been detrended through the PDC-SAP pipeline and a flat offset was the most suitable. For the \ngts\ data for \objonengts\ and \objtwo, and the ASTEP data for \objtwo, and all SPOC 120s cadence data, a hybrid spline was more appropriate. All other photometric data required the use of a GP Matern-3/2 model. For this, red noise values were estimated for the out of transit data before using ``in-transit'' data for the final fits, where ``in-transit'' is a region centered on mid transit with a width set by $\simeq3 \times T_\mathrm{tot}$: 0.5 days in duration for \objonengts, 0.4 days for \objtwo\ and 0.3 days for \objthreengts. These values, listed in Table~\ref{tab:GPpriors}, were used as normal priors when fitting using \allesfitter~\citep{allesfitter,allesfittercode}. The fitted photometry data with chosen models subtracted can be found in Figures~\ref{fig:TIC147Phot}, \ref{fig:TIC237Phot} and \ref{fig:TIC333Phot}.

To estimate priors for the quadratic limb darkening coefficients, we start by calculating $u_1$ and $u_2$ using the package \ldtk~\citep{ldtk}. \ldtk\ calculates both stellar limb darkening profiles and model specific coefficients, where the model is based on the \textsc{Phoenix} stellar spectrum library of \cite{phoenixhusser} and uses $T_\mathrm{\rm{eff}}$, $\log{g}$ and $[\mathrm{Fe/H]}$ from Table~\ref{tab:stellarparams} for each star as inputs. We input the \ngts\ filter~\citep{Wheatley} (which has been multiplied by an atmospheric contribution;~\citealt{ngts3ab}), the \tess\ filter~\citep{TESS} and the red \astep\ filter~\citep{astep4,astep5,astep6} to allow it to find limb darkening parameters specific to each instrument. We then convert these limb darkening parameters to $q_1$ and $q_2$ form~\citep[as parameterized by][]{kipping}. The results of this (listed in Table~\ref{tab:ldpriors}) are directly input into the initial parameters file for \allesfitter\ as normal priors and we allow \allesfitter\ to fit these parameters based on the in-transit photometry data.\\
\begin{table}
	\centering
	\caption{Estimated limb darkening priors using \ldtk~\protect\citep{ldtk}.}
	\label{tab:ldpriors}
	\begin{tabular}{lccc}
		\hline
		Parameter & \ngts\ & \tess\ & \astep\ \\
		\hline
		\textbf{\objonengts}&&\\
		$q_{1}$  & $0.3526\pm0.0018$ & $0.2988\pm0.0015$ &\\
		$q_{2}$ & $0.384\pm0.019$ & $0.376\pm0.019$ &\\
		\hline
		\textbf{\objtwo}&&\\
		$q_{1}$  & $0.3930\pm0.0019$ & $0.3375\pm0.0017$ & $0.2918\pm0.0014$\\
		$q_{2}$ & $0.414\pm0.020$ & $0.402\pm0.020$ & $0.392\pm0.019$\\
		\hline
		\textbf{\objthreengts}&&\\
		$q_{1}$  & $0.4614\pm0.0037$ & $0.4007\pm0.0032$ &\\
		$q_{2}$ & $0.441\pm0.028$ & $0.426\pm0.027$ &\\
		\hline
	\end{tabular}
\end{table}

\tess\ has a large pixel scale of 21", whereas \ngts\ has a smaller pixel scale of 5". This can lead to some contamination of the lightcurves by additional neighbouring stars, so we need to account for dilution, especially in the data from TESS. Dilution is defined in \allesfitter\ as
\begin{equation}
\label{dil}
 D_\mathrm{instrument} = 1 - \frac{F_\mathrm{target}}{F_\mathrm{target}+F_\mathrm{blend}}
\end{equation}
where $D_\mathrm{instrument}=0$ implies an uncontaminated aperture for that instrument. For \objtwo\ and \objthreengts, we can fix the \ngts\ dilution to zero, since there are no neighbouring stars in \gaia~DR3~\citep{gaiapaper,gaiadr3} within 35" for \objtwo\ and within 45" for \objthreengts, which is far larger than the aperture size for \ngts\ (3 pixels radius, or 15"). However, \objonengts\ has a nearby star at 8.2" (\gaia\ Source ID 5387844676776121472), which we account for by calculating a prior for $D$ using \pysynphot~\citep[][]{pysynphot}. First, model spectra are generated for both the target and the neighbour using the \textsc{Phoenix}~\citep{phoenixhusser} model and the \ariadne\ outputs for the target and neighbouring stars for $T_\mathrm{\rm{eff}}$, $\log{g}$ and $\mathrm{[Fe/H]}$ from Section~\ref{sec:ariadne}. To estimate the flux received from each star, we imitate an observation by interpolating each spectrum over the same \ngts\ bandpass used for \ldtk. In equation~\eqref{dil}, this leads to $D =  0.051253\pm0.000311$, which we set as a normal prior.

As discussed in Section~\ref{sec:tess}, the TESS-SPOC and SPOC data used are PDC-SAP flux, so in theory the effective dilution we fit for in \allesfitter\ should be zero as well. However, we allow for the possibility that the pipeline has miscalculated dilution and that additional correction is required. Similar to \cite{toi530}, we set the dilution prior for TESS-SPOC and SPOC to a normal prior with mean 0.00 and standard deviation 0.05.

\subsubsection{Radial Velocity}
\label{sec:rv}
\begin{figure}
     \centering
     \begin{subfigure}[b]{\columnwidth}
         \centering
         \includegraphics[width=0.85\columnwidth]{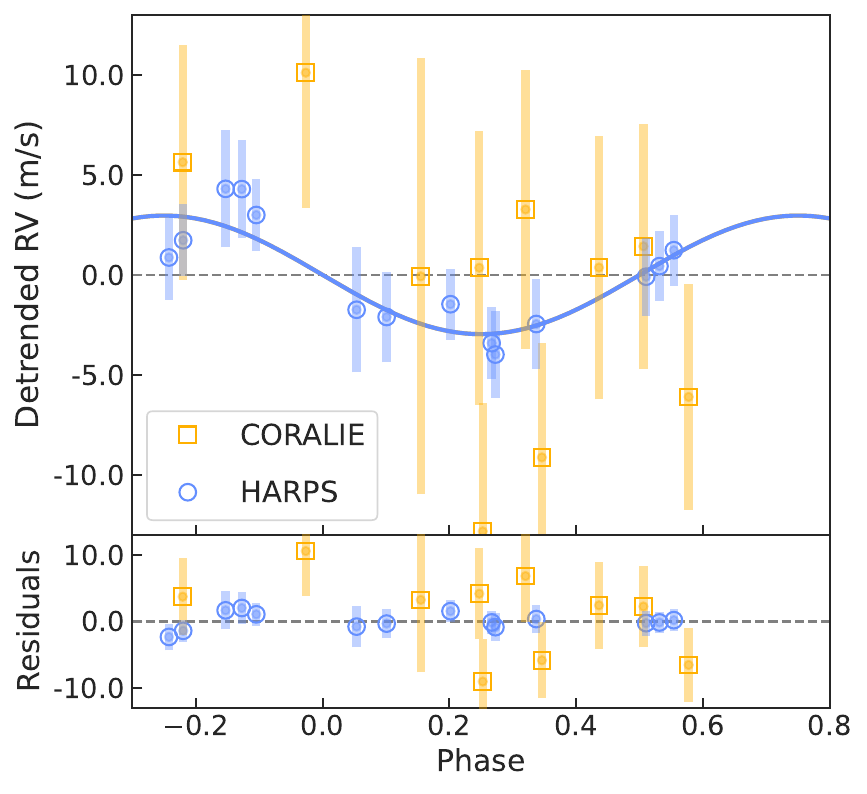}
     \end{subfigure}
     \hfill
     \begin{subfigure}[b]{\columnwidth}
         \centering
         \includegraphics[width=0.85\columnwidth]{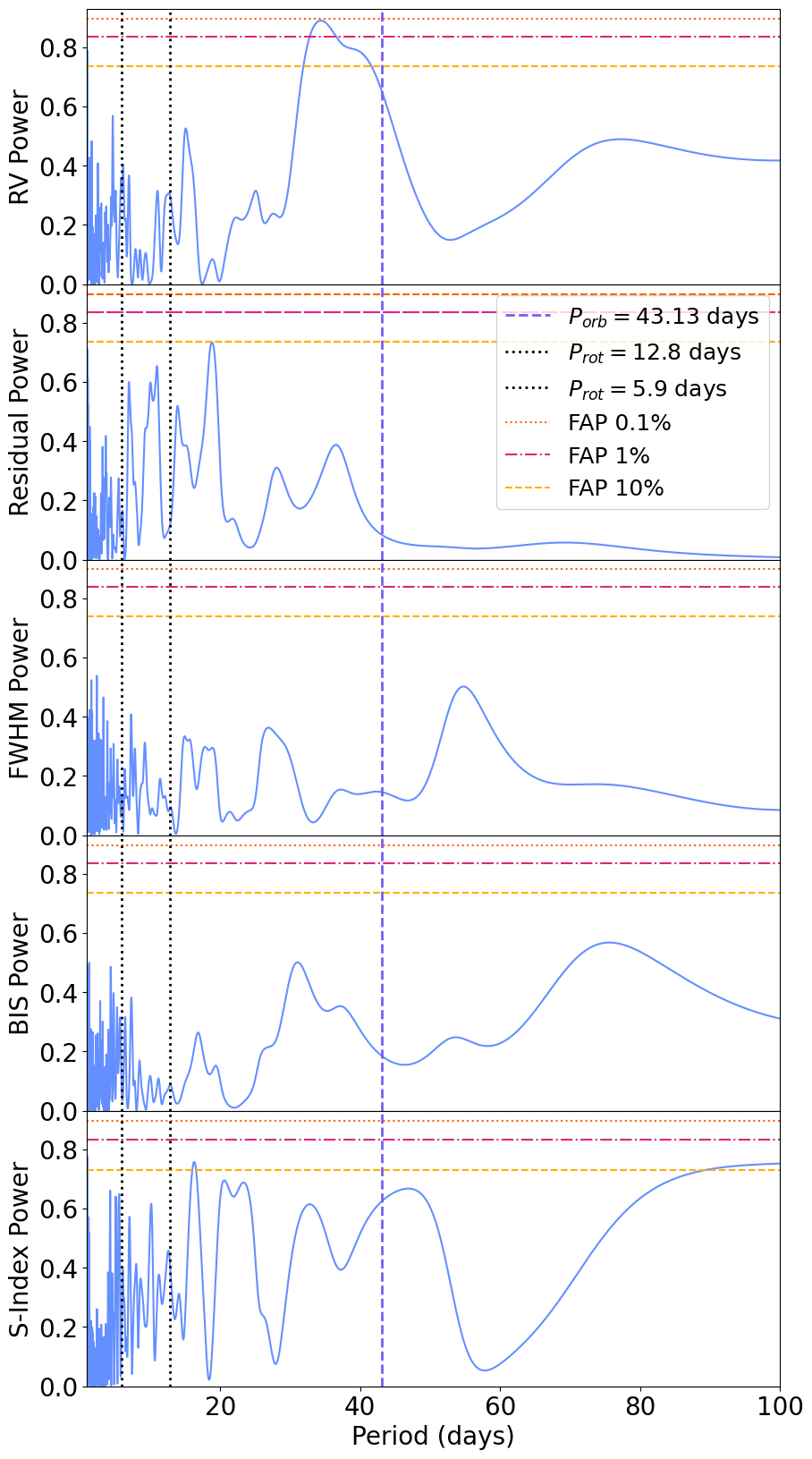}
     \end{subfigure}
     \caption{\textit{Top:} Radial velocity observations of \objonengts\ with residuals, from both \coralie~(yellow squares) and \harps~(blue circles), phase-folded on the orbital period from the joint fit ($P = \objonep~$days). The red noise models subtracted were a hybrid spline for \harps\ and a flat offset of $13.9597\pm0.0019~\mathrm{km/s}$ for \coralie. The solid line is the posterior median model from the joint fit. \textit{Bottom:} Normalized Lomb-Scargle Periodograms for \harps\ radial velocities, residuals, FWHM, BIS and $s_{MW}$ from top to bottom. The orbital period from the joint fit is the purple dashed line, and the estimated rotation period is the black dotted line. 0.1, 1 and 10 per cent FAP are shown as orange dotted, pink dash dot and yellow dashed lines.}
    \label{fig:TIC147rv}
    \end{figure}
    
\begin{figure}
     \centering
     \begin{subfigure}[b]{\columnwidth}
         \centering
         \includegraphics[width=0.85\columnwidth]{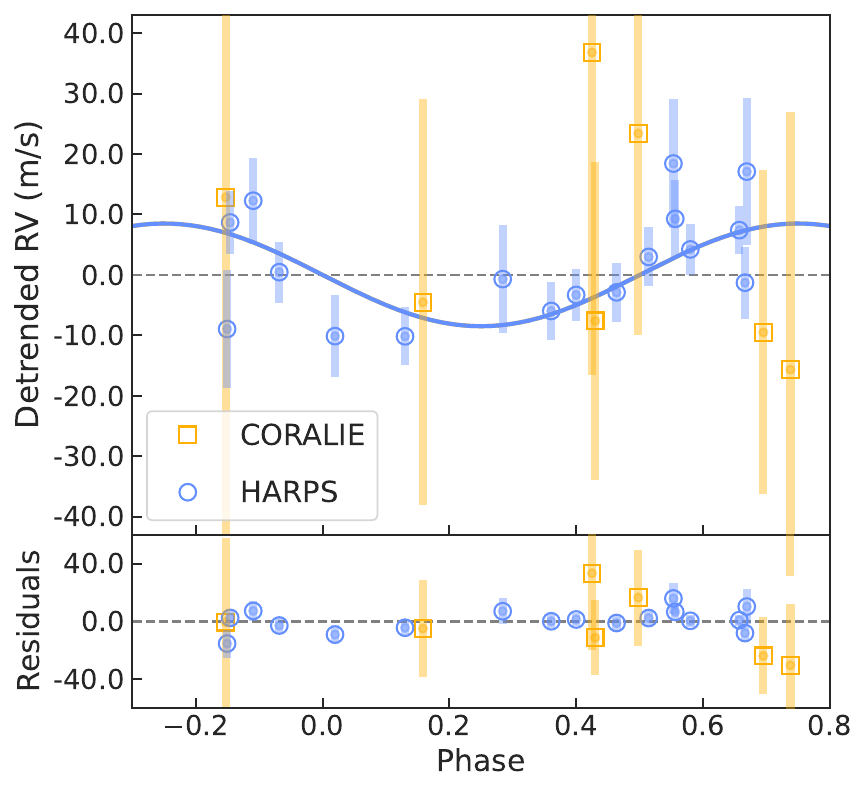}
     \end{subfigure}
     \hfill
     \begin{subfigure}[b]{\columnwidth}
         \centering
         \includegraphics[width=0.85\columnwidth]{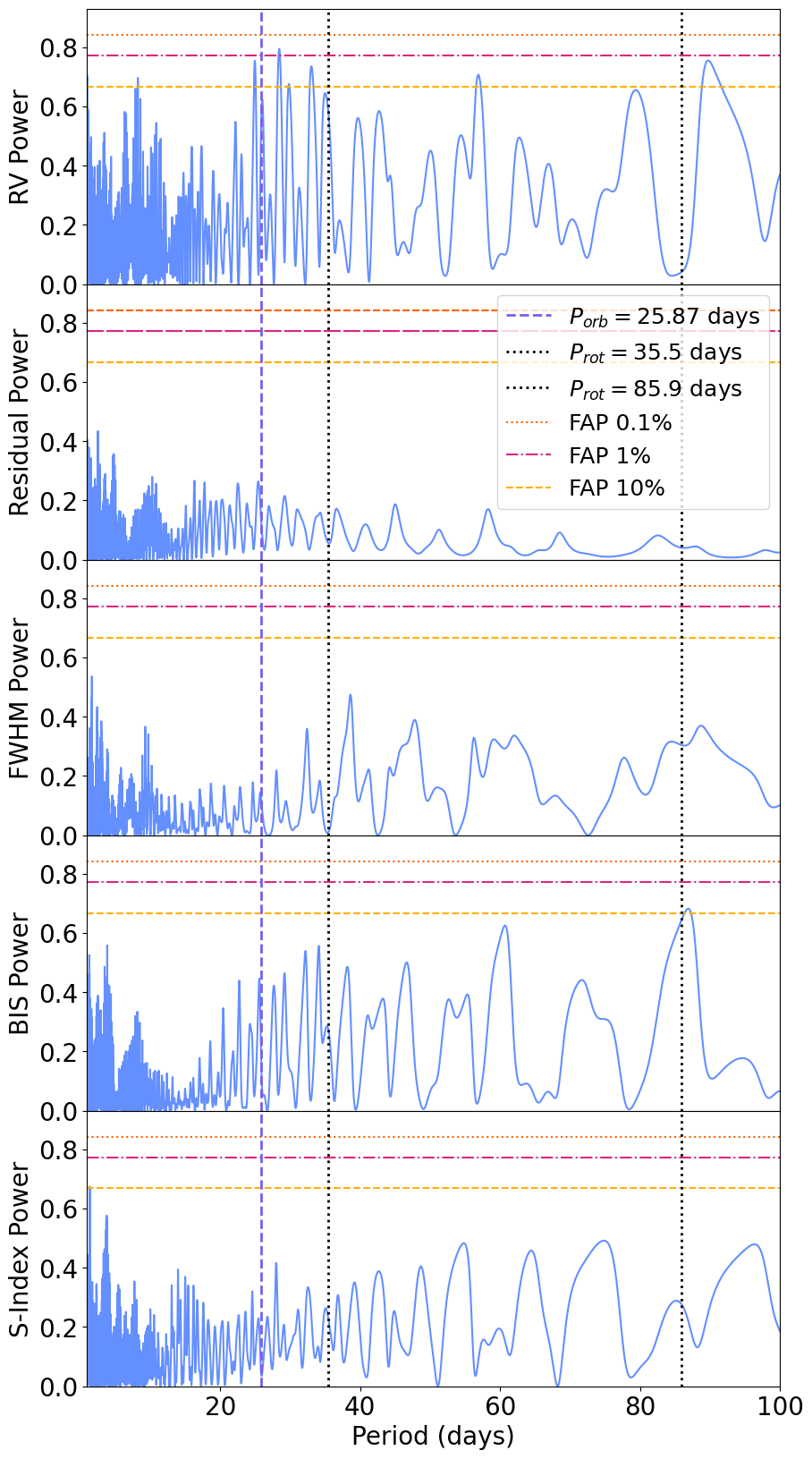}
     \end{subfigure}
     \caption{\textit{Top:} Radial velocity observations of \objtwo\ with residuals, from both \coralie~(yellow squares) and \harps~(blue circles), phase-folded on the orbital period from the joint fit ($P = \objtwop~$days). The red noise models subtracted were a hybrid spline for \harps\ and a flat offset of $41.188\pm0.013~\mathrm{km/s}$ for \coralie. The solid line is the posterior median model from the joint fit. \textit{Bottom:} Normalized Lomb-Scargle Periodograms for \harps\ radial velocities, residuals, FWHM, BIS and $s_{MW}$ from top to bottom. The orbital period from the joint fit is the purple dashed line, and the estimated rotation period is the black dotted line. 0.1, 1 and 10 per cent FAP are shown as orange dotted, pink dash dot and yellow dashed lines.}
     \label{fig:TIC237rv}
    \end{figure}

\begin{figure}
     \centering
     \begin{subfigure}[b]{\columnwidth}
         \centering
         \includegraphics[width=0.85\columnwidth]{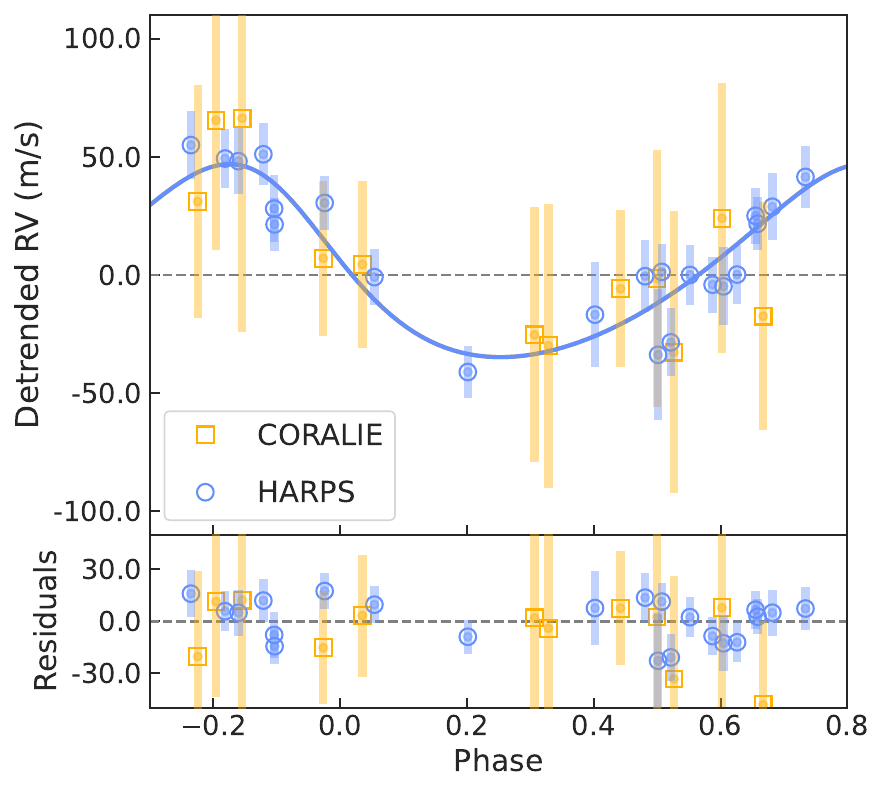}
     \end{subfigure}
     \hfill
     \begin{subfigure}[b]{\columnwidth}
         \centering
         \includegraphics[width=0.85\columnwidth]{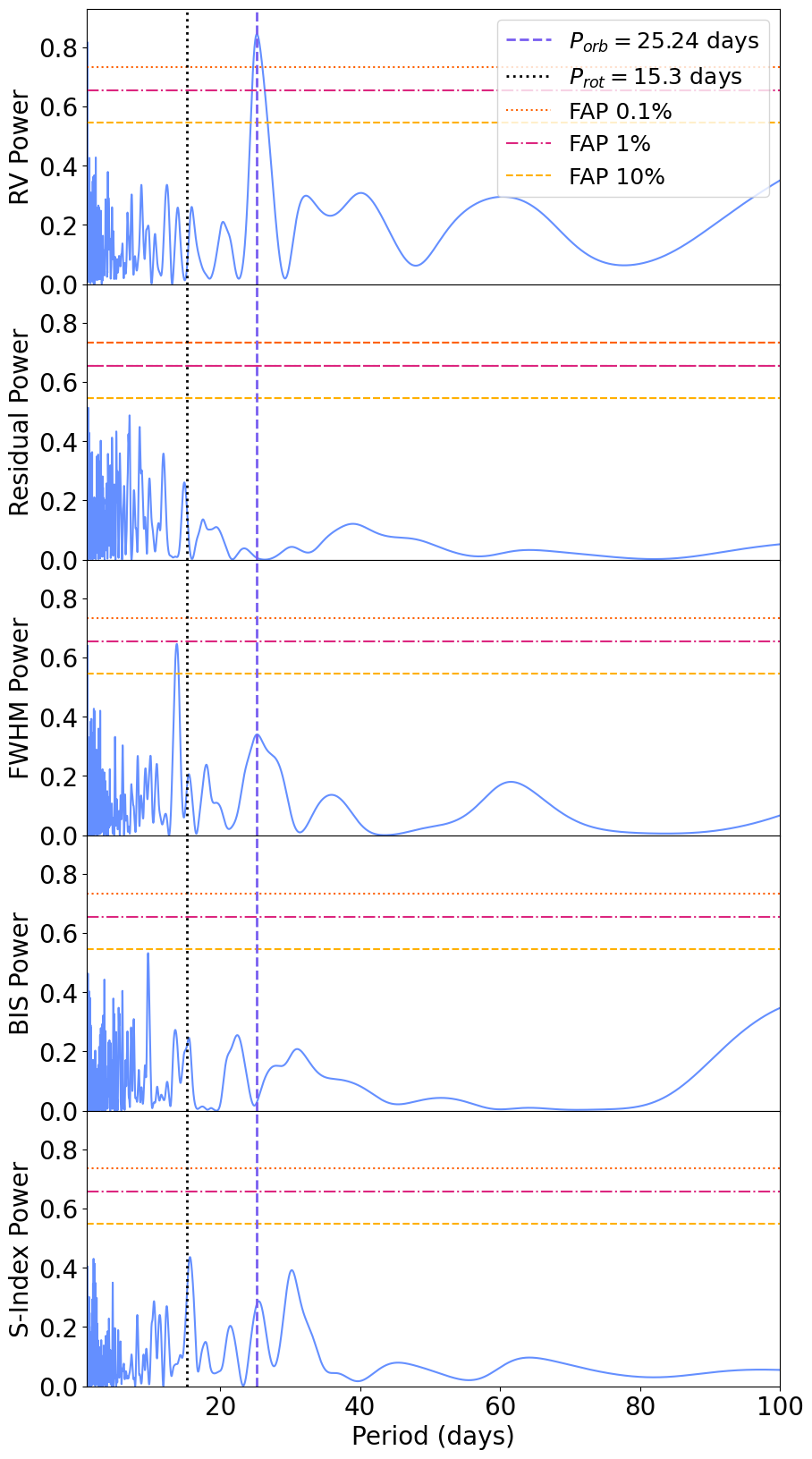} 
     \end{subfigure}
     \caption{\textit{Top:} Radial velocity observations of \objthreengts\ with residuals, from both \coralie~(yellow squares) and \harps~(blue circles), phase-folded on the orbital period from the joint fit ($P = \objthreep~$days). The red noise models subtracted were flat offsets ($37.9606\pm0.0029~\mathrm{km/s}$ for \harps, $37.972\pm0.013~\mathrm{km/s}$ for \coralie) and the solid line is the posterior median model from the joint fit. \textit{Bottom:} Normalized Lomb-Scargle Periodograms for \harps\ radial velocities, residuals, FWHM, BIS and $s_{MW}$ from top to bottom. The orbital period from the joint fit is the purple dashed line, and the estimated rotation period is the black dotted line. 0.1, 1 and 10 per cent FAP are shown as orange dotted, pink dash dot and yellow dashed lines.}
     \label{fig:TIC333rv}
    \end{figure}
Before using the radial velocity data for each object in the fit, we assess whether it is necessary to detrend the data to remove spurious signals and trends introduced by stellar activity. We can use activity indicators, such as the Full Width Half Maximum (FWHM) of the cross correlation function (CCF), the Bisector Span (BIS) or the S-Index ($s_{MW}$), to try to interpret or lessen effects on radial velocity measurements from stellar activity. 

For each object, we calculate a weighted Pearson correlation coefficient with a p-value derived from 2-tailed permutation tests for radial velocity against each of the activity indicators previously mentioned using the \harps\ data for each object. For \objtwo\ and \objthreengts, we find weak to no correlation for all activity indicators with p-values greater than 34 per cent, which suggests that these correlations are not statistically significant. The strongest correlations for these objects are $|R| = 0.35$ with a p-value of 42 per cent for $s_{MW}$ for \objtwo\,b and $|R| = 0.27$ with a p-value of 34 per cent for FWHM for \objthreengts. For \objonengts, there is very weak correlation for FWHM and $s_{MW}$ with p-values greater than 43 per cent, but the BIS span shows a moderate correlation of $|R| = 0.48$ with a p-value of 9 per cent. However, this is not smaller than commonly used critical values of 5 or 1 per cent. A heatmap of correlation for each object's \harps\ radial velocities and activity indicators can be found in the supplementary material.

We search for periodic trends in the radial velocity and activity indicators for each object from \harps\ through producing a normalised generalised Lomb-Scargle periodogram of each dataset (Figures~\ref{fig:TIC147rv}, \ref{fig:TIC237rv}, and \ref{fig:TIC333rv}). Using a fine grid taking 50000 uniformly spaced points between 1 and 100 days, we do not find any significant periodic signals in the activity indicators at less than 1 per cent false-alarm probability (FAP).

For the radial velocity periodogram for \objonengts\,b, we find the strongest peak at about 34.4 days at less than 0.15 per cent FAP, staying above 5 per cent over a wide range of days. For \objonengts\,b, the fitted orbital period is inside this wide peak, where it dips below 10 per cent FAP at longer periods. This unclear signal could be due to having a smaller semi-amplitude signal compared to the noise in the radial velocities and having few points over which to perform periodogram analysis. We do not see signals under 5 per cent FAP in the residuals or activity indicators. The strongest peaks are a 6.1 per cent FAP peak in the S-Index at 16.3~days and a 9.4 per cent FAP peak in the residuals at 18.8~days. However, we see no clear evidence either of the estimated rotation periods from Section~\ref{sec:activity}.

The strongest peak is at 28.4 days for \objtwo\,b at 0.5 per cent FAP. For \objtwo\,b, the orbital period is a non-significant peak amongst a cluster of peaks between 25 and 35 days. This discrepancy is likely due to having fewer radial velocity points spread over a longer timescale, detailed in Table~\ref{tab:TIC237timeseries}. We see a wide peak in the BIS at 86.9~days and 7.6 per cent FAP, which may contain the estimated rotation period at 85.9~days despite the range of stellar metallicities included for the coefficients. We do not see any clear evidence of a rotation period at 35.5~days.

We find one strong peak at 25.3 days for \objthreengts\,b at less than 0.001 per cent FAP. From sufficient data points and a higher semi-amplitude, this well-defined peak period corresponds with the fitted period in Table~\ref{tab:planetparams}. Across all three planets, the most significant peak in activity indicators is in the FWHM for \objthreengts\,b. This peak at 13.8 days and 1.2 per cent FAP does not correspond with any equivalent significant peak in the radial velocity periodogram, but it is however close to half the orbital period, so activity may have had some impact on the radial velocity signal. Additionally, there is a low significance peak in the BIS at 9.7~days and 12.9 per cent FAP. The tallest peak for the S-Index is at 15.7~days at 50.3 per cent FAP, which is non-significant but may correspond with the $\log{R'_{hk}}$ estimated rotation period of 15.3~days.

Combining this analysis with the previous activity analysis in Section~\ref{sec:activity}, and considering whether the number of radial velocity points available would allow us to accurately and meaningfully detrend the data, we do not undertake any additional more complex detrending for the radial velocity points such as including an additional stellar variability function common between instruments on top of the red noise model. However, we note that stellar activity may have had some impact on the final mass derivations.

\subsubsection{Eccentricity \& RV red noise models}
\label{sec:joint}
\begin{figure}
     \centering
     \begin{subfigure}[b]{\columnwidth}
         \centering
         \includegraphics[width=0.9\columnwidth]{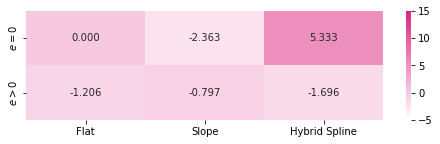}
         \includegraphics[width=0.9\columnwidth]{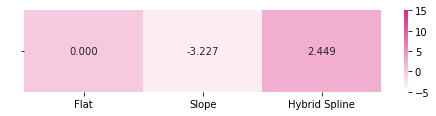}
         \caption{$\ln{R}$ for joint fits of \objonengts\,b.}
         \label{fig:tic147lnZ}
     \end{subfigure}
     \hfill
     \begin{subfigure}[b]{\columnwidth}
         \centering
         \includegraphics[width=0.9\columnwidth]{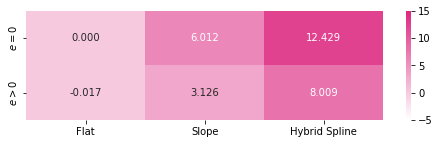}
         \includegraphics[width=0.9\columnwidth]{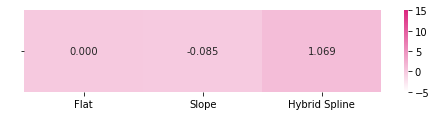}
         \caption{$\ln{R}$ for joint fits of \objtwo\,b.}%
         \label{fig:tic237lnZ}
     \end{subfigure}
     \hfill
     \begin{subfigure}[b]{\columnwidth}
         \centering
         \includegraphics[width=0.9\columnwidth]{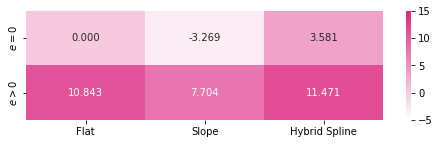}
         \includegraphics[width=0.9\columnwidth]{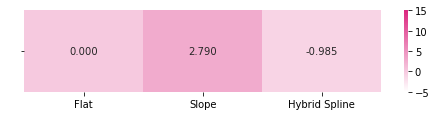}
         \caption{$\ln{R}$ for joint fits of \objthreengts\,b.}
         \label{fig:tic333lnZ}
     \end{subfigure}
        \caption{Heatmaps of $\ln{R}$ values for models used for fitting the three planets in this paper, where the upper left square of each panel is the simplest model. For each object, the first plot includes just \harps\ data, where the top row is assuming a circular orbit, the bottom row allows fitting of parameters such as $\sqrt{e}\cos{\omega}$ and $\sqrt{e}\sin{\omega}$, and the columns refer to the \harps\ red noise model. The second plot includes both \harps\ and \coralie\ data, using the \harps\ red noise model and eccentricity determined from the first plot for each fit. The columns refer to the \coralie\ red noise model. Each fit uses the photometry red noise models discussed in Section~\ref{sec:phot}.}
        \label{fig:lnz}
\end{figure}

For choosing an appropriate radial velocity model, we first perform fits using photometry and only the higher \nicefrac{S}{N} \harps\ data. To determine whether to include eccentricity parameters $\sqrt{e}\cos{\omega}$ and $\sqrt{e}\sin{\omega}$ in the fitted model, or fix the orbit as circular, we first use equation (22) from \citet{lucysweeney}. For \objonengts\,b, both the flat offset and sloped eccentric models produce an eccentricity and associated error that just passes this test. For \objtwo\,b, only the hybrid model satisfies this condition to produce a significant eccentricity. For \objthreengts\,b, each of the fits allowing eccentricity have $e>2.45\sigma_e$ by large margins, easily passing the test, implying an eccentric fit would be highly appropriate for this system. 

To further investigate this and make an informed decision, we can treat each circular model as the simple null model $(M_1)$ and each eccentric model as the more complex model $(M_2)$, then calculate logarithmic Bayesian evidence ($\ln{Z}$) for each model. To decide whether to opt for the more complex model, the logarithmic Bayes Factor $\ln{R}=\ln{Z_{M_2}}-\ln{Z_{M_1}}$ is calculated, shown in Figure~\ref{fig:lnz} with the flat offset circular model as the simplest model. We accept the more complex model if $\ln{R}>4.6$ based on an interpretation of Bayes Factors from \cite{jeffreys} and \cite{kass}. For \objonengts\,b, $\ln{R}$ is only in favour of non-zero eccentricity for the sloped model in Figure~\ref{fig:tic147lnZ}. However, the difference is only $1.57\pm0.25$, which can be interpreted as ``substantial'' evidence but not enough to use the more complex model~\citep{jeffreys,kass,allesfitter}. Considering all red noise models, we adopt the simple circular model and fix eccentricity parameters to zero. For \objtwo\,b, considering Figure~\ref{fig:tic237lnZ}, $\ln{R}$ is in favour of a circular orbit for all red noise models. Again, we choose not to fit for eccentricity and use the simple circular orbit model. However, for \objthreengts\,b, the bottom row is greater than the upper row across all of Figure~\ref{fig:tic333lnZ}. We find $\ln{R}\geq7.89\pm0.23$ when comparing models with the same radial velocity red noise model setting, which is ``decisive'' evidence for the more complex model~\citep{jeffreys,kass,allesfitter}, confirming that eccentricity should be considered for this orbit, therefore we allow \allesfitter\ to fit for $\sqrt{e}\cos{\omega}$ and $\sqrt{e}\sin{\omega}$. 

To ensure the best fit to the data has been performed in terms of red noise models used, since there are few radial velocity points, we take the same logarithmic Bayesian evidence approach as for eccentricity. In terms of model complexity, we trialled a fitted flat offset model for each instrument ($M_1$), a sloped linear model ($M_2$) and a hybrid spline model($M_3$). We did not consider including a GP red noise model in order to not to overfit, since there are at most 14 points per telescope for \objonengts, 17 points for \objtwo, and 22 points for \objthreengts. For \objonengts, when we compare the hybrid spline against the flat offset model after choosing zero eccentricity, we find $\ln{R} = 5.33\pm0.24$ which provides ``decisive'' evidence that selecting a hybrid spline model is the most appropriate model~\citep{jeffreys,kass,allesfitter}. For \objtwo, after choosing a circular model, the hybrid spline version produces a ``decisively'' improved Bayes Factor over both the flat model at $12.43\pm0.19$ and the sloped model by $6.42\pm0.27$, shown in Figure~\ref{fig:tic237lnZ}, hence we choose this model. For \objthreengts, after choosing to fit for eccentricity, only the hybrid spline model surpasses the flat offset Figure~\ref{fig:tic333lnZ}, but only by $0.63\pm0.24$, which is not significant, thus we choose a flat offset. This introduces a caveat for choosing to fit for eccentricity, since \cite{hara2022} show that eccentricity can be exaggerated by the model to account for red noise. However, no significant signals remain in the residuals in Figure~\ref{fig:TIC333rv} and the fitted \harps\ jitter term from this model is far smaller than $K$.

After making the choices above, we perform three further fits for each object, including both \coralie\ and \harps\ data to determine the most appropriate \coralie\ red noise model. At this step, the simplest models in Figure~\ref{fig:lnz} use the previously determined settings for eccentricity and \harps\ red noise model plus a flat offset red noise model for \coralie. Comparing the logarithmic Bayesian evidence in the same way as above, we choose the simpler flat offset for each object since the more complex models do not meet the requirement for ``decisive'' evidence. A full list of priors for the final chosen models can be found in the supplementary material.


\section{Results \& Discussion}
\label{sec:results}
\begin{table*}
	\centering
	\caption{Fitted and derived parameters from joint fits for \objonengts\,b, \objtwo\,b and \objthreengts\,b using \allesfitter. $T_0$ and $P$ are fixed from previous fits of photometric data only.}
	\label{tab:planetparams}
	\begin{tabular}{lccc}
		\hline
		\textit{Fitted Parameters} & \objonengts & \objtwo & \objthreengts\\
		\hline
        $(R_\star + R_p) / a$ & $0.0253_{-0.0012}^{+0.0015}$ & $0.0344_{-0.0018}^{+0.0023}$ & $0.02496_{-0.00096}^{+0.0011}$\\ \vspace{1pt plus 1pt minus 1pt}         
        $R_p / R_\star$ & $0.0245\pm0.0010$ & $0.0521\pm0.0015$ & $0.1317\pm0.0023$\\ \vspace{1pt plus 1pt minus 1pt}          
        $\cos{i}$ & $0.0197_{-0.0015}^{+0.0018}$ & $0.0267_{-0.0022}^{+0.0026}$ & $0.0207_{-0.0016}^{+0.0021}$\\ \vspace{1pt plus 1pt minus 1pt}         
        $T_{0}$ ($\mathrm{BJD}$) & $2459282.7925_{-0.0016}^{+0.0024}$ & $2459265.3032_{-0.0013}^{+0.0014}$ & $2459334.02963_{-0.00056}^{+0.00053}$\\ \vspace{1pt plus 1pt minus 1pt}
        $P$ ($\mathrm{d}$) & $\objonep$ & $\objtwop$ & $\objthreep$\\ \vspace{1pt plus 1pt minus 1pt}          
        $\sqrt{e}\cos{\omega}_p$ & 0 (fixed) & 0 (fixed) & $0.262\pm0.095$ \\ \vspace{1pt plus 1pt minus 1pt}         
        $\sqrt{e}\sin{\omega}_p$ & 0 (fixed) & 0 (fixed) & $0.342_{-0.080}^{+0.065}$\\ \vspace{1pt plus 1pt minus 1pt}         
        $K$ ($\mathrm{m/s}$) & $2.97\pm0.66$ & $8.5_{-3.2}^{+3.0}$ & $40.8\pm4.2$\\ \vspace{1pt plus 1pt minus 1pt}         
        $q_{1;\mathrm{NGTS}}$ & $0.3527\pm0.0018$ & $0.3928\pm0.0019$ & $0.4619\pm0.0035$\\ \vspace{1pt plus 1pt minus 1pt}         
        $q_{2;\mathrm{NGTS}}$ & $0.385_{-0.018}^{+0.017}$ & $0.414\pm0.020$ & $0.449_{-0.027}^{+0.025}$\\ \vspace{1pt plus 1pt minus 1pt}         
        $q_{1;\mathrm{ASTEP}}$ & & $0.2919\pm0.0013$ & \\ \vspace{1pt plus 1pt minus 1pt} 
        $q_{2;\mathrm{ASTEP}}$ & & $0.394\pm0.018$ & \\ \vspace{1pt plus 1pt minus 1pt}         
        $q_{1;\mathrm{TESS-SPOC~1800s}}$ & $0.2988\pm0.0014$ & $0.3377\pm0.0016$ & $0.3999\pm0.0030$\\ \vspace{1pt plus 1pt minus 1pt}         
        $q_{2;\mathrm{TESS-SPOC~1800s}}$ & $0.375\pm0.017$ & $0.401\pm0.019$ & $0.425\pm0.025$ \\ \vspace{1pt plus 1pt minus 1pt}         
        $q_{1;\mathrm{TESS-SPOC~600s}}$ & $0.2986_{-0.0014}^{+0.0015}$ & $0.3374\pm0.0016$ & \\ \vspace{1pt plus 1pt minus 1pt}         
        $q_{2;\mathrm{TESS-SPOC~600s}}$ & $0.375\pm0.018$ & $0.403\pm0.020$ & \\ \vspace{1pt plus 1pt minus 1pt}         
        $q_{1;\mathrm{TESS-SPOC~200s}}$ & $0.2987\pm0.0014$ &  & \\ \vspace{1pt plus 1pt minus 1pt}         
        $q_{2;\mathrm{TESS-SPOC~200s}}$ & $0.378\pm0.017$ &  & \\ \vspace{1pt plus 1pt minus 1pt}         
        $q_{1;\mathrm{SPOC 120s}}$ & & $0.3376\pm0.0017$ & $0.4008\pm0.0030$\\ \vspace{1pt plus 1pt minus 1pt}         
        $q_{2;\mathrm{SPOC 120s}}$ & & $0.402\pm0.019$ & $0.428\pm0.024$ \\ \vspace{1pt plus 1pt minus 1pt}         
        $D_{\mathrm{NGTS}}$ & $0.05126\pm0.00029$ & 0 (fixed) & 0 (fixed) \\ \vspace{1pt plus 1pt minus 1pt}          
        $D_{\mathrm{ASTEP}}$ & & 0 (fixed) &  \\ \vspace{1pt plus 1pt minus 1pt}          
        $D_{\mathrm{TESS-SPOC~1800s}}$ & $-0.007\pm0.046$ & $0.017\pm0.042$ & $0.017\pm0.037$\\ \vspace{1pt plus 1pt minus 1pt}          
        $D_{\mathrm{TESS-SPOC~600s}}$ & $0.016\pm0.045$ & $0.024\pm0.040$ & \\ \vspace{1pt plus 1pt minus 1pt}         
        $D_{\mathrm{TESS-SPOC~200s}}$ & $0.002\pm0.045$ &  & \\ \vspace{1pt plus 1pt minus 1pt}         
        $D_{\mathrm{SPOC 120s}}$ & & $-0.016\pm0.043$  & $-0.018_{-0.035}^{+0.032}$\\ 
        $\mathrm{Offset_{\coralie}}$ ($\mathrm{km/s}$) & $13.9597\pm0.0019$ & $41.188\pm0.013$ & $37.972\pm0.013$ \\ 
        $\mathrm{Offset_{\harps}}$ ($\mathrm{km/s}$)& &  & $37.9606\pm0.0029$ \\ 
        \hline
        {\textit{Derived parameters}}&&&\\ 
        \hline
        $R_\mathrm{p}$ ($\mathrm{R_{\oplus}}$) & $\objoner$ & $\objtwor$ & $\objthreer$\\ \vspace{1pt plus 1pt minus 1pt}         
        $a$ (AU) & $0.257\pm0.018$ & $0.165\pm0.012$ & $0.160\pm0.011$ \\ \vspace{1pt plus 1pt minus 1pt}         
        $i$ (deg) & $88.874_{-0.11}^{+0.086}$ & $88.47_{-0.15}^{+0.13}$ & $88.816_{-0.12}^{+0.091}$\\ \vspace{1pt plus 1pt minus 1pt}          
        $T_\mathrm{tot}$ (h) & $5.255_{-0.086}^{+0.078}$ & $4.291_{-0.080}^{+0.087}$ & $2.958\pm0.047$\\ \vspace{1pt plus 1pt minus 1pt}         
        $T_\mathrm{full}$ (h) & $4.587_{-0.10}^{+0.090}$ & $3.11_{-0.17}^{+0.13}$ & $1.153\pm0.098$\\ \vspace{1pt plus 1pt minus 1pt}         
        $e$ & & & $\objthreee$\\ \vspace{1pt plus 1pt minus 1pt}         
        $\omega$ (deg) & & & $38_{-15}^{+16}$\\ \vspace{1pt plus 1pt minus 1pt}         
        $b$ & $0.796\pm0.026$ & $0.817\pm0.027$ & $0.812_{-0.012}^{+0.010}$\\ \vspace{1pt plus 1pt minus 1pt}          
        $q$ & $0.000046_{-0.000010}^{+0.000011}$ & $0.000123\pm0.000046$ & $0.000582_{-0.000068}^{+0.000076}$\\ \vspace{1pt plus 1pt minus 1pt}          
        $M_\mathrm{p}$ ($\mathrm{M_{\oplus}}$) & $\objonem$ & $\objtwom$ & $\objthreem$\\ \vspace{1pt plus 1pt minus 1pt}          
        $T_\mathrm{eq}$ (K) & $624_{-16}^{+19}$ & $643_{-18}^{+21}$ & $\objthreet$\\ \vspace{1pt plus 1pt minus 1pt}          
        $\delta_\mathrm{tr; undil;NGTS}$ (ppt) & $0.582\pm0.044$ & $2.54_{-0.11}^{+0.13}$ & $15.98_{-0.52}^{+0.54}$ \\ \vspace{1pt plus 1pt minus 1pt}         
        $\delta_\mathrm{tr; dil;NGTS}$ (ppt) & $0.552\pm0.042$ & $2.54_{-0.11}^{+0.13}$ & $15.98_{-0.52}^{+0.54}$ \\ \vspace{1pt plus 1pt minus 1pt}         
        $\delta_\mathrm{tr; undil;ASTEP}$ (ppt) &  & $2.58_{-0.12}^{+0.13}$ &  \\ \vspace{1pt plus 1pt minus 1pt}         
        $\delta_\mathrm{tr; dil;ASTEP}$ (ppt) &  & $2.58_{-0.12}^{+0.13}$ &  \\ \vspace{1pt plus 1pt minus 1pt}         
        $\delta_\mathrm{tr; undil;TESS-SPOC 1800s}$ (ppt) & $0.581_{-0.049}^{+0.052}$ & $2.56_{-0.16}^{+0.18}$ & $16.15_{-0.79}^{+0.86}$\\ \vspace{1pt plus 1pt minus 1pt}  
        $\delta_\mathrm{tr; dil;TESS-SPOC 1800s}$ (ppt) & $0.585\pm0.043$ & $2.52_{-0.11}^{+0.12}$ & $15.84_{-0.52}^{+0.62}$\\ \vspace{1pt plus 1pt minus 1pt}    
        $\delta_\mathrm{tr; undil;TESS-SPOC 600s}$ (ppt) & $0.583_{-0.052}^{+0.058}$  & $2.57_{-0.15}^{+0.16}$ &\\ \vspace{1pt plus 1pt minus 1pt}         
        $\delta_\mathrm{tr; dil;TESS-SPOC 600s}$ (ppt) & $0.574_{-0.045}^{+0.049}$ & $2.51_{-0.12}^{+0.11}$  &\\ \vspace{1pt plus 1pt minus 1pt}         
        $\delta_\mathrm{tr; undil;TESS-SPOC 200s}$ (ppt) & $0.583_{-0.056}^{+0.062}$ &  &\\ \vspace{1pt plus 1pt minus 1pt}         
        $\delta_\mathrm{tr; dil;TESS-SPOC 200s}$ (ppt) & $0.582_{-0.050}^{+0.053}$ &  &\\ \vspace{1pt plus 1pt minus 1pt}         
        $\delta_\mathrm{tr; undil;SPOC 120s}$ (ppt) & & $2.56_{-0.15}^{+0.17}$ & $16.14\pm0.65$\\ \vspace{1pt plus 1pt minus 1pt}          
        $\delta_\mathrm{tr; dil;SPOC 120s}$ (ppt) & & $2.60\pm0.12$ & $16.44\pm0.38$\\    
        \hline
	\end{tabular}
\end{table*}

The fitted and derived parameters for each planet from the global fits with the above chosen eccentricities and red noise models can be found in Table~\ref{tab:planetparams}. The radii have been determined to better than 5 per cent; and the masses to within 25 per cent for \objonengts\,b, 40 per cent for \objtwo\,b, and 15 per cent for \objthreengts\,b. The median models resulting from these fits for the \ngts, \tess\ and \astep\ photometry data with the red noise models subtracted for each system are shown on the phase-folded lightcurves in Figures~\ref{fig:TIC147Phot}, \ref{fig:TIC237Phot} and \ref{fig:TIC333Phot}. Similarly, Figures~\ref{fig:TIC147rv},  \ref{fig:TIC237rv} and \ref{fig:TIC333rv} show the radial velocity median models for each target on the phase-folded \coralie\ and \harps\ data, again with red noise subtracted. Posterior distribution corner plots for the fitted parameters for each fit can be found in the supplementary material.

\subsection{Three warm giants}
\label{sec:allthree}
\begin{figure*}
     \includegraphics[width=0.8\linewidth]{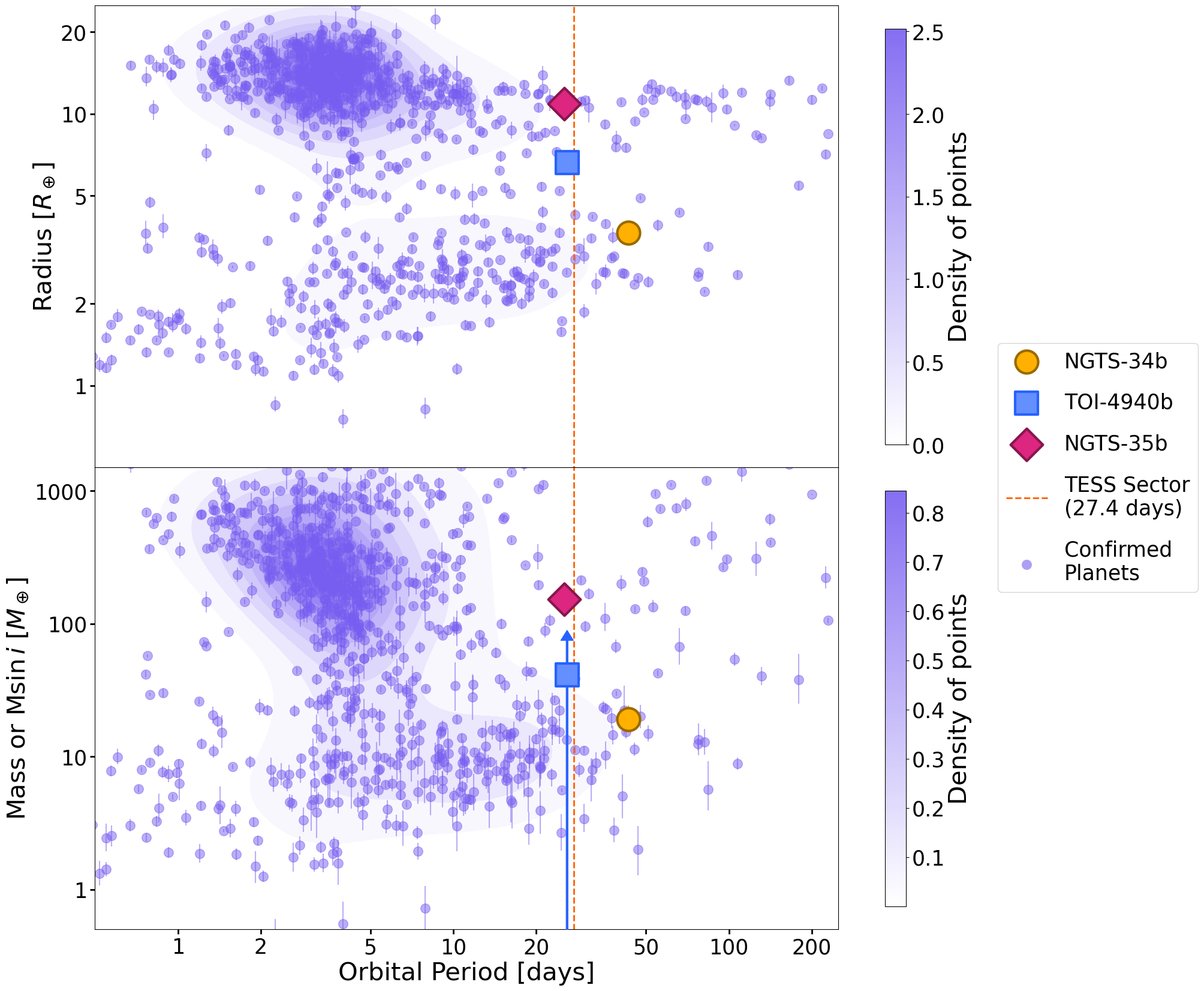}
    \caption{Distributions of known exoplanets between 0.5 and 250 days, and $0.5-25~R_\oplus$ (top) or $0.5-1500~M_\oplus$ (bottom). The purple dots represent confirmed planets listed on the NASA Exoplanet Archive\protect\footref{nasaarchive}, limited to transiting planets with periods defined to better than 6 hours, radii defined to better than 20 per cent, and masses determined via radial velocities to better than 50 per cent. The purple colour bar and associated contours represent the number of planets per element of the meshgrid used ($500\times500$), which has been spaced uniformly within the log equivalent of the dimensions listed above for each subfigure. The yellow circle represents \objonengts\,b, the blue square \objtwo\,b with the arrow representing the upper mass limit, and the pink diamond \objthreengts\,b. The parameters for each planet can be found in full in Table~\ref{tab:planetparams}. The dashed orange line represents the length of a \tess\ sector. }
        \label{fig:RMvsP}
\end{figure*}

We compare \objonengts\,b, \objtwo\,b and \objthreengts\,b to confirmed transiting exoplanets from the NASA Exoplanet Archive\footnote{\label{nasaarchive}\url{https://exoplanetarchive.ipac.caltech.edu/} (accessed 06/03/2025)} with well-defined orbital periods ($\sigma_P < 6~$hours), radii ($\sigma_R < 20~$per cent) and masses ($\sigma_M < 50~$per cent which have been measured through radial velocities to ensure accurate mass determination) in Figure~\ref{fig:RMvsP}. Most of the planets that appear in the discovered exoplanet population occupy two higher density shorter period regions: Hot Jupiters and Neptunes/Sub-Neptunes. Together, these three planets occupy a more sparsely populated region of R-P and M-P space, spread between the populations in terms of R or M and at longer periods than these populations.

\begin{figure*}
     \centering
         \includegraphics[width=0.85\linewidth]{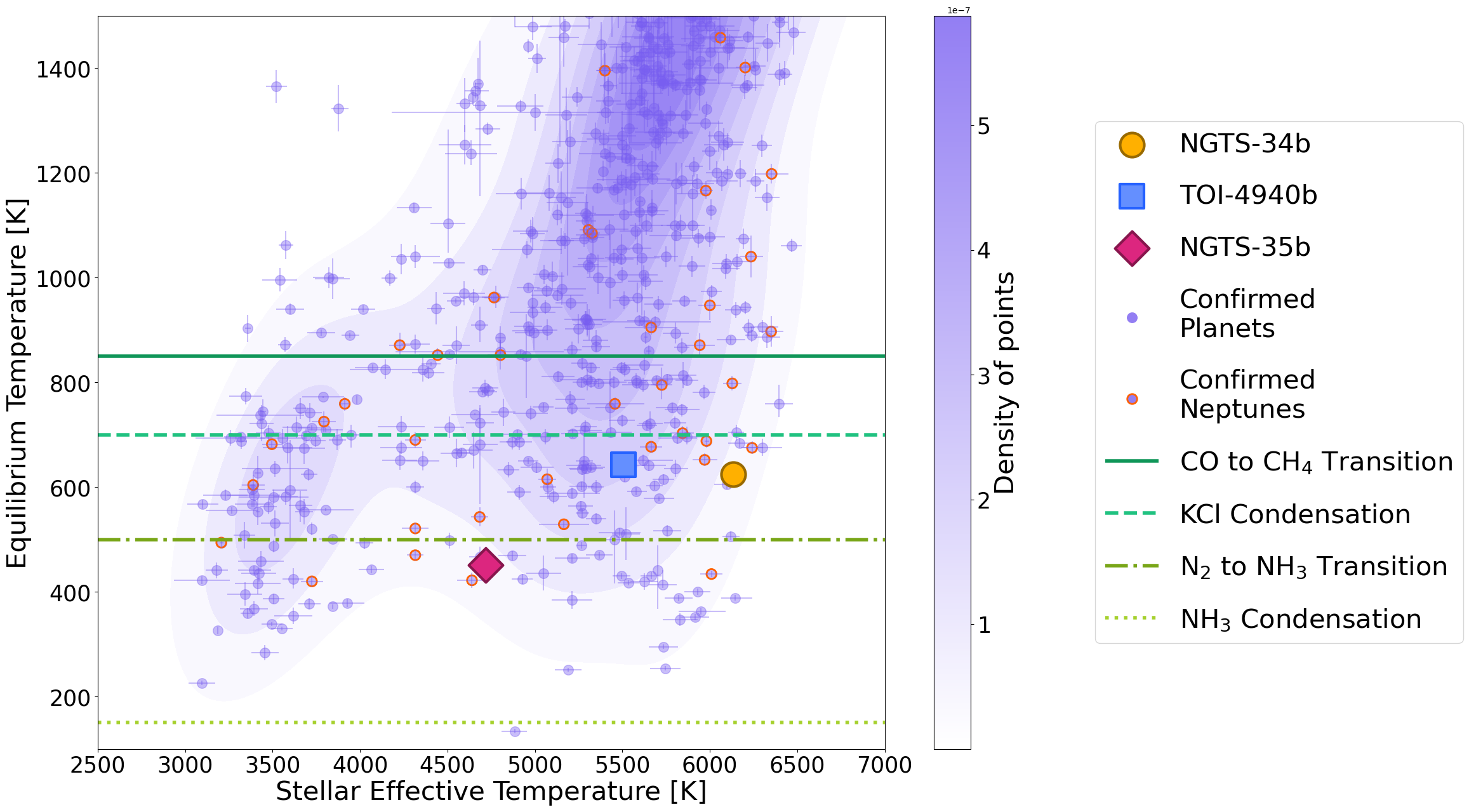}
         \caption{Distribution of known exoplanets with $T_\mathrm{eq,p}$ between 100 and 1500~K, around stars with $T_\mathrm{\rm{eff}}$ between 2500 and 7000~K, limited to transiting planets with periods defined to better than 6 hours, radii defined to better than 20 per cent, and masses determined via radial velocities to better than 50 per cent. The purple dots represent confirmed planets listed on the NASA Exoplanet Archive\protect\footref{nasaarchive}, with Neptunes($3~\mathrm{R_{\oplus}}$ - $0.5~\mathrm{R_{J}}$ and $10~\mathrm{M_{\oplus}}$ - $50~\mathrm{M_{\oplus}}$) circled in orange. The purple colour bar and associated contours representing the number of planets per element of the meshgrid used ($500\times500$), which has been uniformly spaced within the parameter space. In order of temperature, transition lines are plotted as: CO to CH\textsubscript{4} at 850~K as a solid line, KCl condensation at 700~K as a dashed line, N\textsubscript{2} to NH\textsubscript{3} at 500~K as a dash-dotted line, and NH\textsubscript{3} condensation at 150K as a dotted line~\citep{transitions}. Again, the yellow circle represents \objonengts\,b, the blue square \objtwo\,b, and the pink diamond \objthreengts\,b, with parameters listed in full in Table~\ref{tab:planetparams}.}
         \label{fig:TvsST}
\end{figure*}

A key derived parameter for these three planets is their relatively cool equilibrium temperatures under 700~K, especially \objthreengts\,b which is discussed further in Section~\ref{sec:TIC333}. This renders them assets for cool atmosphere studies, since they are all in the region where CH\textsubscript{4} dominates over CO and KCl condenses, shown in Figure~\ref{fig:TvsST}. In this temperature regime, they could be missing the mineral clouds that can mute signals for HJs, allowing investigation into links between these planets' atmospheres and our own Solar System. However, note that \allesfitter\ estimates equilibrium temperature through
\begin{equation}
\label{eq:teq}
    T_\mathrm{eq,p} = T_{\mathrm{\rm{eff}},\star} \cdot \frac{(1-A)^{1/4}}{E} \cdot \sqrt{\frac{R_{\star}}{2a}}
\end{equation}
assuming a bond albedo $A = 0.3$ (comparable to gas giants in our own Solar System ie. \citet{jupiteralbedo,neptunealbedo}) and an emissivity $E = 1$~\citep[as is default in][]{allesfittercode,allesfitter}, which may not necessarily be the case.

\subsection{\objonengts\,b}
\label{sec:TIC147}
\begin{figure}
         \centering
         \includegraphics[width=0.9\columnwidth]{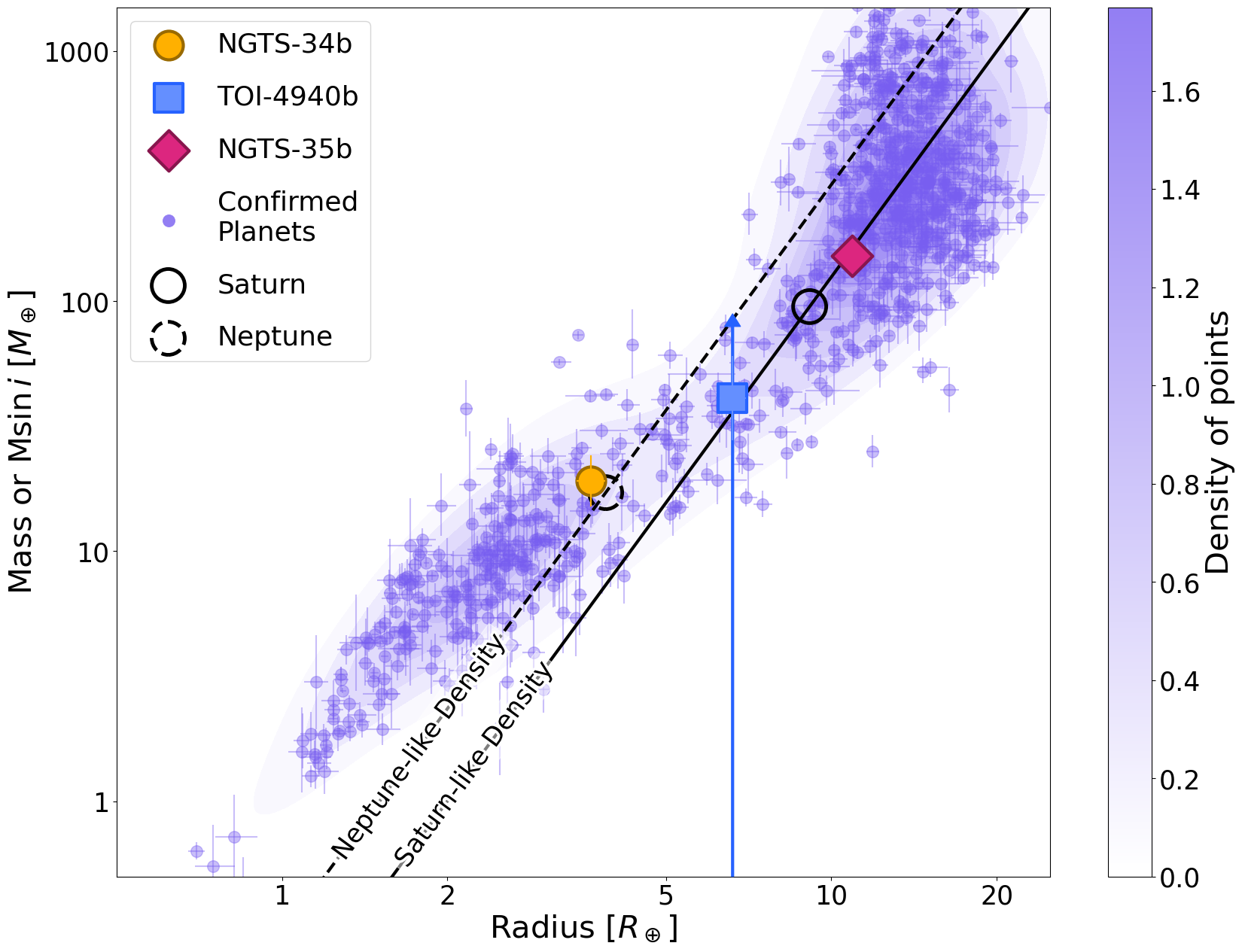}
         \caption{Distributions of known exoplanets between $0.5-25~R_\oplus$ and $0.5-1500~M_\oplus$. The purple dots represent confirmed planets listed on the NASA Exoplanet Archive\protect\footref{nasaarchive}, limited to transiting planets with periods defined to better than 6 hours, radii defined to better than 20 per cent, and masses determined via radial velocities to better than 50 per cent. The purple colour bar and associated contours represent the number of planets per element of the meshgrid used ($500\times500$), which has been spaced uniformly within the log equivalent of the dimensions listed above. The yellow circle represents \objonengts\,b, the blue square \objtwo\,b with the arrow representing the upper mass limit, and the pink diamond \objthreengts\,b, with parameters listed in full in Table~\ref{tab:planetparams}. The solid black ring represents Saturn and the solid black line is a constant density equal to Saturn's density. The dashed black ring and line are the equivalents for Neptune.}
         \label{fig:densityplot}
\end{figure}

\begin{figure*}
     \centering
    \includegraphics[width=0.8\linewidth]{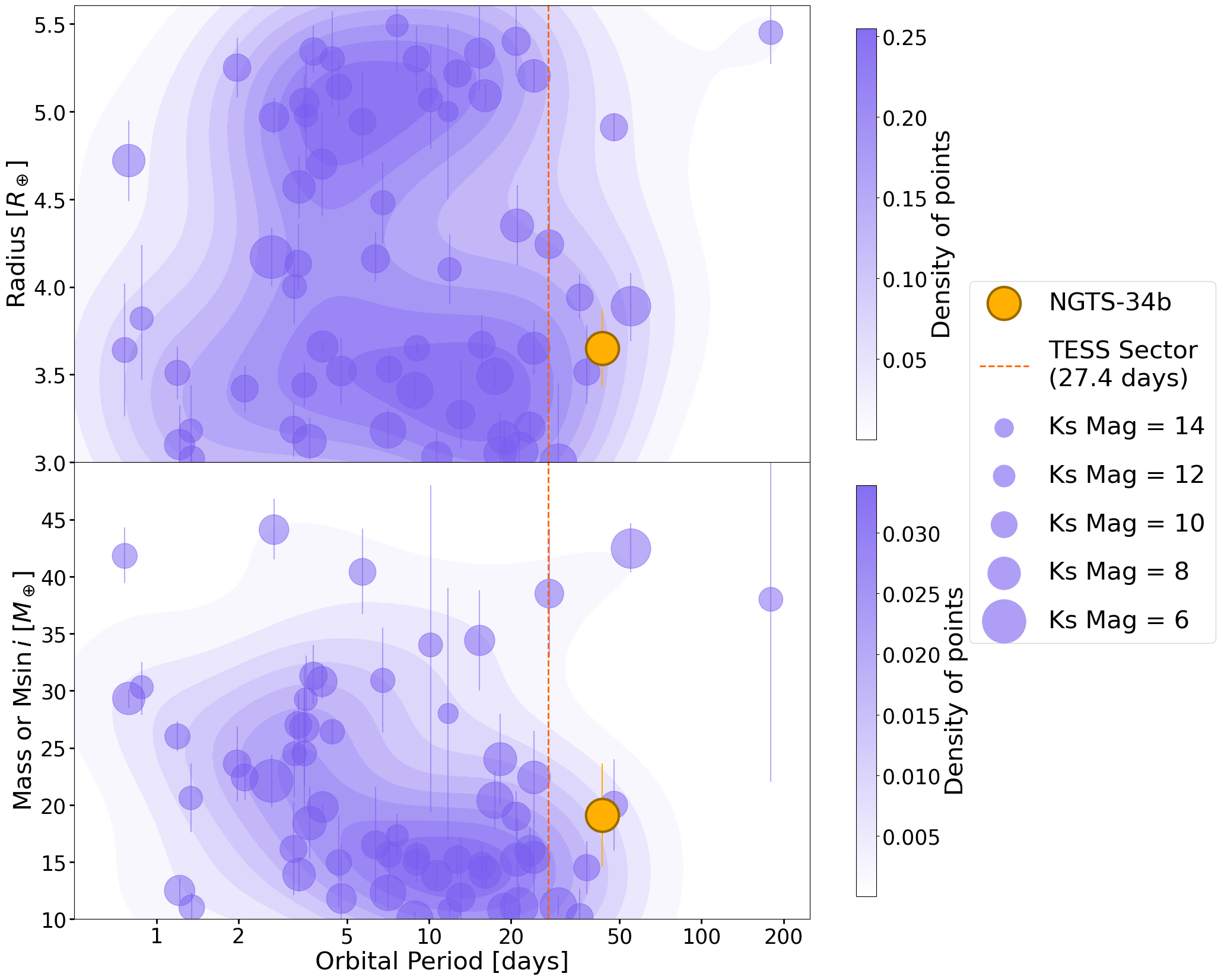}
    \caption{Similar to Figure~\ref{fig:RMvsP}, restricted to Neptunes($3~\mathrm{R_{\oplus}}$ - $0.5~\mathrm{R_{J}}$ and $10~\mathrm{M_{\oplus}}$ - $50~\mathrm{M_{\oplus}}$). The yellow circle with errorbars is again \objonengts\,b, with parameters listed in full in Table~\ref{tab:planetparams}. The size of each point is scaled by the inverse square of the host star's 2MASS $K_s$ magnitude, such that systems with brighter stars are larger.}
    \label{fig:TIC147rmp}
\end{figure*}
     
The joint fit of \objonengts\,b, the smallest and lightest of the three, determines it to be a Neptune with radius $\objoner~\mathrm{R_{\oplus}}$. It has the longest period in this paper at $\objonep~$days. 

It has been fitted with a circular orbit due to having a significantly higher $\ln{R}$ than any other combination. The radial velocity semi-amplitude is $2.97\pm0.66~\mathrm{m/s}$ from the global fit, which leads to a derived mass of $\objonem~\mathrm{M_{\oplus}}$. This means it occupies a very similar parameter space as Neptune, to the point of having a density within $1\sigma$ of Neptune's, which can be seen in Figure~\ref{fig:densityplot}. The hybrid spline red noise model chosen for the \harps\ radial velocities could indicate the presence of an additional planet, some uncharacterized stellar activity or other noise effect from the telescope.

If we consider only known transiting planets with well-defined parameters as before, \objonengts\,b is one of a handful of the longest period transiting Neptunes. This can be seen in Figure~\ref{fig:TIC147rmp}, where we have applied additional cuts by defining a Neptune as having $3~\mathrm{R_{\oplus}} \leq R_p \leq 0.5~\mathrm{R_{J}}$ and $10~\mathrm{M_{\oplus}} \leq M_p \leq 50~\mathrm{M_{\oplus}}$. With these cuts applied, it is in the coolest 25 per cent of this group at an equilibrium temperature of $624_{-16}^{+19}~$K, shown highlighted in orange in Figure~\ref{fig:TvsST}.

Since it orbits a bright star, \objonengts\,b is also amenable to atmospheric spectroscopy. Using values in Table~\ref{tab:stellarparams} and Table~\ref{tab:planetparams} in the framework laid out in \cite{kemptontsm}, we find a transmission spectroscopy metric of 22.5, which is high considering how cool and how far from the host star \objonengts\,b is. With a 2MASS\textit{Ks} magnitude of $7.9250 \pm 0.0210$, it is also one of the larger points in Figure~\ref{fig:TIC147rmp}. This makes it ideal for observing with infrared instruments that can observe chemical compositions such as MIRI~\citep{miri} on JWST or Ariel~\citep{ariel}. As a well-defined transiting cool Neptune on such a long orbit far from a bright host star, \objonengts\,b is a valuable candidate for both cool atmosphere studies and for better understanding cool giant formation, migration and evolution.

\subsection{\objtwo\,b}
\label{sec:TIC237}
Fitting \objtwo\,b shows it is heavier and larger than \objonengts\,b, but still between Neptune and Saturn. It has a radius $\objtwor~\mathrm{R_{\oplus}}$ and a period $\objtwop~$days. The mass from the joint fit is $\objtwomwhole~\mathrm{M_{\oplus}}$ from the radial velocity semi-amplitude $8.5_{-3.2}^{+3.0}~\mathrm{m/s}$. Owing to the large uncertainty, we interpret this as a $3\sigma$ upper mass limit of $\objtwom~\mathrm{M_{\oplus}}$. The hybrid spline model chosen for the higher \nicefrac{S}{N} \harps\ data could again represent some otherwise uncharacterized stellar activity, instrumental noise or an additional companion.

At these masses and radii, \objtwo\,b exists in a sparser area of M-R space for well-characterised transiting planets, sat between two denser regions seen in Figure~\ref{fig:densityplot}. The mass also places it near the middle of the mass range of the desert in \citet{idalin1,idalin2} as discussed in Section~\ref{sec:intro}, and is an example of longer period transiting planet searches filling in the period valley caused by selection effects in this intermediate mass range~\citep{wittenmyerperiodvalley}. For this planet, we calculate a transmission spectroscopy metric of 20.8~\citep{kemptontsm}, meaning it is still feasible to observe with JWST. However, using the upper mass limit, this could be as low as 9.6.

\subsection{\objthreengts\,b / \objthreetoi\,b}
\label{sec:TIC333}
The largest and most massive planet in this paper is \objthreengts\,b, Jupiter sized at $R_p=0.972\pm0.058~R_J$, but less dense at around a Saturn mass with $M_p=0.477_{-0.060}^{+0.068}~M_J$, derived from radial velocity semi-amplitude $K=40.8\pm4.2~\mathrm{m/s}$. After choosing the flat red noise model, there are no significant peaks left in the residuals periodogram in Figure~\ref{fig:TIC333rv}. With a period as long as $\objthreep~$days, this is far from the population of transiting HJs in Figure~\ref{fig:RMvsP}. \objthreengts\,b is also on a moderately eccentric orbit at $e=\objthreee$, which suggests a more dynamic migration mechanism such as high eccentricity migration is involved.  

Shown in Figure~\ref{fig:TvsST}, we can see \objthreengts\,b is approximately 200~K colder than the other two planets in this paper at $T_\mathrm{eq} = \objthreet~$K. In fact, it is in the coolest 6 per cent(i.e. top 45) of known transiting exoplanets with well-defined parameters as described earlier. With this colder equilibrium temperature, just below where NH\textsubscript{3} begins to dominate over N\textsubscript{2}, but above NH\textsubscript{3} condensation, \objthreengts\,b exists in the region where nitrogen abundance is measurable as ammonia in Figure~\ref{fig:TvsST}. For this planet, we determine a transmission spectroscopy metric of 41.4~\citep{kemptontsm}, which is high for a cool near-Jupiter sized planet, especially with such a low equilibrium temperature. As an eccentric very cool Saturn, this is a promising candidate to follow-up with JWST observations, where measurable nitrogen abundance as part of the C/N/O ratio may provide constraints on the distance at which this planet formed.

\subsection{Interior Modelling}
Weakly irradiated transiting exoplanets are ideal objects to explore their interior structure. Whilst we only derive an upper mass limit for \objtwo\,b, we find that \objonengts\,b and \objthreengts\,b are well-characterized exoplanets with precise planetary mass and radius measurements and with some constraints on the age of the host star. Given their equilibrium temperature ranging from 450 to 620~K, we assume that the planets are not inflated and thus there is no degeneracy between the heavy element content and the amount of radius inflation. We use the planetary evolution code \texttt{completo}~\citep{mordasini2012} to model the core and the envelope of the planet along with a semi-grey atmospheric model. We build a grid of evolution models assuming no bloating of the planet and we couple this grid with a Bayesian inference model to estimate the heavy element content. We allow the planetary core mass to vary between 0 and $\rm 10\,M_{\oplus}$. The envelope is composed of hydrogen and helium in which heavy elements, modelled as water, are homogeneously mixed. Water is modelled following the equation of state AQUA2020~\citep{haldemann2020}. Hydrogen and helium are modelled with the equation of state from \cite{chabrierdebras2021}. We find that the heavy elements mass ($M_{Z}$) is equal to $17\pm4$ $M_{\oplus}$ for \objonengts\,b, and $37^{+13}_{-14}$ $M_{\oplus}$ for \objthreengts\,b. We estimate the overall stellar metallicity from the scaling law with the iron abundance ($\rm Z_{\star} = 0.0142 \times 10^{[Fe/H]}$, from \cite{asplund2009}, \cite{millerandfortney}). We then find that the heavy element enrichment ($\rm Z_{p} / Z_{\star} $, where $\rm Z_{p} = \frac{M_{Z}}{M_{P}} $,) is $37\pm10$ for \objonengts\,b, and $13 ^{+5} _{-6}$ for \objthreengts\,b. The less massive planet \objonengts\,b is significantly metal enriched compared to the host star, while the composition of \objthreengts\,b is not significantly enriched.


\section{Conclusions}
\label{sec:conc}
We report three warm giants with equilibrium temperatures below 700K: one of the longest period Neptunes, a planet in the mass desert with a similar density to Saturn and a significantly cool larger Saturn on an eccentric orbit. These three planet confirmations are the result of combining data from several instruments. They were first detected as candidates with few transits in \tess\ data, then supported with ground-based follow-up photometry from the \ngts\ Long Period Planets Programme and \astep\ to locate additional transits to either confirm or refine the orbital period and radial velocity measurements taken by the \harps\ and \coralie\ spectrographs to confirm their planetary nature.

To characterise the host stars of these systems, both spectral fitting of the combined spectra and SED fitting to broadband photometry were used. This revealed \objonengts\ to be an F-type star with radius $1.367\pm0.061~R_\odot$ and mass $1.266_{-0.081}^{+0.063}~M_\odot$, \objtwo\ to be a G-type with radius $1.163_{-0.053}^{+0.054}~R_\odot$ and mass $1.012_{-0.047}^{+0.040}~M_\odot$ and \objthreengts\ to be a K-type with radius $0.753_{-0.053}^{+0.051}~R_\odot$ and mass $0.789_{-0.048}^{+0.035}~M_\odot$. Performing joint modelling on the photometry and radial velocity data derived precise parameters for the three planets. \objonengts\,b has radius $\objoner~R_\oplus$ and mass $\objonem~M_\oplus$, on a circular orbit of 43.1 days with equilibrium temperature of $624_{-16}^{+19}~$K. \objtwo\,b has radius $\objtwor~R_\oplus$ and upper mass limit of $\objtwom~M_\oplus$, on a circular orbit of 25.9 days with equilibrium temperature of $643_{-18}^{+21}~$K. \objthreengts\,b has radius $\objthreer~R_\oplus$ and mass $\objthreem~M_\oplus$, on a orbit of 25.2 days with moderate eccentricity $\objthreee$ and an especially cold equilibrium temperature of $\objthreet~$K.

These three new planets add to the growing number of warm giants, which will help solidify theories of formation and evolution aided by future observations with JWST. Long period planet searches continue to produce interesting candidates and results, with several other recently published papers such as \cite{faithduo} which presents \objthreengts\,b as part of 85 duotransit candidates, or TOI-2447\,b/NGTS-29\,b~\citep{toi2447} and TOI-4862\,b/NGTS-30\,b~\citep{toi4862} which also have equilibrium temperatures below 500~K. 

\section*{Acknowledgements}
This paper includes public data collected by the \tess\ mission. Funding for the \tess\ mission is provided by NASA's Science Mission Directorate. We acknowledge the use of public \tess\ data from pipelines at the \tess\ Science Office and at the \tess\ Science Processing Operations Center. Resources supporting this work were provided by the NASA High-End Computing (HEC) Program through the NASA Advanced Supercomputing (NAS) Division at Ames Research Center for the production of the SPOC data products. This research has made use of the Exoplanet Follow-up Observation Program website, which is operated by the California Institute of Technology, under contract with the National Aeronautics and Space Administration under the Exoplanet Exploration Program.

This work includes data collected under the NGTS project at the ESO La Silla Paranal Observatory. The NGTS facility is operated by the consortium institutes with support from the UK Science and Technology Facilities Council (STFC) under projects ST/M001962/1, ST/S002642/1 and ST/W003163/1.

This work makes use of observations from the ASTEP telescope. ASTEP benefited from the support of the French and Italian polar agencies IPEV and PNRA in the framework of the Concordia station program and from OCA, INSU, Idex UCAJEDI (ANR- 15-IDEX-01) and ESA through the Science Faculty of the European Space Research and Technology Centre (ESTEC). This research also received funding from the European Research Council (ERC) under the European Union's Horizon 2020 research and innovation program (grant agreement No. 803193/BEBOP), from the Science and Technology Facilities Council (STFC; grant No. ST/S00193X/1, ST/W002582/1 and ST/Y001710/1), and from the ERC/UKRI Frontier Research Guarantee programme (CandY/ EP/Z000327/1).

Based on observations obtained at the Southern Astrophysical Research (SOAR) telescope, which is a joint project of the Minist\'{e}rio da Ci\^{e}ncia, Tecnologia e Inova\c{c}\~{o}es (MCTI/LNA) do Brasil, the US National Science Foundation’s NOIRLab, the University of North Carolina at Chapel Hill (UNC), and Michigan State University (MSU).

Some of the observations in this paper made use of the High-Resolution Imaging instrument Zorro and were obtained under Gemini LLP Proposal Number: GN/S-2021A-LP-105. Zorro was funded by the NASA Exoplanet Exploration Program and built at the NASA Ames Research Center by Steve B. Howell, Nic Scott, Elliott P. Horch, and Emmett Quigley. Zorro was mounted on the Gemini South telescope of the international Gemini Observatory, a program of NSF’s NOIR Lab, which is managed by the Association of Universities for Research in Astronomy (AURA) under a cooperative agreement with the National Science Foundation. on behalf of the Gemini partnership: the National Science Foundation (United States), National Research Council (Canada), Agencia Nacional de Investigación y Desarrollo (Chile), Ministerio de Ciencia, Tecnología e Innovación (Argentina), Ministério da Ciência, Tecnologia, Inovações e Comunicações (Brazil), and Korea Astronomy and Space Science Institute (Republic of Korea).

This research has made use of the NASA Exoplanet Archive, which is operated by the California Institute of Technology, under contract with the National Aeronautics and Space Administration under the Exoplanet Exploration Program.

This research made use of Lightkurve, a Python package for Kepler and TESS data analysis ~\citep{lightkurve}.

This publication makes use of The Data \& Analysis Center for Exoplanets (DACE), which is a facility based at the University of Geneva (CH) dedicated to extrasolar planets data visualisation, exchange and analysis. DACE is a platform of the Swiss National Centre of Competence in Research (NCCR) PlanetS, federating the Swiss expertise in Exoplanet research. The DACE platform is available at \url{https://dace.unige.ch}.

AK is supported by a STFC studentship (ST/T506242/1). The contributions at the Mullard Space Science Laboratory by E.M.B. have been supported by STFC through the consolidated grant ST/W001136/1. D.D. acknowledges support from the TESS Guest Investigator Program grant 80NSSC22K0185, and from the NASA Exoplanet Research Program grant 18-2XRP18\_2-0136. B.S.G. was supported by the Thomas
Jefferson Chair for Discovery and Space Exploration at the Ohio State University. MK acknowledges the support of the Natural Sciences and Engineering Research Council of Canada (NSERC), RGPIN-2024-06452. Cette recherche a été financée par le Conseil de recherches en sciences naturelles et en génie du Canada (CRSNG), RGPIN-2024-06452. JSJ greatfully acknowledges support by FONDECYT grant 1240738 and from the ANID BASAL project FB210003. ML acknowledges support of the Swiss National Science Foundation under grant number PCEFP2\_194576. The contribution of ML has been carried out within the framework of the NCCR PlanetS supported by the Swiss National Science Foundation under grants 51NF40\_182901 and 51NF40\_205606.

 For the purpose of open access, the author has applied a Creative Commons Attribution (CC BY) licence to the Author Accepted Manuscript version arising from this submission. This research used the ALICE High Performance Computing facility at the University of Leicester.
\section*{Data Availability}
This paper includes data collected by the \tess\ mission that are publicly available from the Mikulski Archive for Space Telescopes (MAST, \url{https://mast.stsci.edu/portal/Mashup/Clients/Mast/Portal.html}). Public \harps\ data are available from the ESO archive (\url{http://archive.eso.org/cms.html}).
Radial velocity measurements and photometry from \ngts\ and \astep\ will be available from the VizieR archive server hosted by the Université de Strasbourg (\url{cdsarc.u-strasbg.fr}) and the ExoFOP pages for each object (\url{https://exofop.ipac.caltech.edu/tess/}).

\allesfitter, \ariadne, \ldtk, \lightkurve, \specmatch\ and \pysynphot\ are all open-source and public software.

\section*{Affiliations}
\textit{$^{1}$School of Physics and Astronomy, University of Leicester, University Road, Leicester, LE1 7RH, UK\\
$^{2}$Leiden Observatory, Leiden University, P.O. Box 9513, 2300 RA Leiden, The Netherlands\\
$^{3}$Observatoire de Genève, Université de Genève, Chemin Pegasi, 51, 1290 Versoix, Switzerland\\
$^{4}$Space Research and Planetary Sciences, Physics Institute, University of Bern, Gesellschaftsstrasse 6, 3012 Bern, Switzerland\\
$^{5}$Department of Physics, University of Warwick, Gibbet Hill Road, Coventry CV4 7AL, UK\\
$^{6}$Centre for Exoplanets and Habitability, University of Warwick, Gibbet Hill Road, Coventry CV4 7AL, UK\\
$^{7}$Instituto de Astronomía, Universidad Católica del Norte, Angamos 0610, 1270709, Antofagasta, Chile\\
$^{8}$Mullard Space Science Laboratory, University College London, Holmbury St Mary, Dorking, RH5 6NT, UK\\
$^{9}$Institute of Physics, Laboratory of Astrophysics, Ecole Polytechnique Fédérale de Lausanne (EPFL), Observatoire de Sauverny, 1290 Versoix, Switzerland\\
$^{10}$Department of Physics and Astronomy, University of New Mexico, 210 Yale Blvd, Albuquerque, NM 87131, USA\\
$^{11}$NASA Goddard Space Flight Center, 8800 Greenbelt Rd, Greenbelt, MD 20771, USA\\
$^{12}$Department of Physics \& Astronomy, University of Minnesota – Duluth, Duluth, MN 55812, USA\\
$^{13}$U.S. National Science Foundation National Optical-Infrared Astronomy Research Laboratory (NSF NOIRLab), Tucson, AZ, USA\\
$^{14}$Department of Astronomy, The Ohio State University, Columbus, OH, USA\\
$^{15}$Xingming Observatory, Urumqi 830011, China\\
$^{16}$Observatoire de la Côte d'Azur, Université Côte d’Azur, CNRS, Laboratoire Lagrange, Bd de l'Observatoire, CS 34229, 06304 Nice cedex 4, France\\
$^{17}$School of Physics \& Astronomy, University of Birmingham, Edgbaston, Birmingham B15 2TT, United Kingdom\\
$^{18}$Department of Physics, Engineering and Astronomy, Stephen F. Austin State University, 1936 North St, Nacogdoches, TX 75962, USA\\
$^{19}$Department of Physics and Astronomy, The University of North Carolina at Chapel Hill, Chapel Hill, NC 27599-3255, USA\\
$^{20}$NASA Ames Research Center, Moffett Field, CA 94035, USA\\
$^{21}$Caltech/IPAC, Mail Code 100-22, Pasadena, CA 91125, USA\\
$^{22}$Bay Area Environmental Research Institute, Moffett Field, CA 94035, USA\\
$^{23}$SETI Institute, 339 Bernardo, Suite 200, Mountain View, CA 94043 USA\\
$^{24}$Department of Physics and Astronomy, University of British Columbia, 6224 Agricultural Rd., Vancouver, BC V6T 1Z1, Canada\\
$^{25}$Royal Astronomical Society, Burlington House, Piccadilly, London W1J 0BQ, UK\\
$^{26}$Department of Physics and Kavli Institute for Astrophysics and Space Research, Massachusetts Institute of Technology, Cambridge, MA 02139, USA\\
$^{27}$Departmento de Astronomía, Universidad de Chile, Casilla 36-D, Santiago, Chile\\
$^{28}$Centro de Excelencia en Astrofísica y Tecnologías Afines (CATA), Camino El Observatorio 1515, Las Condes, Santiago, Chile\\
$^{29}$Astronomy Unit, Queen Mary University of London, G.O. Jones Building, Bethnal Green, London E1 4NS, United Kingdom\\
$^{30}$European Space Agency (ESA), European Space Research and Technology Centre (ESTEC), Keplerlaan 1, 2201 AZ Noordwijk, The Netherlands\\
$^{31}$Instituto de Estudios Astrofísicos, Facultad de Ingeniería y Ciencias, Universidad Diego Portales, Av. Ejército Libertador 441, Santiago, Chile\\
$^{32}$University Observatory, Faculty of Physics, Ludwig-Maximilians-Universit{\"a}t M{\"u}nchen, Scheinerstr. 1, 81679 Munich, Germany\\
$^{33}$Department of Physics and Astronomy, McMaster University, 1280 Main St W, Hamilton, ON, L8S 4L8, Canada\\
$^{34}$Astrophysics Research Centre, School of Mathematics and Physics, Queen’s University Belfast, Belfast, BT7 1NN, UK}


\bibliographystyle{mnras}
\bibliography{example} 



\newpage
\subfile{supplementary.tex}

\bsp	
\label{lastpage}
\end{document}

%% file: supplementary.tex
\appendix
\onecolumn
\begin{figure*}
\section{Raw Data}
     \centering
         \centering
         \includegraphics[width=1\linewidth]{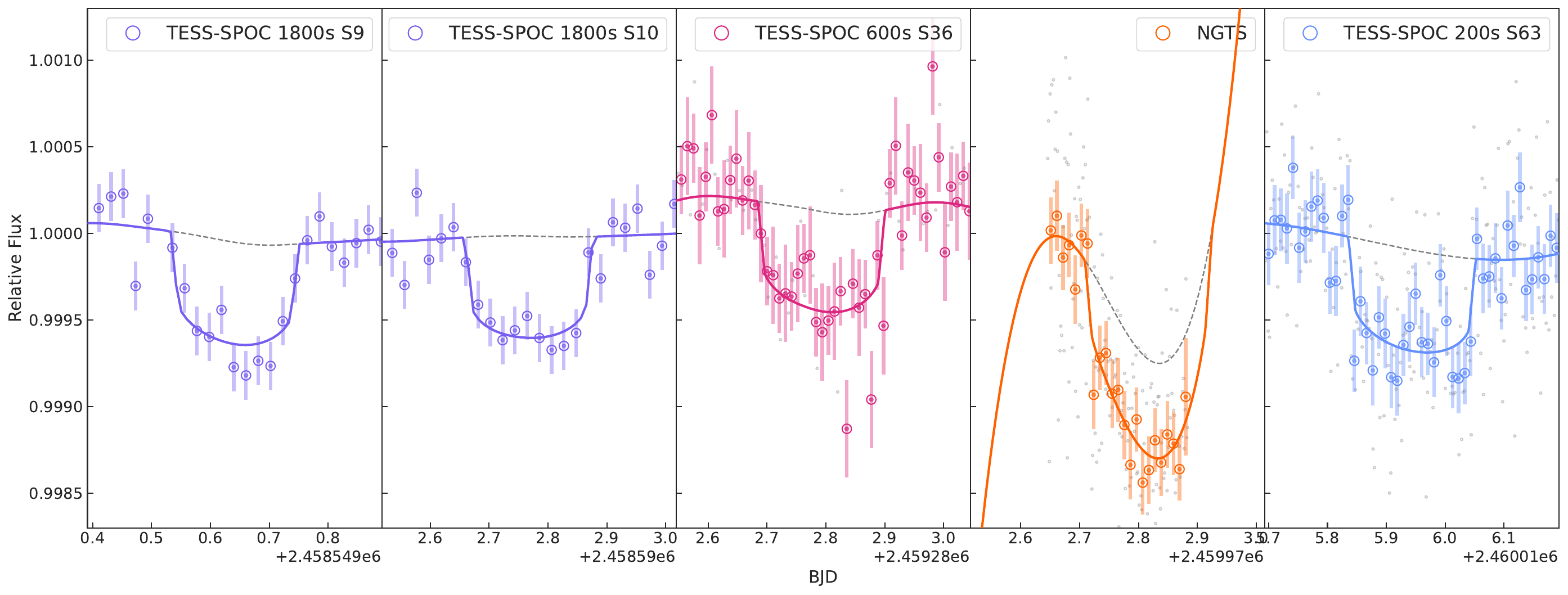}
        \caption{Normalised raw photometry data per transit in chronological order for \objonengts. From left to bottom: TESS-SPOC 1800s (Sectors 9 then 10) in purple where the baseline was a GP (Matern-3/2); TESS-SPOC 600s (Sector 36) in pink where the baseline was a GP (Matern-3/2); \ngts\ binned to 2 minutes in grey and binned to 15 minutes in orange, where the baseline was a hybrid spline; and TESS-SPOC 200s (Sector 63) in grey and binned to 15 minutes in blue, where the baseline was a hybrid spline. The grey dashed lines are the red noise models per instrument/cadence and the solid lines are the posterior median models for each instrument/cadence from the joint fit from \allesfitter.}
        \label{fig:147rawphot}
\end{figure*}

\begin{figure*}
     \centering
         \centering
         \includegraphics[width=1\linewidth]{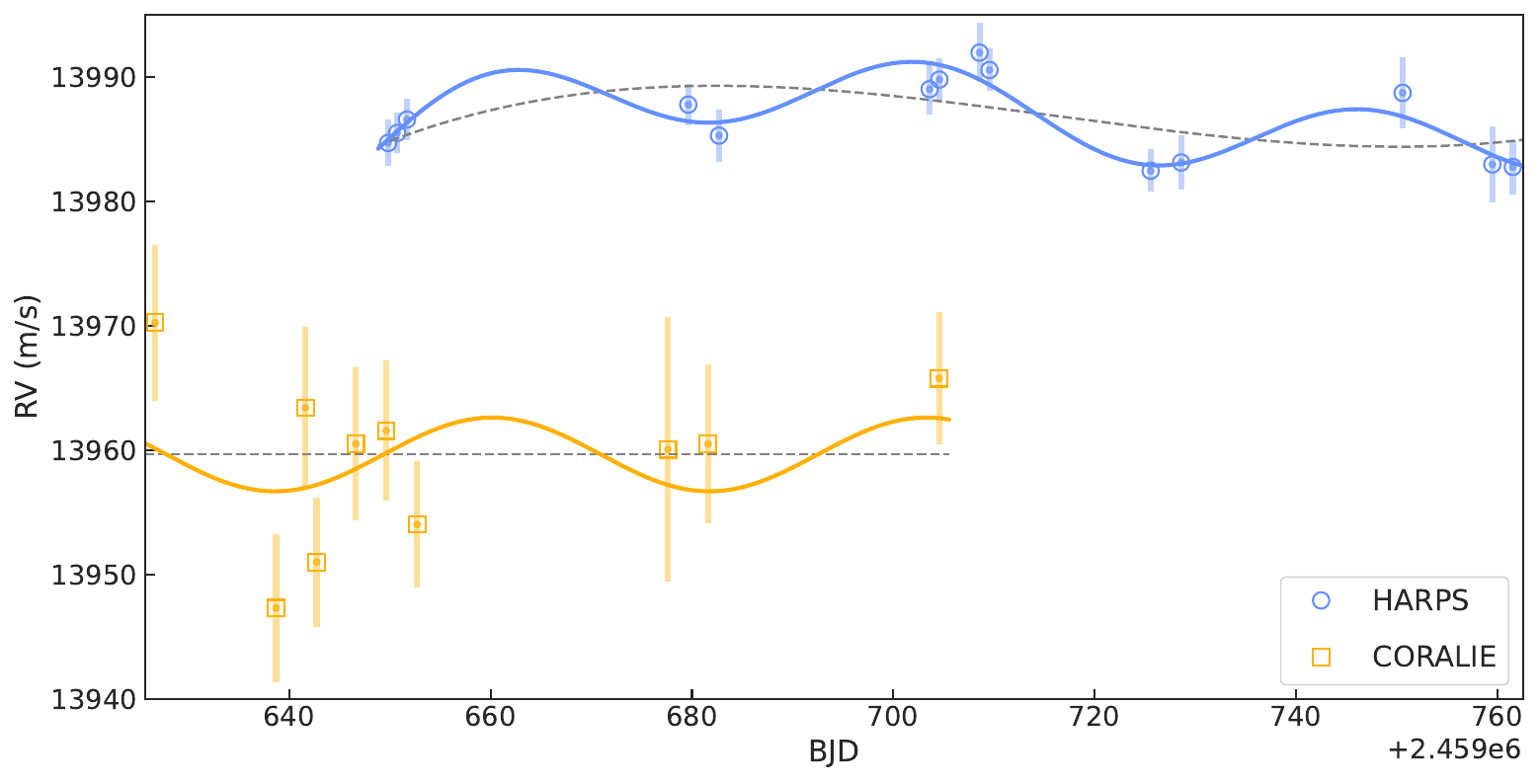}
        \caption{Raw RV data for \objonengts, from both \coralie~(yellow squares) and \harps~(blue circles). The red noise models (dashed grey) subtracted for each instrument were a hybrid spline for \harps\ and a flat offset of $13.9597\pm0.0019~\mathrm{km/s}$ for \coralie. The solid lines are the posterior median models for each instrument from the joint fit.}
        \label{fig:147rawrv}
\end{figure*}

\begin{figure*}
     \centering
         \centering
         \includegraphics[width=0.95\linewidth]{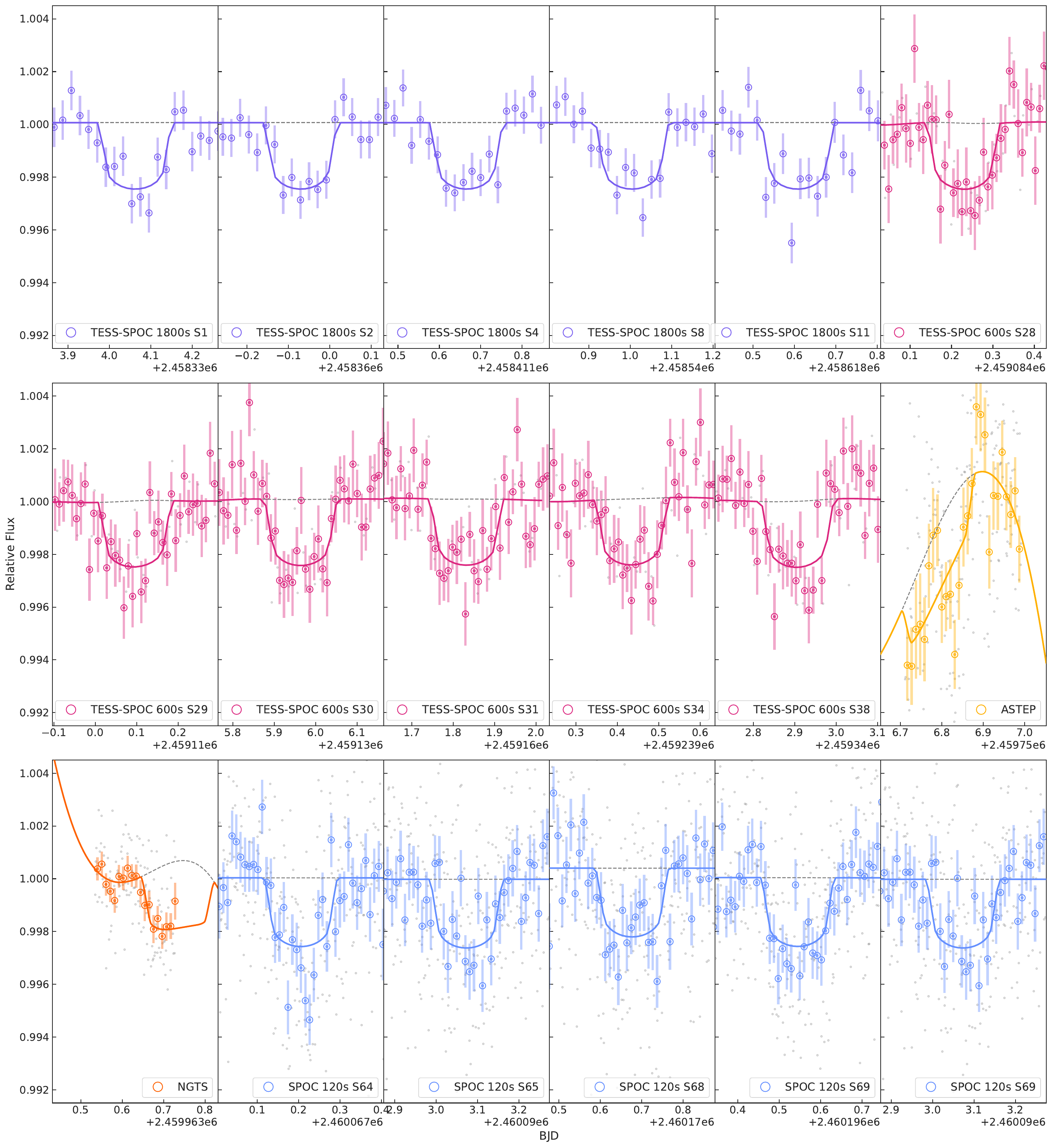}
        \caption{Normalised raw photometry data per transit for \objtwo\ in chronological order. From top left to bottom right: TESS-SPOC 1800s (Sectors 1, 2, 4, 8 and 11) in purple where the baseline was a flat offset; TESS-SPOC 600s (Sectors 28, 29 ,30, 31, 34 and 38) in grey and binned to 15 minutes within the phase-folded lightcurve in pink, where the baseline was a GP (Matern-3/2); \astep\ in 100s exposures in grey and binned to 15 minutes in yellow, where the baseline was a hybrid spline; \ngts\ binned to 2 minutes in grey and binned to 15 minutes in orange, where the baseline was a hybrid spline; and SPOC 120s (2 in Sector 64, and one each in 65, 68, and 69) in grey and binned to 15 minutes within the phase-folded lightcurve in blue, where the baseline was a hybrid spline. The grey dashed lines are the red noise models per instrument/cadence and the solid lines are the posterior median models for each instrument/cadence from the joint fit from \allesfitter.}
        \label{fig:237rawphot}
\end{figure*}

\begin{figure*}
     \centering
         \centering
         \includegraphics[width=1\linewidth]{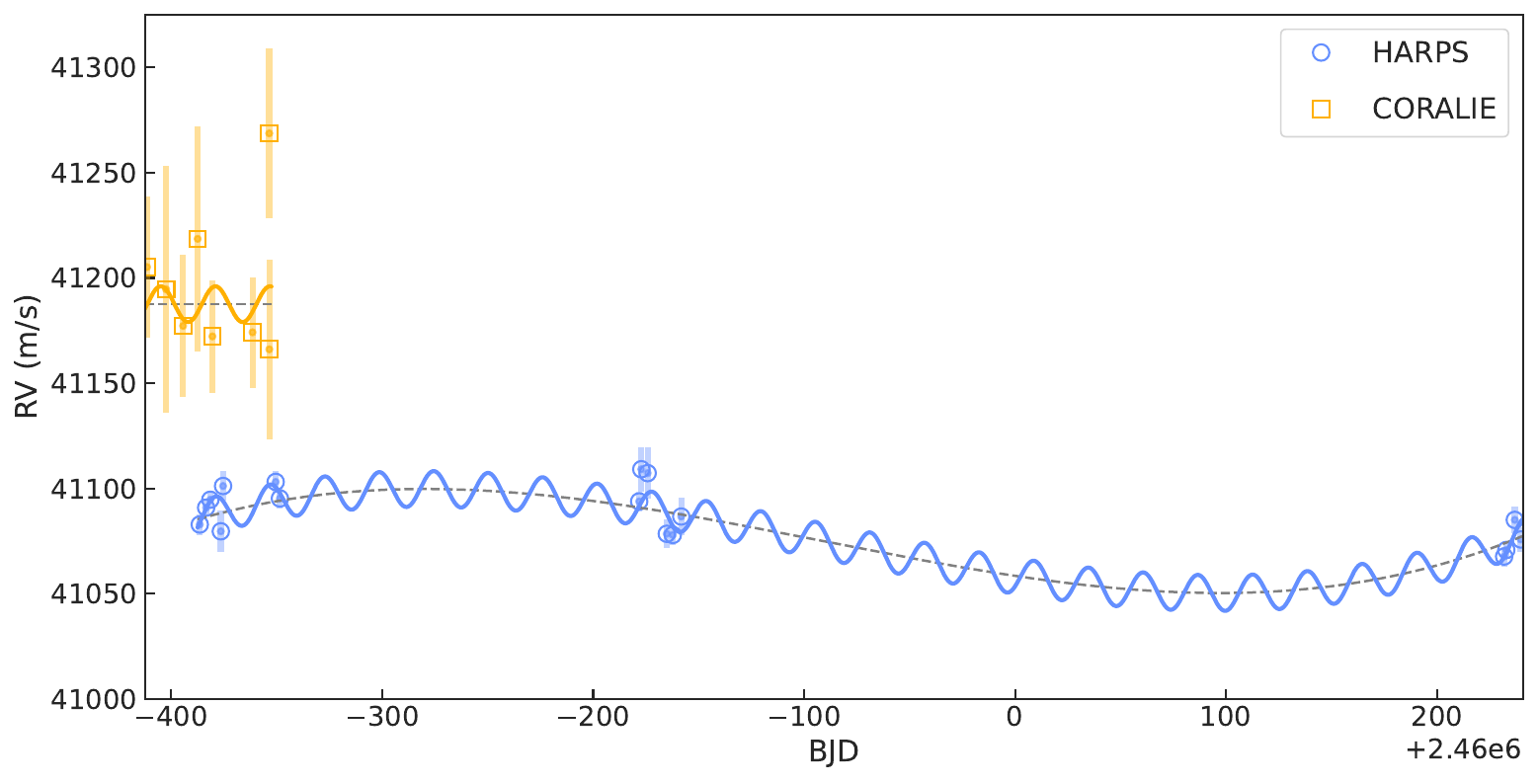}
        \caption{Raw RV data for \objtwo, from both \coralie~(yellow squares) and \harps~(blue circles). The red noise models (dashed grey) subtracted for each instrument were a hybrid spline for \harps\ and a flat offset of $41.188\pm0.013~\mathrm{km/s}$ for \coralie. The solid lines are the posterior median models for each instrument from the joint fit.}
        \label{fig:237rawrv}
\end{figure*}

\begin{figure*}
     \centering
         \centering
         \includegraphics[width=1\linewidth]{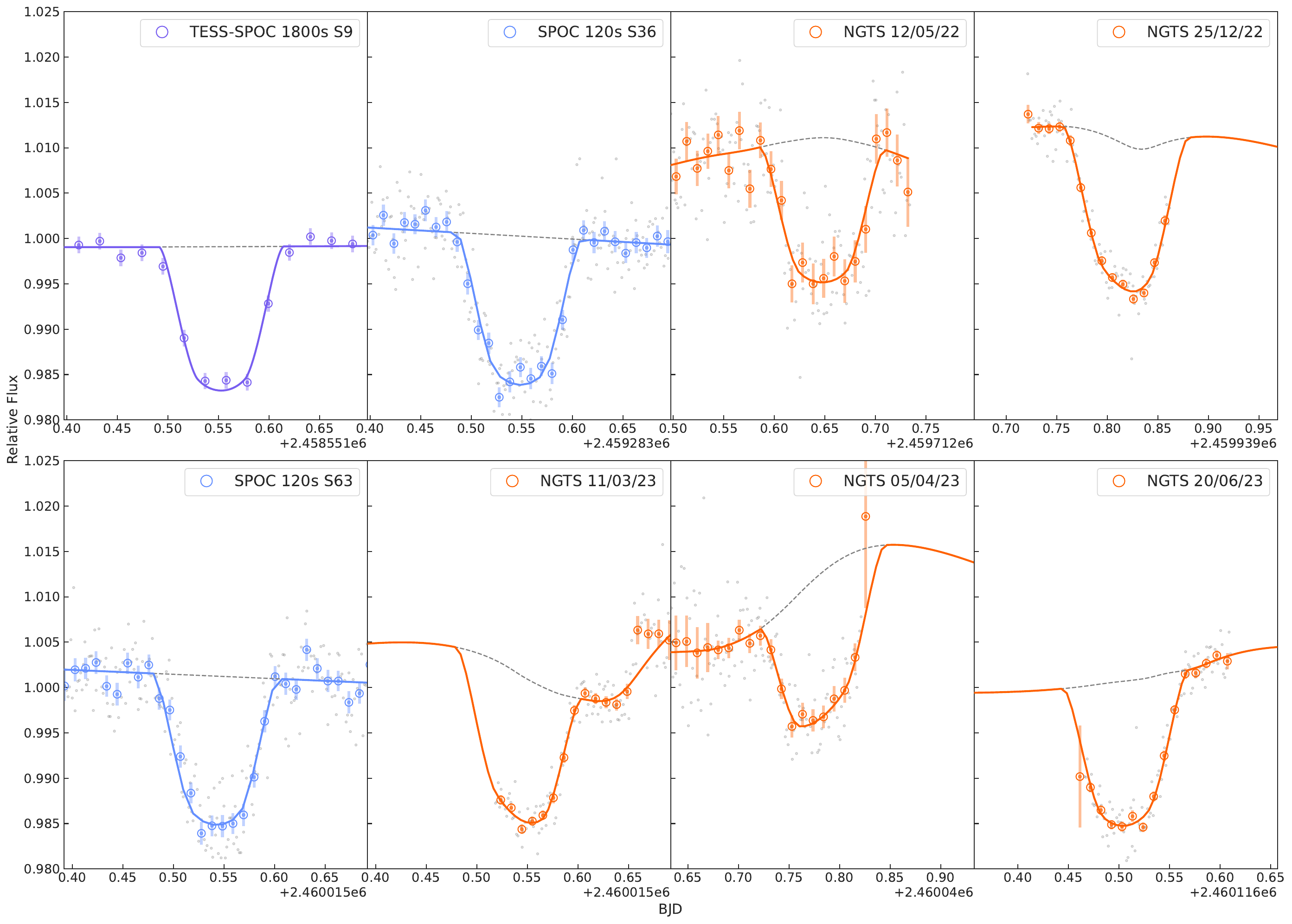}
        \caption{Normalised raw photometry data per transit for \objthreengts. From top left to bottom right: TESS-SPOC 1800s (Sector 9) in purple where the baseline was a GP (Matern-3/2); SPOC 120s (Sector 36) in blue, where the baseline was a hybrid spline; \ngts\ on nights 12\textsuperscript{th} May and 25\textsuperscript{th} December 2022 binned to 2 minutes in orange, where the baseline was a GP (Matern-3/2); SPOC 120s (Sector 63) again in blue with a hybrid spline), and \ngts\ on nights 11\textsuperscript{th} March, 5\textsuperscript{th} April and 20\textsuperscript{th} June 2023, again binned to 2 minutes in orange with a GP (Matern-3/2). The grey dashed lines are the red noise models per instrument/cadence and the solid lines are the posterior median models for each instrument/cadence from the joint fit from \allesfitter.}
        \label{fig:333rawphot}
\end{figure*}

\begin{figure*}
     \centering
         \centering
         \includegraphics[width=1\linewidth]{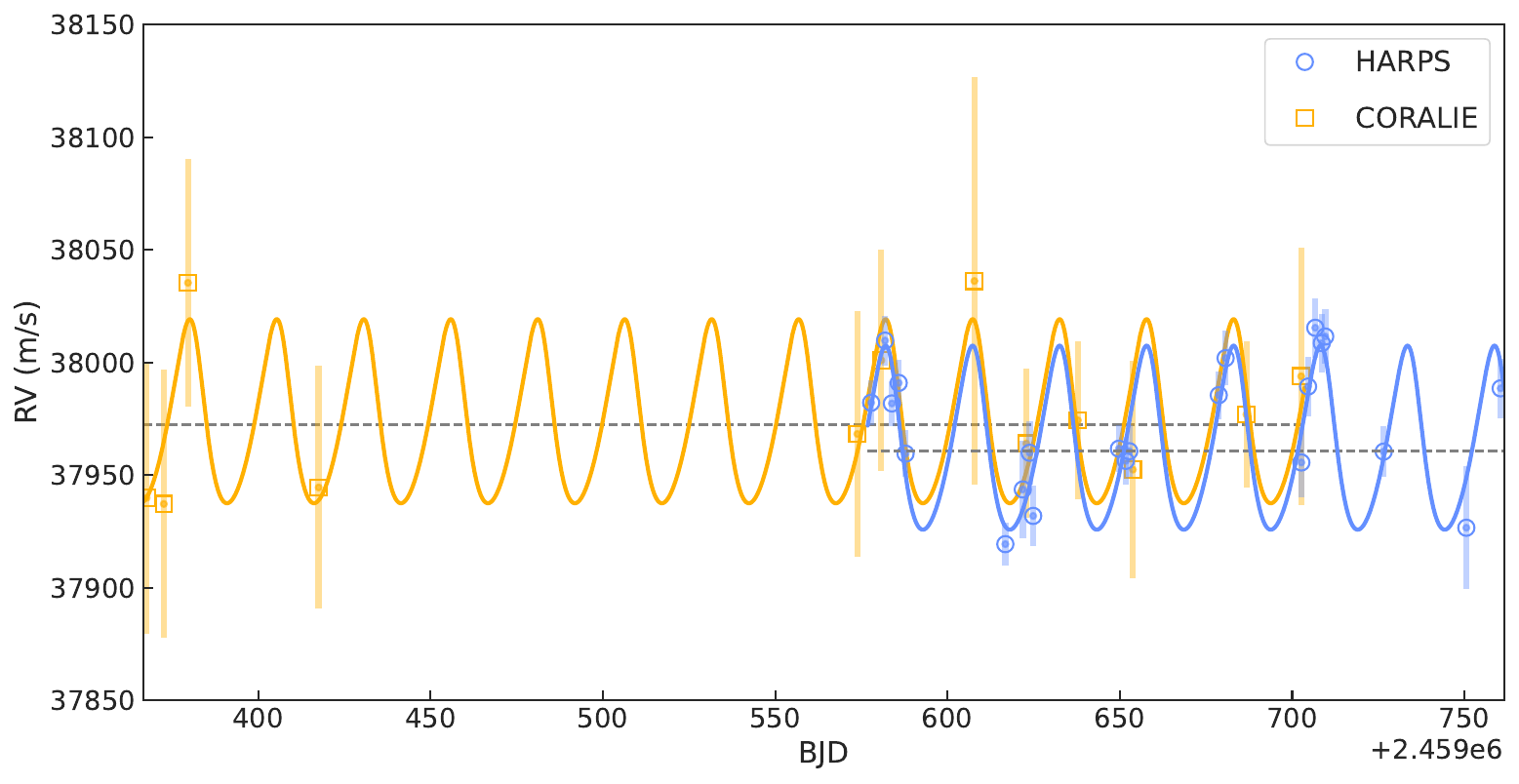}
        \caption{Raw RV data for \objthreengts, from both \coralie~(yellow squares) and \harps~(blue circles). The red noise models (dashed grey) subtracted for each instrument were flat offsets of $37.972\pm0.013~\mathrm{km/s}$ and $37.9606\pm0.0029~\mathrm{km/s}$ respectively, and the solid lines are the posterior median models for each instrument from the joint fit.}
        \label{fig:333rawrv}
\end{figure*}

\clearpage

\begin{figure*}
\section{Timeline of Data}
	\includegraphics[width=1\columnwidth]{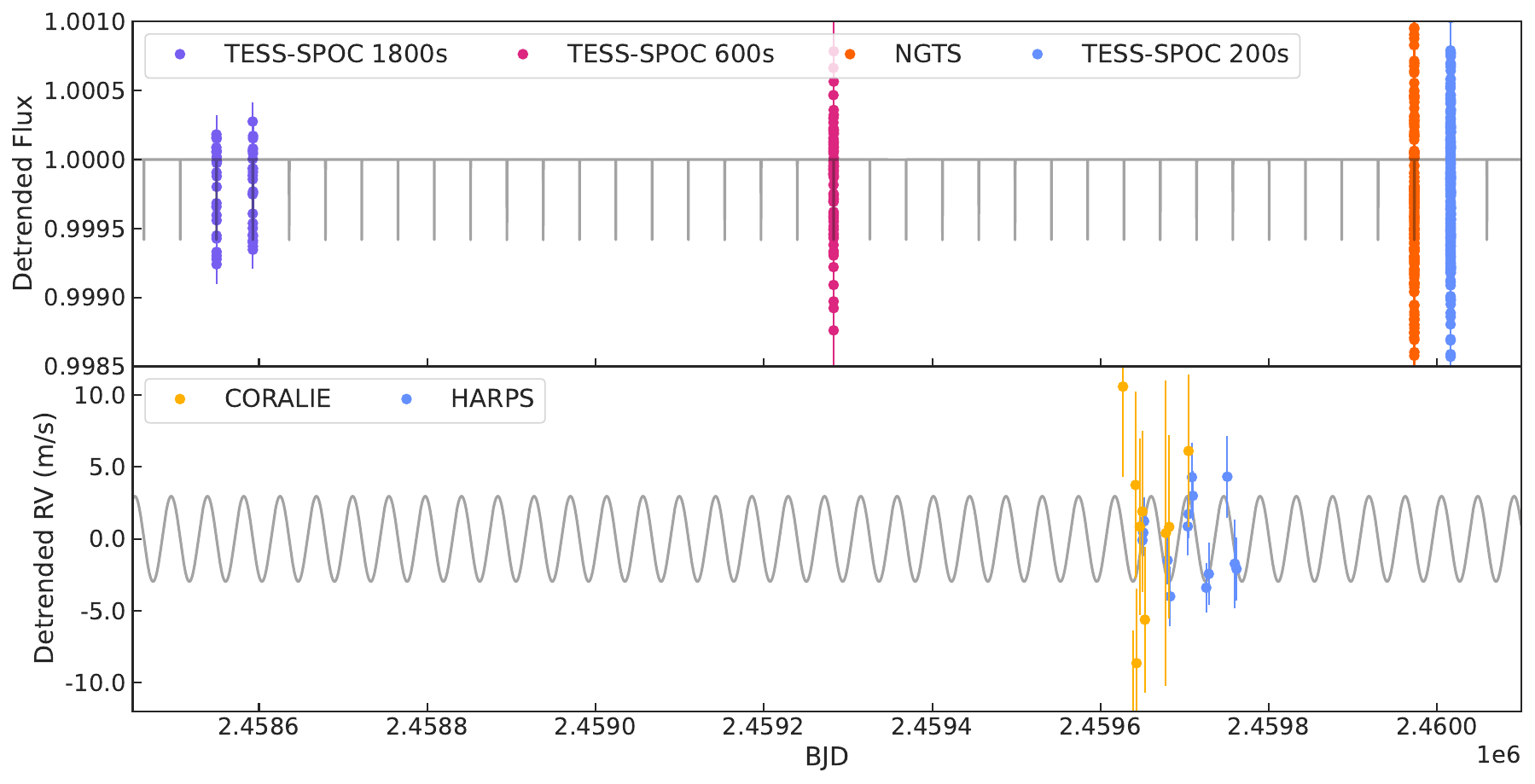}
    \caption{Time series of in-transit photometry and radial velocity observations of \objonengts\ grouped by instrument and cadence. From left to right in the top panel: TESS-SPOC 1800s (Sectors 9 and 10) in purple where the red noise was modelled by a GP (Matern-3/2); TESS-SPOC 600s (Sector 36) in pink where the red noise was modelled by a GP (Matern-3/2); \ngts\ binned to 2 minutes in orange, where the red noise was modelled by a hybrid spline; and TESS-SPOC 200s (Sector 63) in blue, where the red noise was modelled by a hybrid spline. From left to right in the bottom panel: \coralie\ in yellow then \harps\ in blue where the red noise was modelled by a hybrid spline for \harps\ and a flat offset of $13.9597\pm0.0019~\mathrm{km/s}$ for \coralie. The solid grey lines are the posterior median models for the most recent data sets (TESS-SPOC 200s and \harps) from the joint fit from \allesfitter.}
    \label{fig:TIC147timeseries}
\end{figure*}
\begin{figure*}
	\includegraphics[width=0.9\columnwidth]{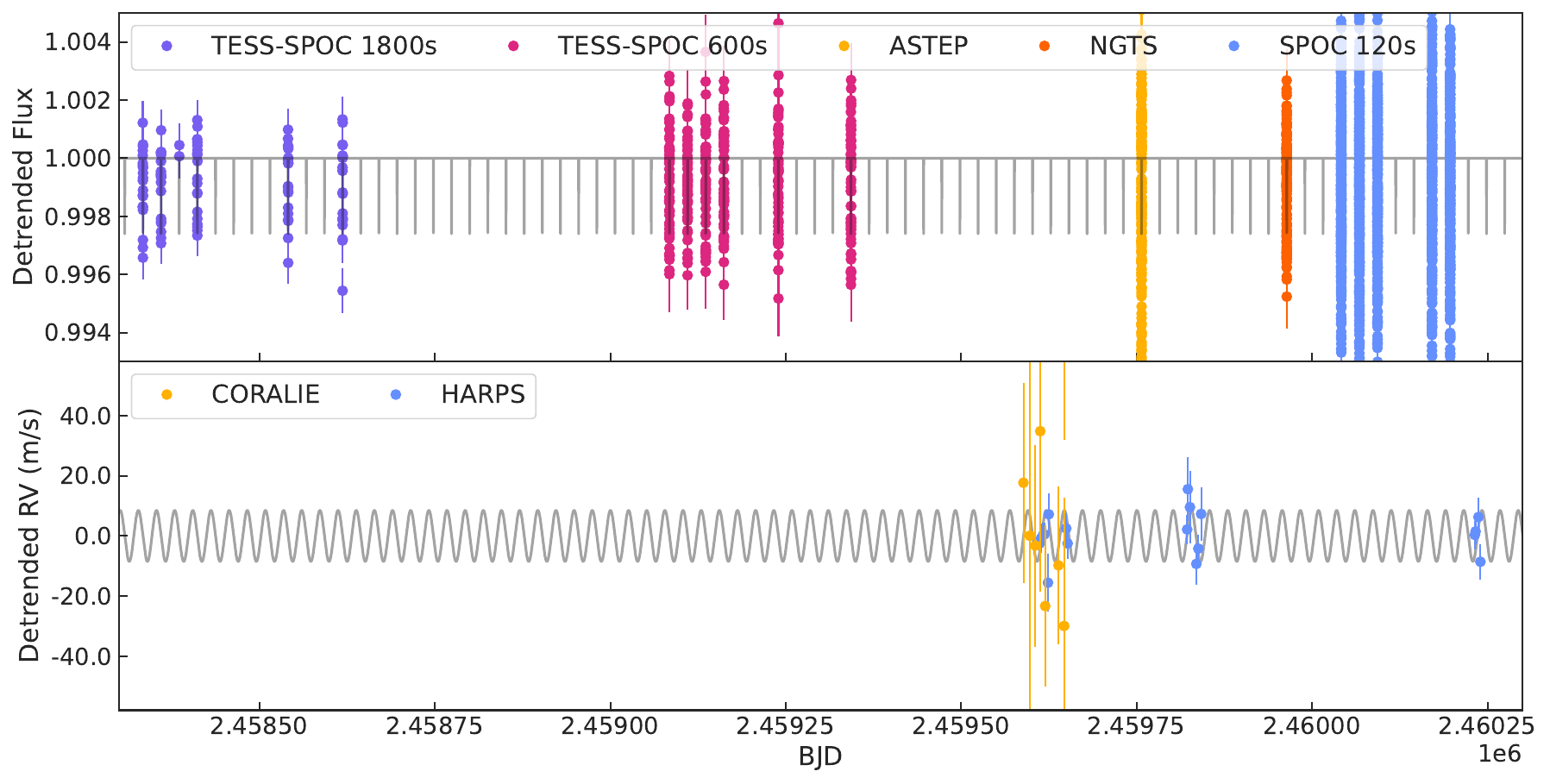}
   \caption{Time series of photometry and radial velocity observations of \objtwo\ grouped by instrument and cadence. From left to right in the top panel: TESS-SPOC 1800s (Sectors 1-4, 8 and 11) in purple where the red noise was modelled by a flat offset; TESS-SPOC 600s (Sectors 28-31, 34 and 38) in pink, where the red noise was modelled by a GP (Matern-3/2); \astep\ in 100s exposures in yellow, where the red noise was modelled by a hybrid spline; \ngts\ binned to 2 minutes in orange, where the red noise was modelled by a hybrid spline; and SPOC 120s (Sectors 61, 62, 64, 65, 68 and 69) in blue, where the red noise was modelled by a hybrid spline. From left to right in the bottom panel: \coralie\ in yellow then \harps\ in blue where the red noise was modelled by a hybrid spline for \harps\ and a flat offset of $41.188\pm0.013~\mathrm{km/s}$ for \coralie. The solid grey lines are the posterior median models for the most recent data sets (SPOC 120s and \harps) from the joint fit from \allesfitter.}
    \label{fig:TIC237timeseries}
\end{figure*}
\begin{figure*}
	\includegraphics[width=0.9\columnwidth]{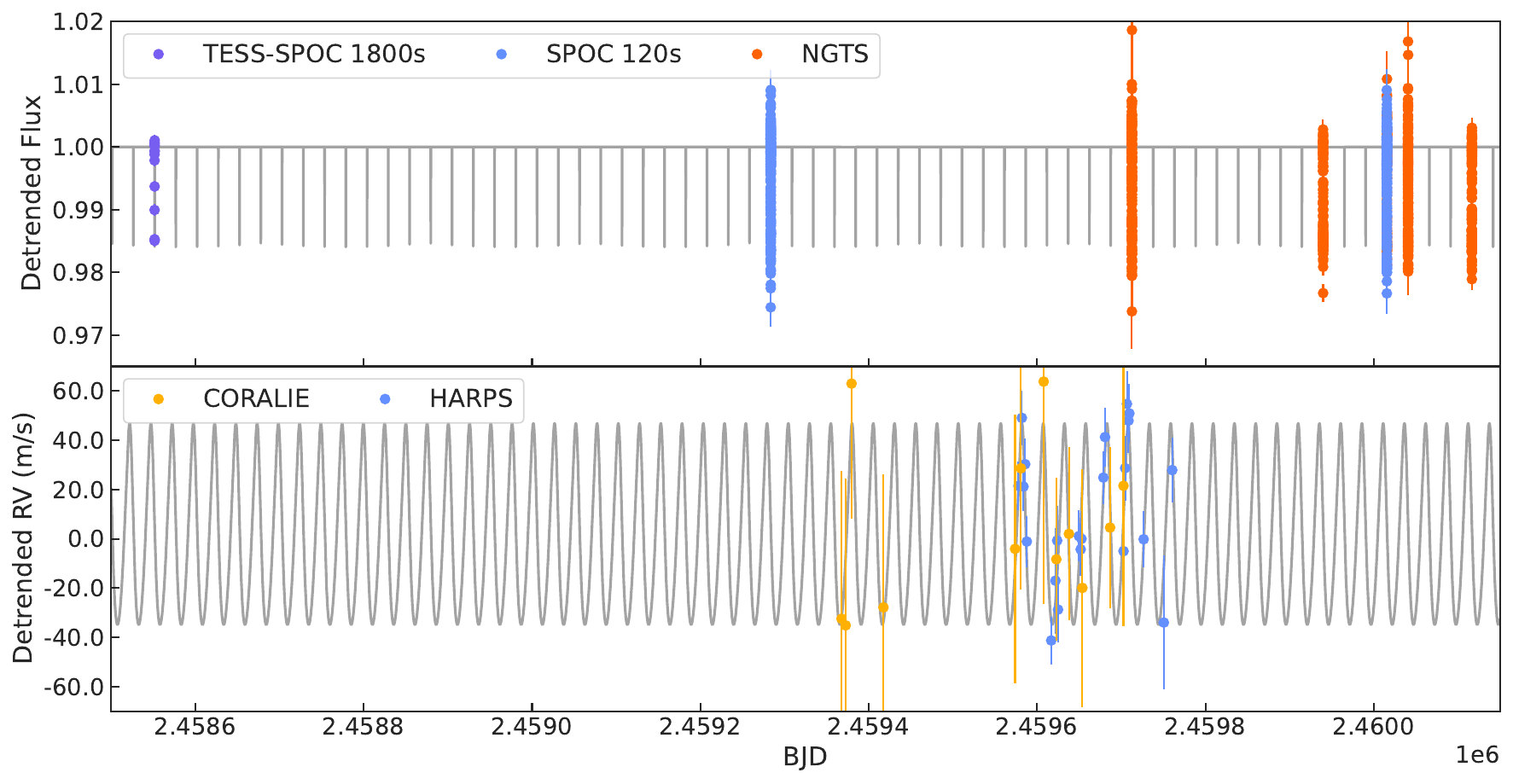}
    \caption{Time series of photometry and radial velocity observations of \objthreengts\ grouped by instrument and cadence. From left to right in the top panel: TESS-SPOC 1800s (Sector 9) in purple where the red noise was modelled by a GP (Matern-3/2); SPOC 120s (Sector 36 and later Sector 63) in blue, where the red noise was modelled by a hybrid spline; and \ngts\ binned to 2 minutes in orange, where the red noise was modelled by a GP (Matern-3/2). From left to right in the bottom panel: \coralie\ in yellow then \harps\ in blue where the red noise was modelled by flat offsets ($37.9606\pm0.0029~\mathrm{km/s}$ for \harps, $37.972\pm0.013~\mathrm{km/s}$ for \coralie). The solid grey lines are the posterior median models for the most recent data sets (\ngts\ and \harps) from the joint fit from \allesfitter.}
   \label{fig:TIC333timeseries}
\end{figure*}
\newpage
\begin{table}
\centering
\section{Global Fit Priors}
\label{sec:priors}
	\caption{Priors for chosen global fit of photometric and radial velocity data. $T_{0}$ and $P$ are fixed from previous fits of photometric data only. Uniform: (value, lower bound, upper bound), Normal: (mu,sigma)}
	\label{tab:priors}
	\begin{tabular}{lcccc}
		\hline
		\textit{Fitted Parameters} & Distribution &\objonengts & \objtwo & \objthreengts\\
		\hline
        $(R_\star + R_p) / a$ & Uniform & (0.0255, 0.00000, 0.0500) & (0.0344, 0.00000, 0.0500) &  (0.0228, 0.00000, 0.0500)\\
        $R_p / R_\star$ & Uniform & (0.0244, 0.0000, 0.5000) & (0.0524, 0.0000, 0.5000) & (0.1323, 0.0000, 0.5000)\\ 
        $\cos{i}$ & Uniform & (0.0199, 0.0000, 0.1000) & (0.0266, 0.00000, 0.1000) &  (0.0166, 0.0000, 0.1000)\\
        $T_{0}$ ($\mathrm{BJD}$) & Fixed & $2459282.7925_{-0.0016}^{+0.0024}$ & $2459265.3032_{-0.0013}^{+0.0014}$ & $2459334.02963_{-0.00056}^{+0.00053}$ \\
        $P$ ($\mathrm{d}$) & Fixed & $\objonep$ & $\objtwop$ & $\objthreep$\\ 
        $\sqrt{e}\cos{\omega}_p$ & Uniform & 0 (Fixed) & 0 (Fixed) & (0, -1, 1) \\
        $\sqrt{e}\sin{\omega}_p$ & Uniform & 0 (Fixed) & 0 (Fixed) & (0, -1, 1) \\  
        $K$ ($\mathrm{km/s}$) & Uniform & (0.00300, 0.00000, 0.10000) & (0.0124, 0.0000, 0.1000) & (0.0381, 0.0000, 0.1000)\\ 
        $q_{1;\mathrm{NGTS}}$ & Normal & (0.3526, 0.0018) & (0.3930, 0.0019) & (0.4614, 0.0037)\\
        $q_{2;\mathrm{NGTS}}$ & Normal & (0.384, 0.019) & (0.414, 0.020) & (0.441, 0.028)\\
        $q_{1;\mathrm{ASTEP}}$ & Normal & & (0.2918, 0.0014) & \\
        $q_{2;\mathrm{ASTEP}}$ & Normal & & (0.392, 0.019) &\\
        $q_{1;\mathrm{TESS-SPOC~1800s}}$ & Normal & (0.2988, 0.0015) & (0.3375, 0.0017) & (0.4007, 0.0032)\\
        $q_{2;\mathrm{TESS-SPOC~1800s}}$ & Normal & (0.376, 0.019) & (0.402, 0.020) & (0.426, 0.027)\\
        $q_{1;\mathrm{TESS-SPOC~600s}}$ & Normal & (0.2988, 0.0015) & (0.3375, 0.0017) & \\
        $q_{2;\mathrm{TESS-SPOC~600s}}$ & Normal & (0.376, 0.019) & (0.402, 0.020) & \\
        $q_{1;\mathrm{TESS-SPOC~200s}}$ & Normal & (0.2988, 0.0015) &  & \\
        $q_{2;\mathrm{TESS-SPOC~200s}}$ & Normal & (0.376, 0.019) &  & \\
        $q_{1;\mathrm{SPOC 120s}}$ & Normal & & (0.3375, 0.0017) & (0.4007, 0.0032)\\
        $q_{2;\mathrm{SPOC 120s}}$ & Normal & & (0.402, 0.020) & (0.426, 0.027)\\
        $D_{\mathrm{NGTS}}$ & Normal & (0.051253, 0.000311) & 0 (Fixed) & 0 (Fixed) \\ 
        $D_{\mathrm{ASTEP}}$ & Normal &  & 0 (Fixed) &  \\ 
        $D_{\mathrm{TESS-SPOC~1800s}}$ & Normal & (0, 0.05) & (0, 0.05) & (0, 0.05) \\ 
        $D_{\mathrm{TESS-SPOC~600s}}$ & Normal & (0, 0.05) & (0, 0.05) & \\
        $D_{\mathrm{TESS-SPOC~200s}}$ & Normal & (0, 0.05) &  & \\
        $D_{\mathrm{SPOC 120s}}$ & Normal & & (0, 0.05) & (0, 0.05)\\
        $\ln{\sigma_\mathrm{NGTS}}$ & Uniform & (-7.3, -10.0, -1.0) & (-6.7, -10.0, -1.0) & (-5.2, -10.0, -1.0)\\
        $\ln{\sigma_\mathrm{ASTEP}}$ & Uniform & & (-5.5, -10.0, -1.0) &  \\
        $\ln{\sigma_\mathrm{TESS-SPOC~1800s}}$ & Uniform & (-8.7, -10.0, -1.0) & (-7.3, -10.0, -1.0) & (-7.1, -10.0, -1.0)\\
        $\ln{\sigma_\mathrm{TESS-SPOC~600s}}$ & Uniform & (-8.2, -10.0, -1.0) & (-6.7, -10.0, -1.0) & \\
        $\ln{\sigma_\mathrm{TESS-SPOC~200s}}$ & Uniform & (-7.8, -10.0, -1.0) &  & \\
        $\ln{\sigma_\mathrm{SPOC 120s}}$ & Uniform & & (-5.9, -10.0, -1.0) & (-5.7, -10.0, -1.0) \\
        $\ln{\sigma_\mathrm{jitter}}(RV_\mathrm{\coralie})$ ($\mathrm{km/s}$) & Uniform & (-5.0, -7.6, -1.6) & (-5.0, -7.6, -1.6) & (-5.0, -7.6, -1.6)\\
        $\ln{\sigma_\mathrm{jitter}}(RV_\mathrm{\harps})$ ($\mathrm{km/s}$) & Uniform & (-6.0, -7.6, -1.6) & (-6.0, -7.6, -1.6) & (-6.0, -7.6, -1.6)\\
        $\mathrm{Offset_{\coralie}}$ ($\mathrm{km/s}$) & Uniform & (13.96, 13.95, 13.97) & (41.2, 41.3, 41.4) & (37.971, 37.921, 38.021)\\
        $\mathrm{Offset_{\harps}}$ ($\mathrm{km/s}$) & Uniform & & & (37.960, 37.950, 37.970)\\
        $\mathrm{Offset_{NGTS}}$ & Normal &  &  & (0.00, 0.01)\\
        $\mathrm{GP~}\ln{\sigma}_\mathrm{NGTS}$ & Normal &  &  & (-4.82, 0.13)\\
        $\mathrm{GP~}\ln{\rho}_\mathrm{NGTS}$ & Normal &  &  & (-1.24, 0.27)\\
        $\mathrm{Offset_{TESS-SPOC~1800s}}$ & Normal & (0.00, 0.01) & (0.00, 0.01) & (0.00, 0.01)\\
        $\mathrm{GP~}\ln{\sigma}_\mathrm{TESS-SPOC~1800s}$ & Normal & (-9.32, 0.08) &  & (-7.00, 0.20) \\
        $\mathrm{GP~}\ln{\rho}_\mathrm{TESS-SPOC~1800s}$ & Normal & (-1.62, 0.16) &  & (0.13, 0.25)\\
        $\mathrm{Offset_{TESS-SPOC~600s}}$ & Normal & (0.00, 0.01) & (0.00, 0.01) & \\
        $\mathrm{GP~}\ln{\sigma}_\mathrm{TESS-SPOC~600s}$ & Normal & (-9.38, 0.11) & (-8.65, 0.10) & \\
        $\mathrm{GP~}\ln{\rho}_\mathrm{TESS-SPOC~600s}$ & Normal & (-2.17, 0.26) & (-2.21, 0.34) & \\
        \hline
	\end{tabular}
\end{table}

\twocolumn
\begin{figure*}
\section{Radial Velocity Activity Indicators\label{activityappendix}}
     \centering
     \begin{subfigure}[p]{\columnwidth}
         \centering
         \includegraphics[width=0.9\columnwidth]{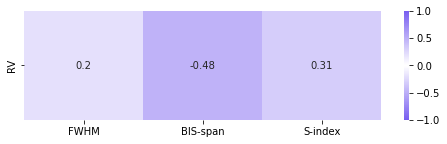}
         \caption{Correlation matrix for \objonengts.}
         \label{fig:tic147corr}
     \end{subfigure}
     \hfill
     \begin{subfigure}[p]{\columnwidth}
         \centering
         \includegraphics[width=0.9\columnwidth]{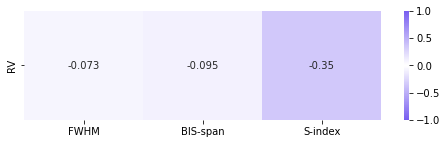}
         \caption{Correlation matrix for \objtwo.}
         \label{fig:tic237corr}
     \end{subfigure}
     \hfill
     \begin{subfigure}[p]{\columnwidth}
         \centering
         \includegraphics[width=0.9\columnwidth]{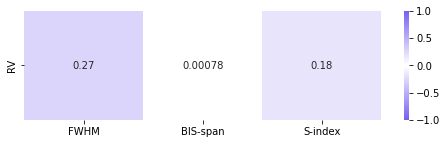}
         \caption{Correlation matrix for \objthreengts.}
         \label{fig:tic333corr}
     \end{subfigure}
        \caption{Weighted Pearson's correlation between \harps\ radial velocity data and activity indicators: Full Width Half Maximum, BIS-Span and S-index, where a darker colour indicates a stronger correlation between radial velocity and each indicator.}
        \label{fig:corr}
\end{figure*}

\onecolumn
\begin{figure*}
\section{Global Fit Posteriors}
\label{sec:corners}
    \includegraphics[width=1\columnwidth]{TIC147_Figures/TIC147nscorner.pdf}
    \caption{Posterior distributions of fitted parameters for \objonengts\,b.}
    \label{fig:TIC147corner}
\end{figure*}
\begin{figure*}
    \includegraphics[width=1\columnwidth]{TIC237_Figures/TIC237nscorner.pdf}
    \caption{Posterior distributions of fitted parameters for \objtwo\,b.}
    \label{fig:TIC237corner}
\end{figure*}
\begin{figure*}
    \includegraphics[width=1\columnwidth]{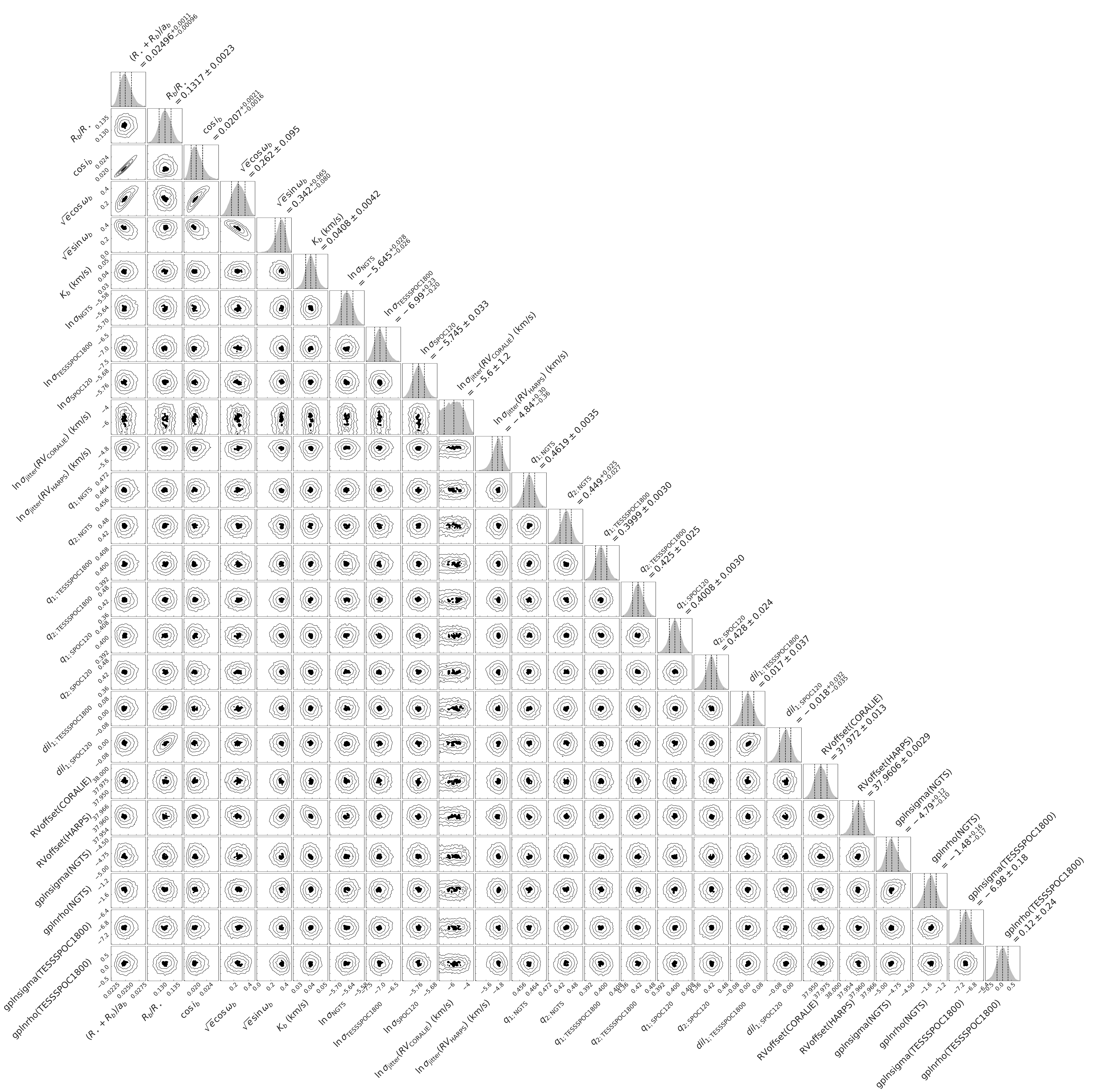}
    \caption{Posterior distributions of fitted parameters for \objthreengts\,b.}
    \label{fig:TIC333corner}
\end{figure*}